\documentclass[twocolumn]{aa}
\usepackage{graphicx}
\begin{document}

\authorrunning{Nagao et al.}
\titlerunning{The Evolution of the BLR}

\title{The Evolution of the Broad-Line Region among SDSS Quasars}

\author{
        Tohru Nagao            \inst{1, 2},
        Alessandro Marconi     \inst{1}, \and
        Roberto Maiolino       \inst{1}
       }

\offprints{T. Nagao}

\institute{
           INAF -- Osservatorio Astrofisico di Arcetri,
           Largo Enrico Fermi 5, 50125 Firenze, Italy\\
           \email{tohru@arcetri.astro.it,
                  marconi@arcetri.astro.it,
                  maiolino@arcetri.astro.it}
           \and
           National Astronomical Observatory of Japan,
           2-21-1 Osawa, Mitaka, Tokyo 151-8588, Japan
          }

\date{Received ;  accepted }

\abstract{
   Based on 5344 quasar spectra taken from the SDSS Data Release 2,
   the dependences of various emission-line flux ratios on redshift 
   and quasar luminosity are investigated in the ranges 
   $2.0 \leq z \leq 4.5$ and $-24.5 \geq M_B \geq -29.5$.
   We show that the emission lines in the composite spectra are 
   fitted better with power-law profiles than with double Gaussian
   or modified Lorentzian profiles, and in particular we show that
   the power-law profiles are more appropriate
   to measure broad emission-line fluxes than other methods.
   The composite spectra show that there are statistically
   significant correlations between quasar luminosity and various
   emission-line flux ratios, such as N{\sc v}/C{\sc iv} and
   N{\sc v}/He{\sc ii}, while there are only marginal correlations
   between quasar redshift and emission-line flux ratios.
   We obtain detailed photoionization models to interpret the
   observed line ratios. The correlation of line ratios with 
   luminosity is interpreted in terms of higher gas metallicity in 
   more luminous quasars. For a given quasar luminosity, there is 
   no metallicity evolution for the redshift range 
   $2.0 \leq z \leq 4.5$. The typical metallicity of BLR gas clouds 
   is estimated to be $Z \sim 5 Z_\odot$, although the inferred 
   metallicity depends on the assumed BLR cloud properties, such as 
   their density distribution function and their radial distribution.
   The absence of a metallicity evolution up to $z \sim 4.5$
   implies that the active star-formation epoch of quasar host
   galaxies occurred at $z \ga 7$.
}

\maketitle

\section{Introduction}

Quasars are among the most luminous objects in the universe
and therefore they can be detected and
investigated in detail even at very high redshifts,
up to $z>6$ (Fan et al. 2001, 2003, 2004). As a consequence,
they have been frequently used as a tool to pursue the
exploration of the distant universe, such as the investigation
of the intergalactic matter (see, e.g., 
Rauch 1998 for a review), the cosmic re-ionization history 
(e.g., Fan et al. 2002; Gnedin 2004) and the metal-enrichment 
history in the universe. Since the gas in the broad-line region 
(BLR) of quasars is most likely photoionized, as suggested by 
reverberation mapping observations (e.g., Peterson 1993), it is
possible to investigate the chemical composition of gas clouds 
in BLRs by comparing spectroscopic data with photoionization 
model calculations. The gas-phase elemental abundances 
are determined by the star-formation history of the galaxies, 
therefore studies on the metallicity of BLR in distant quasars 
are highly insightful of the metal-enrichment 
history and the galaxy formation in the very early universe
(e.g., Matteucci \& Padovani 1993; Hamann \& Ferland 1993, 1999; 
Venkatesan et al. 2004). 
Note that the analysis on the gas metallicity of high-$z$
galaxies not hosting AGN are extremely difficult and 
time-consuming, because 
even the brightest galaxies are much fainter than quasars at the 
same redshift (but see, e.g., Teplitz et al. 2000; 
Kobulnicky \& Koo 2000; Pettini et al. 2001).

It has been often claimed that the gas metallicity of the BLR 
in quasars is higher than solar (e.g., Baldwin 
\& Netzer 1978; Hamann \& Ferland 1992; Ferland et al. 1996; 
Dietrich et al. 1999, 2003). The inferred metallicity is
sometimes very high, several times solar, and for the most 
extreme case QSO 0353--383 it reaches as much as 
$Z_{\rm BLR} \sim 15 Z_\odot$ (Baldwin et al. 2003). 
Since such high metallicities require
deep gravitational potentials and intense star-forming activity
in the host galaxies, they provide strong constraints on the
evolutionary scenarios of quasar host galaxies
(e.g., Hamann \& Ferland 1993, 1999; Di Matteo et al. 2004).
Another surprising result is that the BLR metallicity does not 
appear to decrease at the highest redshift proved so far
(e.g., Pentericci et al. 2002;
Dietrich et al. 2003; Maiolino et al. 2003). 
This finding gives tight constraints on the 
first epoch of star formation in the host galaxies,
especially when the minimum timescale required for the
enrichment of some metals (C, Si, Fe and so on) is taken
into account. On the other hand, some observations suggest that
quasars at higher-$z$ show higher metallicity than 
lower-$z$ quasars (e.g., Hamann \& Ferland 1992, 1993, 1999). 
It is also recognized that the BLR metallicity tends to be 
higher in more luminous quasars (e.g., Hamann \& Ferland 
1993, 1999). This trend may suggest a connection between the 
BLR metallicity of quasars and evolutionary processes of the 
host galaxies. However, since higher-luminosity quasars tend 
to be selectively observed at higher-$z$, it is not clear 
how the BLR metallicity depends on the luminosity and 
redshift, individually.

To investigate the BLR metallicity and understand the 
chemical evolution of quasars and their host galaxies, 
it is thus necessary to observe a large number of 
quasars with sufficiently large ranges of luminosity and 
redshift. Francis \& Koratkar (1995) compared the rest-UV
spectra of quasars at the local universe observed with IUE 
with those at $1.7 < z < 3.3$ obtained by
Large Bright Quasar Survey (LBQS; Foltz et al. 1987)
and found little redshift evolution of the UV 
spectra. Dietrich et al. (2002) investigated the
spectra of a larger sample of quasars at $0 < z < 5$
spanning ~6 orders of magnitude in luminosity, based on 
the data compilation of various observations that were 
performed with IUE, HST, and some ground-based 
observatories. However, to avoid possible systematic 
uncertainties, it is crucially important to use large 
homogeneous samples of quasars obtained by the same 
instrument and manner. Thanks to the public data 
release of the Sloan Digital Sky Survey (SDSS) project 
(York et al. 2000), spectra of more than a few thousands quasars 
are now available,
and therefore a systematic examination of BLR properties 
becomes feasible. Although the signal-to-noise ratio of
each spectrum in the SDSS database is not high enough to
measure emission-line fluxes accurately, 
higher quality spectra can be obtained by the ``composite'' of
numerous individual spectra. Composite spectra are a very
efficient tool not only to increase the data quality, but also
to minimize effects due to individual characteristics of each
quasars. Note that the datasets of the 2dF and 6dF QSO Redshift 
Surveys (e.g., Croom et al. 2004) are also available to examine 
various statistical properties of quasars 
(e.g., Croom et al. 2002); however, most quasars 
found by these surveys are only at $z<3$ and the spectra are not 
calibrated in (relative) flux, making it difficult to obtain 
accurate emission-line flux ratios.
By using composite SDSS spectra, various 
spectroscopic properties of quasars have already been
investigated, such as the spectral energy distribution
(SED) (Vanden Berk et al. 2001) and
broad absorption line (BAL) objects (Reichard et al. 2003a).
The purpose of this paper is to use composite spectra of SDSS 
quasars to derive the dependences of BLR 
emission-line spectra on redshift and luminosity in wide ranges, 
although the luminosity range is narrower for 
high-redshift quasars due to the limited spectroscopic 
sensitivity of the SDSS dataset.

In this paper, we present our making of the composite spectra
of SDSS quasars for various redshift and luminosity intervals. 
The measured emission-line flux ratios and other spectroscopic
properties are discussed in detail. 
By combining these observational results with new extensive
photoionization models, we investigate the metallicity and other 
spectroscopic properties 
of the BLRs. Throughout this paper, we adopt a cosmology with 
($\Omega_{\rm tot}$, $\Omega_{\rm M}$, $\Omega_{\Lambda}$)
=(1.0, 0.3, 0.7) and $H_0$ = 70 km s$^{-1}$ Mpc$^{-1}$ .

\section{Composite spectra}

\subsection{Spectral composition}

The spectroscopic data of SDSS quasars (Richards et al. 2002a)
were obtained from the SDSS archive, Data Release 2 
(DR2; Abazajian et al. 2004)\footnote{The reason why we did 
not use later releases (DR3 or DR4) is because, as discussed 
in the following, we had to inspect individually each 
spectrum to remove broad absorption line quasars, and
DR2 has a number of objects for which this process is feasible.}. 
The spectral resolution of
the SDSS spectroscopic data is $\sim$2000, that corresponds
to $\Delta v \sim 150$ km s$^{-1}$, which is high enough
for our scientific purposes. Only quasars at 
$2.0 \leq z \leq 4.5$ are considered in this paper, because
in this redshift range the wavelength coverage of the SDSS 
spectroscopic data
($3800{\rm \AA} \la \lambda_{\rm obs} \la 9200 {\rm \AA}$) 
includes at least the rest frame wavelength interval from 
Ly$\alpha$ $\lambda$1216 to He {\sc ii} $\lambda$1640.
There are 6181 quasars\footnote{
We regard the SDSS spectroscopic targets as quasars
when the objects are classified as ``quasars''
(\texttt{specClass}=3) or ``high-$z$ quasars'' 
(\texttt{specClass}=4)
by the SDSS classification algorithm.} within this redshift
range and with a redshift confidence level of \texttt{zConf} 
$\geq$ 0.75 in the DR2 archive; while 311 objects are excluded
because not matching this \texttt{zConf} criterion. 109
objects are removed from the 6181 quasars:
84 spectra of them suffer from bad focusing of
the spectrograph collimator and 25 spectra suffer from
leaking light from a LED (see Abazajian et al. 
2004 for more details). Therefore the number of usable
quasar spectra is 6072. Note however that this sample is 
not a complete one in
any sense, because the spectroscopic targets are selected 
heterogeneously: some of them are selected through their
SDSS photometric properties, and others are
selected by cross-identification with radio or X-ray
sources.

The SDSS quasar selection algorithm picks up not only
``normal'' quasars, but also
broad absorption line (BAL) quasars (Richards et al. 2002a).
Indeed Reichard et al. (2003b) found 224 BAL quasars among
3814 quasars in the SDSS early data release
(EDR; Stoughton et al. 2002). BAL quasars
should be removed because the BAL features affect
fluxes and profiles of broad emission lines in the composite 
spectra, therefore we
removed BAL quasars from our sample. Here we did not adopt 
the standard ``Balnicity Index (BI)'' (Weymann et al. 1991) to
identify BAL quasars. This is because BI does not identify
quasars with a strong absorption line close to the
systemic velocity of the quasar 
($\mid \! \! v_{\rm outflow} \! \! \mid < 3000$ km s$^{-1}$), 
by definition. For our purpose, however, quasars with such 
associated absorption lines (see, e.g., Foltz et al. 1986) 
should be also removed to investigate the BLR emission-line 
properties correctly. We checked all of the 6072 quasars by eye
and identified 724 quasars with strong absorption features,
which were removed from our quasar sample.

Each spectrum was then corrected for Galactic reddening 
with the reddening curve of Cardelli et al. (1989),
even though the effect is very small in most cases because 
the SDSS survey area is at high Galactic latitude,
i.e. $E(B-V)<0.05$ mag for $\sim83\%$ and
$E(B-V)<0.10$ mag for $\sim98\%$ of our sample
[the median value is $E(B-V)=0.029$ mag]. Note that the
spectroscopic data in the EDR archive and the DR1 archive are
already corrected for the Galactic reddening, which is different
from the spectroscopic data in the DR2 archive. 
The $k$-correction for each quasar was applied for calculating
the absolute $B$ magnitude. For this purpose, a 
simple power-law SED with a power-law index of $\alpha=0.5$ 
(where $f_\nu \propto \nu^{-\alpha}$) is assumed for the 
intrinsic spectral shape of quasars, following other 
studies on SDSS quasars (e.g., Schneider et al. 2002, 2003).
This assumption seems valid at least at 
$\lambda_{\rm rest} < 5000 {\rm \AA}$,
based on the composite spectrum of the whole SDSS quasar sample
(Vanden Berk et al. 2001).
In this study, the absolute magnitude of quasars was calculated
from the $i^{\prime}_{\rm PSF}$ magnitude, because an
$i^\prime$-band flux is not affected by the Lyman-break 
for quasars at $z \leq 4.5$.

For most of the quasars in our sample, the signal-to-noise ratio
of the individual spectra is not high enough to 
measure the properties of the broad emission lines accurately. 
However, as mentioned above, we can investigate the BLR 
properties as a function of redshift and quasar luminosity by 
constructing composite spectra of quasars in certain parameters 
ranges and by examining the BLR properties in such composite 
spectra. As discussed by Vanden Berk et al. (2001), there are 
mainly two methods to combine the spectroscopic data. One is 
the arithmetic mean method, which preserves the relative fluxes 
of emission features. The other is the geometric mean method, 
which preserves global continuum shapes. Since we are interested 
mainly in broad emission-line properties and not in global 
continuum SED, we adopt the former strategy to combine the 
quasar spectra. However, the choice of the composite
method does not affect results and discussion significantly:
the measured emission-line fluxes vary
less than 10\% if we adopt the geometric mean instead of
the arithmetic mean. The spectral composition was
performed by using the IRAF\footnote{
IRAF (Image Reduction and Analysis Facility) is distributed
by the National Optical Astronomy Observatory, which is operated
by the Association of Universities for Research in Astronomy Inc.,
under corporative agreement with the National Science Foundation.} 
task, \texttt{scombine}, adopting a 5$\sigma$ clipping rejection 
criterion to remove bad pixels.
Before combining the spectral data, it is necessary to shift
the data from the observed frame to the quasar rest frame.
However, the accurate determination of quasar redshifts is
not an easy task, because broad emission lines tend to be
shifted blueward or redward compared to the quasar rest frame
(e.g., Gaskell 1982; Richards et al. 2002b).
Although narrow emission lines are sometimes used as a measure
of systemic recession velocities of quasars (e.g., 
Vanden Berk et al. 2001), we cannot use
narrow emission lines because only the rest-frame ultraviolet
spectral region (which lacks of narrow lines) is available due 
to the redshift range of the sample. Despite this uncertainty,
we simply adopted the redshift assigned by the SDSS reduction 
pipeline. This choice is acceptable for us because 
we are mainly interested in the broad emission-line flux ratios,
but it should be kept in mind that a consequence may be 
non-negligible uncertainties in the velocity profiles of
emission lines in the composite spectra.
We will discuss this issue briefly in \S\S4.1.
After shifting the spectra to the quasar rest frame, the data
were re-binned to a common dispersion of 1 ${\rm \AA}$.
Then each individual spectrum was normalized to the mean flux
density at $1445{\rm \AA} < \lambda_{\rm rest} < 1485{\rm \AA}$.
In this normalization process, 4 objects were removed from the
5348 objects because their spectra show significant problems in
the wavelength range of 
$1445{\rm \AA} < \lambda_{\rm rest} < 1485{\rm \AA}$.
As a consequence the number of quasars used in the following
analysis is 5344. 
Then the quasars are divided into redshift and luminosity bins
with the intervals of $\Delta z = 0.5$ and $\Delta M_B = 1$ mag.
In Table 1 the final number of objects used in our analysis is 
given for each redshift and luminosity bin. Among the
composite spectra, we analyze only those which were created 
by at least 5 individual spectra.
The resulting composite spectra are shown in Figures 1--21.

\subsection{Emission-line measurement: the method}

The measurement of emission-line fluxes in quasar spectra is
a complicated issue, both because some adjoining emission-line 
such as ``Ly$\alpha$ and N{\sc v}$\lambda$1240'' and
``Al{\sc iii}$\lambda$1857, Si{\sc iii}]$\lambda$1892 and
C{\sc iii}]$\lambda$1909'' are heavily blended, and also because
the continuum level is often not easy to estimate.
There are two methods which have been employed mostly.
One is by fitting the detected emission-line feature by some
appropriate function, and the other is by defining a ``local'' 
continuum level for each emission line and integrating the
flux above the adopted continuum level. Zheng et al. (1997)
adopted the former method to measure emission-line fluxes
in the composite HST quasar spectrum through
multiple Gaussian fitting (see also, e.g., Laor et al. 1994). 
Vanden Berk et al. (2001) adopted
the latter strategy and measured line fluxes in the composite
spectrum of 2204 SDSS quasars spanning a wide luminosity and 
redshift range by summing up the line
flux above a defined local continuum level.
Here it should be kept in mind that both methods
have serious difficulties to measure accurate emission-line 
fluxes.
As for the profile-fitting method, it is crucial to
choose appropriate functions to fit emission lines.
A simple Gaussian or 
Lorentzian profile does not work since the broad emission
lines of quasars generally show asymmetric velocity profiles 
(e.g., Corbin 1997; Vanden Berk et al. 2001; Baskin \& Laor 2005).
Although the multi-Gaussian method can achieve
reasonably sufficient profile fit, it requires
many free parameters.
It is also reported that the best-fit profile function may 
depend on the velocity width and other AGN properties
(e.g., Sulentic et al. 2002).
As for the local-continuum method, on the other hand,
it is not clear whether the adopted continuum level is
appropriate or not; this uncertainty may be crucial especially
when blended emission lines are concerned.

In this work, we adopt the profile-fitting method because
we are interested in the fluxes of emission lines including
heavily blended ones, such as N{\sc v}$\lambda$1240, which is
generally regarded as an important metallicity diagnostic
(e.g., Hamann \& Ferland 1993; Hamann et al. 2002).
The following function is adopted to fit the line profiles:
\begin{equation}
  F_\lambda = \cases{
    F_0 \times \left(\frac{\lambda}{\lambda_0}\right)^{-\alpha} 
      \ \ \ {\rm for} \ \lambda > \lambda_0 \cr
    F_0 \times \left(\frac{\lambda}{\lambda_0}\right)^{+\beta}
      \ \ \ {\rm for} \ \lambda < \lambda_0 \cr}
\end{equation}
here the power-law indices ($\alpha$ and $\beta$) are different 
between the blue side and the red side of a given emission 
line (i.e., $\alpha \neq \beta$ generally), 
which allows us to fit asymmetric velocity profiles. 
This function can achieve a better fit than 
double-Gaussian and modified (i.e., asymmetric) 
Lorentzian methods (see \S\S4.1).
Since the emission lines with a different ionization degree
tend to show systematically different velocity profiles
(e.g., Gaskell 1982; Wilkes 1984, 1986; Baskin \& Laor 2005), 
we divide the detected UV emission lines
into two main groups: high-ionization lines (HILs; 
N{\sc v}$\lambda$1240, O{\sc iv}$\lambda$1402, 
N{\sc iv}]$\lambda$1486, C{\sc iv}$\lambda$1549 and
He{\sc ii}$\lambda$1640) and low-ionization lines (LILs;
Si{\sc ii}$\lambda$1263, Si{\sc iv}$\lambda$1398,
O{\sc iii}]$\lambda$1663, Al{\sc ii}$\lambda$1671,
Al{\sc iii}$\lambda$1857, Si{\sc iii}$\lambda$1887
and C{\sc iii}]$\lambda$1909), with the boundary ionization
potential of $\sim 40$ eV. See, e.g., Collin-Souffrin \&
Lasota (1988) for a discussion on this dichotomy. 
For reader's convenience, the ionization potentials of the
concerned ions are given in Table 2. 
Since emission-line profiles (width, asymmetry, and shift)
are correlated with the ionization potential of the
corresponding ions (e.g., McIntosh et al. 1999), emission
lines with similar ionization degree tend to show similar
profiles, approximately. 
We thus assume that all emission lines in the same group 
have the same power-law indices of $\alpha$ and $\beta$, 
and the same wavelength shift of the emission-line center 
($\lambda_0$) with respect to the systemic velocity. 
All the line profiles within the same group were fitted
{\it simultaneously} to infer the best $\alpha$, $\beta$
and $\lambda_0$.
As for the Ly$\alpha$ profile,
we adopt the same red-side index ($\alpha$) as that for HILs 
and the other parameters are left free; this is to
avoid that absorption by intervening IGM on the blue side of
Ly$\alpha$ affects the overall fit
(as we will discuss later on, we are
not interested in the Ly$\alpha$ itself but in 
N{\sc v}$\lambda$1240, which is blended with it).
The model fitting is 
performed in the spectral region at 
$1210{\rm \AA} < \lambda < 2000{\rm \AA}$ except for the
composite spectra of quasars at $4.0 \leq z < 4.5$,
for which the fitting is performed at 
$1210{\rm \AA} < \lambda < 1687{\rm \AA}$ due to
the smaller (rest-frame) spectral coverage.
The slope and the amplitude of the power-law continuum
are initially estimated in two wavelength regions where
the emission-line contribution to the total flux appears
to be small ($1445{\rm \AA} < \lambda < 1455{\rm \AA}$ and
$1973{\rm \AA} < \lambda < 1983{\rm \AA}$) and finally 
determined by the model fitting. 
Since the redder emission-line free region is not available
for the composite spectra of quasars at $4.0 \leq z < 4.5$,
we refer to a spectral region at
$1687{\rm \AA} < \lambda < 1697{\rm \AA}$ instead of
that at $1973{\rm \AA} < \lambda < 1983{\rm \AA}$ for the
initial guess of the parameters for the continuum emission.

The O{\sc i}+Si{\sc ii} composite at 1305${\rm \AA}$ and
C{\sc ii}$\lambda$1335 is measured simply by summing up all of
flux above the continuum level for each line
($1286{\rm \AA} \leq \lambda \leq 1322{\rm \AA}$ for the
O{\sc i}+Si{\sc ii} composite and $1322{\rm \AA} \leq \lambda 
\leq 1357{\rm \AA}$ for C{\sc ii}$\lambda$1335), because their
velocity profiles are different from both HILs and LILs.
We also measure the flux of a ``1600${\rm \AA}$ bump'',
which is clearly seen in the residual spectra (see lower panels
in Figures 1--21). We simply sum up the flux above the continuum 
level ($1570{\rm \AA} \leq \lambda \leq 1631{\rm \AA}$)
also for this unidentified spectral feature.
The nature of the 1600${\rm \AA}$ bump will be discussed in
\S\S4.2. 
All these spectral regions (O{\sc i}+Si{\sc ii} $\lambda$1305,
C{\sc ii}$\lambda$1335, and ``1600${\rm \AA}$ bump'') are 
excluded from the fit.
Also, the wavelength region 
at $1687{\rm \AA} < \lambda < 1833{\rm \AA}$ is excluded from
the fitting process because there are heavily blended emission
lines such as N{\sc iv}$\lambda$1719, Al{\sc ii}$\lambda$1722,
N{\sc iii}]$\lambda$1750 and Fe{\sc ii} multiplets.

\subsection{Emission-line measurement: results}

The fitting results are shown in Figures 1--21. 
The measured emission-line fluxes, normalized to the 
C{\sc iv}$\lambda$1549 flux, are given in Tables 3--7.
Here the errors given in Tables 3--7 contain only
the statistical errors, which are estimated from
the covariance matrix in the standard way.
The measured profile parameters, i.e., the velocity shifts
from the systemic velocity, the $\alpha$ and $\beta$ 
parameters, and the velocity widths in FWHM, are given in 
Table 8. To see how the BLR emission-line properties depend on
redshift and luminosity, some measured emission-line flux 
ratios are plotted as a function of redshift and absolute 
$B$ magnitude in Figures 22 and 23.
For Si{\sc iv}$\lambda$1397 and
O{\sc iv}]$\lambda$1402 only the sum of their flux is considered,
because the wavelength separation
of those two lines is so small that the fitting process 
can hardly deblend them for some composite spectra
(in Table 3 the measured O{\sc iv}]$\lambda$1402 flux is
nearly zero only for the brightest case while the
Si{\sc iv} flux increases suddenly at this luminosity).
For the same reason, only the sum of the O{\sc iii}]$\lambda$1663
and the Al{\sc ii}$\lambda$1671 is plotted in Figures 22 and 23.
As shown in Figure 22, there are no apparent redshift 
dependences in the major emission-line ratios, although
the highest-redshift data may deviate from the lower-redshift
trend in some flux ratios such as 
Si{\sc ii}$\lambda$1263/C{\sc iv}$\lambda$1549,
(O{\sc i}$\lambda$1304+Si{\sc ii}$\lambda$1307)/C{\sc 
iv}$\lambda$1549, and (O{\sc iii}]$\lambda$1663+Al{\sc 
ii}$\lambda$1671)/C{\sc iv}$\lambda$1549.
Figure 22 shows that systematic differences in the 
emission-line flux ratios at different
luminosities are present. This tendency is 
more clear in Figure 23, where significant 
correlations between some flux ratios and luminosity are seen.

In order to examine more quantitatively the significance
of the correlation of emission-line flux ratios with
redshift and luminosity, in Figures
24 and 25 we show the flux ratios normalized to the values
measured from the composite spectra at $2.0 \leq z < 2.5$
or $-26.5 > M_B \geq -27.5$. For these normalized flux ratios,
we apply a linear fit, whose results are also shown
in Figures 24 and 25. The slopes of the best-fit results are
given in Table 9. We also examine whether the flux ratios
are correlated with redshift or absolute $B$ magnitude with a
statistical significance, by applying the Spearman rank-order 
test. The derived Spearman rank-order correlation
coefficients ($r_{\rm S}$) and their statistical significance
$p(r_{\rm S})$, which is the probability of the data being
consistent with the null hypothesis that the flux ratio is
not correlated with redshift or absolute $B$ magnitude,
are also given in Table 9. The results of the Spearman
rank-order test are summarized as follows:
\begin{itemize}
\item There are no statistically significant correlations
      between the examined flux ratios and redshift
      except for (O{\sc i}+Si{\sc ii})/C{\sc iv}, which is
      marginally positively correlated with redshift.
\item The flux ratios of N{\sc v}/C{\sc iv}, 
      (Si{\sc iv}+O{\sc iv})/C{\sc iv}, Al{\sc iii}/C{\sc iv},
      Si{\sc iii}]/C{\sc iv} and 
      N{\sc v}/He{\sc ii} show statistically significant
      positive correlations with absolute $B$ magnitude. On the
      contrary, the flux ratios of (1600${\rm \AA}$ bump)/C{\sc iv} 
      and He{\sc ii}/C{\sc iv} show significant negative 
      correlations with absolute $B$ magnitude.
\end{itemize}
Note that the possible marginal correlations seen in Figure 24 are 
mainly due to the data for the highest redshift quasar 
composite spectra; there are no apparent correlations for those 
flux ratios in the range $2.0 < z < 4.0$.

We also examine the dependences of emission-line shifts and
FWHMs on
redshift and absolute $B$ magnitude. However, as mentioned 
in \S\S2.1, the shape of the emission lines 
may be inaccurate due to the uncertainty in redshift of individual
quasars. Therefore we focus on the relative velocity difference
between HILs and LILs, which is less affected by uncertainties
in the absolute velocity determinations.
The relative differences of HILs and LILs in emission-line
peaks ($\Delta v_{\rm HIL} - \Delta v_{\rm LIL}$) and FWHMs of HILs
and LILs are plotted in Figure 26, as a function of redshift and 
absolute $B$ magnitude. These profile parameters appear to be
strongly correlated with absolute $B$ magnitude while not correlated
with redshift. The Spearman rank-order test results in 
probabilities of the uncorrelation between redshift and
emission-line shift, FWHM of HILs and LILs of
0.46, 0.11 and 0.45 respectively, 
while those between absolute $B$ magnitude
and emission-line shift, FWHM of HILs and LILs are
$2.5 \times 10^{-3}$, $5.3 \times 10^{-5}$ and $2.0 \times 10^{-2}$.
These results indicate that the correlations between absolute $B$ 
magnitude and emission-line shift and  FWHM of HILs are
statistically significant while those between redshift and
emission-line shift, FWHM of HILs and LILs are statistically 
uncorrelated. 
The relations between FWHMs, redshift, and
absolute $B$ magnitude are interesting issues since they
contains information on the growth of supermassive black holes
(SMBHs). 
We do not discuss this issue further since this topic is 
beyond the scope of this paper.

\section{Photoionization models}

\subsection{Method}

To interpret these results quantitatively, it is very
useful to compare emission-line flux ratios with 
photoionization models. However, it is well known
that simple one-zone photoionization models cannot properly
describe BLRs because gas clouds in BLRs span wide ranges in 
densities and/or ionization degrees in general (e.g., 
Davidson 1977; Collin-Souffrin et al. 1982). To investigate the
physical properties of gas clouds in the BLRs of quasars,
Baldwin et al. (1995) proposed the locally optimally emitting 
cloud (LOC) model, which is a multi-zone photoionization model.
In this model, gas clouds with a wide range of physical 
conditions are present at a wide range of distances, 
and thus the net emission-line spectra can be calculated by 
integrating in the parameter space of gas density and 
radius, assuming some distribution functions. 
This model can predict fluxes of both low-ionization emission 
lines and high-ionization emission lines consistently and
simultaneously, and thus it has been sometimes used to investigate
physical and/or chemical properties of ionized gas clouds in
BLRs (e.g., Korista et al. 1998; Korista \& Goad 2000; 
Hamann et al. 2002).

Adopting this LOC model, we carried out photoionization model
calculations by using the photoionization code $Cloudy$
version 94.00 (Ferland 1997). For simplification, we assume  
a plane-parallel geometry and a constant gas density for each 
gas cloud. Dust grains are not included in our calculations 
because gas clouds in BLRs are thought to be in a dust-free 
region (e.g., Netzer \& Laor 1993)
and we have verified that the physical conditions in our grid
of models imply dust sublimation in most cases. Those few cases
which allow dust survival were excluded from the final 
calculation, as briefly mentioned also in \S\S3.2.
Our assumption for the chemical composition is the same as that
of Hamann et al. (2002), in which the relative metal abundances 
scale by keeping solar relative values except for nitrogen, which
scales as the square power of other metal abundances (see 
Table 3 of Hamann et al. 2002). Here the solar elemental 
abundances are taken from Grevesse \& Anders (1989) with 
extensions by Grevesse \& Noels (1993). 
As for the SED of the ionizing photons, two types of SED
are adopted to see possible SED effects on the results: 
one is a SED with a strong UV thermal bump which matches the
quasar template of Scott et al. 2004, and the other is with a 
weak UV thermal bump which matches the HST quasar templates
(Zheng et al. 1998; Telfer et al. 2002; see also Marconi
et al. 2004 for more details). Both SEDs have the same 
optical to X-ray ratio $\alpha_{\rm OX}$ (Zamorani et al. 1981;
Ferland 1997), i.e., $\alpha_{\rm OX} = -1.49$, but different 
slopes in the energy range of 
$9.1{\rm eV} \leq h\nu \leq 35.5{\rm eV}$. See Figure 27 for a 
graphical representation of the two SEDs (see also Nagao et al.
2005). The calculation for each cloud is stopped
when the cloud thickness reaches $10^{23}$ cm$^{-2}$ or
when the cloud temperature drops below 3000 K.
We performed model runs for gas clouds with
a gas density ($n$) in the range of $10^{7-14}$ cm$^{-3}$
with a 0.2 dex step, with a flux of ionizing photons ($\Phi$) in 
the range $10^{17} - 10^{24}$ cm$^{-2}$ s$^{-1}$ also with a 0.2 
dex step, and with metallicities of $Z/Z_\odot$ = 
0.2, 0.5, 1.0, 2.0, 5.0 and 10.0. 
Therefore the number of the performed model runs is 
1296 for each metallicity and SED, giving a total number of 
15552 runs.

Once the calculations are completed, we can obtain the net emission-line
flux by integrating the line emissivity of all clouds; i.e.,
\begin{equation}
  L_{\rm line} =
    \int \!\!\! \int 4\pi r^2 F_{\rm line}(r,n) \ f(r) g(n)\ dn\ dr
\end{equation}
where $f(r)$ and $g(n)$ are the cloud distribution functions for
radius and gas density, respectively. Note that the radius is 
specified by the ionizing photon flux. Baldwin et al. (1995) 
assumed simple power-law functions for both $f(r)$ and $g(n)$;
i.e., $f(r) \propto r^{\Gamma}$ and $g(n) \propto n^{\beta}$.
It has been shown that the observed BLR emission-line spectra are
generally well reproduced by the LOC models with $\Gamma \sim -1$ 
and $\beta \sim -1$ (e.g., Baldwin 1997; Korista \& Goad 2000).

\subsection{Model results}

In Figure 28, we present some examples of calculated line 
emissivities as a function of gas density and ionizing photon 
flux. Contours indicate the predicted equivalent widths
for full geometrical coverage referred to the incident continuum
at $1215{\rm \AA}$, for models with ionizing
SED with a large UV thermal bump and a metallicity 
of $Z/Z_\odot = 1.0$.
It should be mentioned that the model calculations for 
some clouds with certain pair of ($n$, $\Phi$) fail
because of a thermal instability effect of the ionized gas
(see Ferland 1997 for details of this problem). This problem is 
more serious when high metallicity gas clouds are examined.
However even for the highest-metallicity cases (i.e., 
$Z/Z_\odot$ = 10.0), the fraction of the crashed runs is less
than 6\% of the 1296 model runs. 
The line emissivities for the crashed cases are 
estimated by simple interpolations on the $n$-$\Phi$ plane
using the results of the neighboring uncrashed models.

In Figure 29, the net emission-line flux ratios are presented
as a function of gas metallicity, which are obtained by
the integrations as given in the equation (2).
Here we adopt $\Gamma = -1$ and $\beta = -1$, i.e.,
$f(r) \propto r^{-1}$ and $g(n) \propto n^{-1}$.
In the integration procedure, gas clouds with a gas density of
$n < 10^8$ cm$^{-3}$ are excluded because such low-density
clouds are thought to be implausible for BLRs, which is 
inferred by the absence of broad [O {\sc iii}]$\lambda$4363
emission lines in spectra of AGNs (note that the critical
density of the [O {\sc iii}]$\lambda$4363 transition is
$n_{\rm crit} = 3.3 \times 10^7$ cm$^{-3}$). Gas clouds with an
ionizing photon flux of $\Phi < 10^{18}$ cm$^{-2}$ s$^{-1}$
are also excluded, because the energy density temperature of
the incident continuum emission for clouds with such a small
$\Phi$ falls below 1000 K for our
adopted SEDs, at such temperature dust grains may survive
and absorb most of the ionizing photons (as well as most
of any UV line flux which may be produced;
Netzer \& Laor 1993).
The adopted integration range is thus 
$8 \leq {\rm log} \ n \leq 14$ and $18 \leq {\rm log} \ \Phi \leq 24$.
The predicted net emission-line flux ratios are also given
in Table 10.

As apparent in Figure 29, most of emission-line fluxes
normalized to the flux of C{\sc iv}$\lambda$1549 are positively
correlated with the gas metallicity. This is mainly because
the C{\sc iv}$\lambda$1549 transition is an important coolant.
This is especially true in metal-poor environments where the cooling
by other metal lines is less effective, making
C{\sc iv}$\lambda$1549 emission become stronger when the
gas metallicity is lower. The effects of ionizing continuum
SED on the resultant predictions are very small, 
generally far less than a factor of 2. 
Note that some of the predicted flux ratios are sensitive
to the adopted weighting functions, $f(r)$ and $g(n)$,
especially when lines with a different ionization degree are
concerned (e.g., C{\sc ii}/C{\sc iv}). We discuss the effect
of this dependence on our results in \S\S4.2.
Note that Hamann et al. (2002) also presented the
results of the LOC photoionization model calculations
with the same assumption on the relative elemental
abundance ratios. Our results are almost consistent
with those of Hamann et al. (2002). The small
differences may be due to the differences in the
adopted SEDs, to the integration ranges of $n$ and $\Phi$,
to the version of Cloudy, and to the column density of
clouds.

\section{Discussion}

\subsection{Comparison of emission-line fitting methods}

Before comparing the results of photoionization model 
calculations with the measured emission-line flux ratios
of SDSS DR2 quasars, we should discuss whether our 
measurement method is appropriate or not. Our adopted
formula for measuring emission lines [equation (1)] is
different from the widely adopted formulae such as
multi-Gaussian and (modified) Lorentzian. In Figure 30,
we compare the fitting results by adopting equation (1),
double Gaussian (allowing the different central velocity 
for the two Gaussian components), and modified Lorentzian 
that is described by the following formula;
\begin{equation}
F_\lambda = \frac{F_0}
  {1+\left(\frac{\lambda - \lambda_0}{\Delta\lambda}\right)^\alpha}
\end{equation}
($\alpha = 2$ gives a usual Lorentzian function).
Our adopted function provides a much better fit
than the double Gaussian and modified Lorentzian;
this is apparent especially on the blue side of
C{\sc iv}$\lambda$1549 emission and on the red side of 
He{\sc ii}$\lambda$1640 emission.
Note that the number of the free parameter in
our fitting function is four (amplitude, central wavelength, 
$\alpha$ and $\beta$) while those of double Gaussian and 
modified Lorentzian are six and four, respectively.
Although multi-Gaussian methods with three or more components
may achieve better fit for a single line, they require
too many parameters and makes the interpretation
more complex. As a consequence, we conclude that our
power-law fitting is better than the other methods to
describe broad emission-line profiles of quasar composite
spectra.

However, since the power-law formula is not a conventional 
one to describe the BLR emission-line profiles, we should be
careful to judge whether the power-law formula is a really
representative for the BLR emission. In order to investigate
whether the power-law emission-line profiles of our quasar
composite spectra are due to some artificial effect, rather
than resulting from the intrinsic kinematic properties
of the BLR, some individual spectra of SDSS quasars
with a high S/N are also fitted by using the power-law formula in 
the same way as for the composite spectra. 
The fitting results are shown in Figures 31 -- 33, where
SDSS J085417.6+532735 ($z=2.42$, $M_B = -28.6$),
SDSS J080342.0+302254 ($z=2.03$, $M_B = -28.9$), and
SDSS J154359.4+535903 ($z=2.37$, $M_B = -28.5$) are 
investigated. It is clear that the power-law formula 
describes properly individual spectra of quasars, 
and not only
quasar composite spectra. This suggests that the power-law
profile is really representative of the BLR emission and
it should be insightful to investigate kinematic properties
and the geometrical configuration of gas clouds in BLRs.
We do not discuss these issues further because these are
beyond the scope of this paper.

In \S\S2.1, it was mentioned that possible uncertainties
in the redshift assigned to each object may introduce
artifacts in the emission-line profiles of composite spectra. 
To check how much this effect might be significant,
we investigate the difference in redshift determined by a
specific emission line and the redshift assigned for each 
object by the SDSS pipeline (i.e., the average of various
spectral features).
What really matters is not the absolute redshift difference
between a specific line and the average redshift from
other line, but the dispersion of such a difference.
As for the C{\sc iv}$\lambda$1549 emission of quasars with 
$-26.5 > M_B \geq -27.5$ at $2.0 \leq z < 2.5$,
the average difference $z_{\rm CIV} - z_{\rm QSO}$
and its standard deviation are
$< \!\! z_{\rm CIV} - z_{\rm QSO} \!\! > = -0.007 \pm 0.006$.
This standard deviation 
($\sigma_{< \!\! z_{\rm CIV} - z_{\rm QSO} \!\! >} = 0.006$) 
corresponds to a velocity dispersion
of $\sim 1800$ km s$^{-1}$. Therefore we should be aware that
velocity structures on scales less than this velocity interval 
of the C{\sc iv}$\lambda$1549 emission-line profile
in the composite spectra may be affected by some artificial
effects; more global velocity structures are thought to be
free from such effects.

As mentioned in \S\S2.2, there is another measurement method
which has often been adopted, that is the local-continuum
method. It is useful to compare the results of our
measurements with the values measured by adopting local
continuum levels, as in Vanden Berk et al. (2001).
In Figure 34, we show the estimated local continuum levels
for the wavelength regions around N{\sc v}$\lambda$1240,
C{\sc iv}$\lambda$1549 and C{\sc iii}]$\lambda$1909. Here
we adopt roughly the same $\lambda_{\rm lo}$ and 
$\lambda_{\rm hi}$ as those defined by Vanden Berk et al. 
(2001). The local continuum is linear for the wavelength 
region around C{\sc iv}$\lambda$1549, while for the 
wavelength regions around N{\sc v}$\lambda$1240 and 
C{\sc iii}]$\lambda$1909 the local continua are determined
by third polynomial fitting by using the wavelength parts 
outside the emission lines. For this comparison, the 
composite quasar spectrum at $2.0 \leq z < 2.5$ and 
$-25.5 > M_B \geq -26.5$ is used, because most of quasars 
used by Vanden Berk et al. (2001) are at lower redshift and 
are less luminous than ours. The flux measurement results 
are presented in Table 11 and compared with the values of 
Vanden Berk et al. (2001) and with the results of our 
method in \S\S2.2. The fluxes reported by Vanden Berk et al. 
(2001) are similar to the values measured by us by adopting 
the local-continuum method, but are very different from the 
values given in Table 3a obtained by fitting with equation 
(1). This indicates that the flux measurement for several 
of the BLR emission lines highly depends on the adopted 
measurement method. Which method is more appropriate? To 
tackle this problem, emission-line profiles are very 
useful, because emission lines with similar ionization 
potentials are thought to arise in similar regions in BLRs, 
and therefore should have similar emission-line profiles. 
In our fitting method, HILs (N{\sc v}$\lambda$1240, 
C{\sc iv}$\lambda$1549 and He{\sc ii}$\lambda$1640) have 
the same velocity profile and the same velocity shift from 
the systemic velocity by definition. The emission-line 
width and skewness for each line reported by 
Vanden Berk et al. (2001) are, on the other hand, very 
different among these three emission lines; the reported 
width and skewness are ($\sigma$, Skew) =
(2.71${\rm \AA}$, $-0.21$), (14.33${\rm \AA}$, $-0.04$) and
(4.43${\rm \AA}$, $-0.22$) for N{\sc v}$\lambda$1240, 
C{\sc iv}$\lambda$1549 and He{\sc ii}$\lambda$1640, 
respectively. The difference of the line skewness should 
correspond to the difference in the kinematic status of the 
line-emitting clouds, implying a strong segregation of the 
line emitting regions among HILs. These line widths 
correspond to the velocity widths of 655 km s$^{-1}$, 
2779 km s$^{-1}$ and 810 km s$^{-1}$, respectively. 
The difference of a factor of 3--4 in the velocity width 
corresponds to the difference of a factor of $\sim$10 in 
the radius from the nucleus, if the BLR motions are 
dominated by the gravitational potential of the SMBH. 
As presented in Figure 28, the photoionization model 
suggests that the emitting region of 
N{\sc v}$\lambda$1240 and C{\sc iv}$\lambda$1549 are not
segregated with a factor of 10 in radius from the nucleus. 
Reverberation mapping observations also suggest that HILs 
arise from a similar region, and actually 
He{\sc ii}$\lambda$1640 sometimes shows even more rapid 
time variations than C{\sc iv}$\lambda$1549 (e.g., 
Clavel et al. 1991; Korista et al. 1995; 
Peterson \& Wandel 1999). Taking all of the above matters 
into account, we conclude that our measurement method is 
better than the local continuum method to measure the 
emission-line fluxes accurately.

\subsection{Comparison of observed line ratios and trends with models}

Our analysis on the SDSS DR2 composite spectra 
clearly indicates that some emission-line flux ratios
(N{\sc v}/C{\sc iv}, Si{\sc ii}/C{\sc iv}, 
(Si{\sc iv}+O{\sc iv})/C{\sc iv}, Al{\sc iii}/C{\sc iv} and
N{\sc v}/He{\sc ii}) positively correlate with absolute $B$ 
magnitude with a high statistical significance, but are 
independent of redshift. As presented in \S\S3.2, photoionization
models suggest that these flux ratios positively correlate 
with the gas metallicity. This means that the dependences of
emission line ratios on absolute $B$ magnitude are caused by the
dependence of the BLR gas metallicity on the luminosity.
In other words, {\it the BLR gas metallicity positively depends on
the quasar luminosity, but independent of the quasar redshift}.
This conclusion is also suggested by some earlier studies. 
As for Seyfert 1 galaxies at the local universe, the 
C{\sc iii}]/C{\sc iv} flux ratio depends strongly on the 
luminosity (V\'{e}ron-Cetty et al. 1983), which suggests the 
dependence of the gas metallicity on the luminosity. 
The correlation of the BLR metallciity with the quasar luminosity
has been reported by, e.g., Hamann \& Ferland (1999),
Warner et al. (2004), and Shemmer et al. (2004).
Warner et al. (2004) also reported the correlation of
the BLR metallciity with the mass of SMBHs
(see also Shemmer et al. 2004).

The dependence of the gas metallicity on the quasar luminosity 
is expected, since (1) the quasar luminosity 
should positively correlate with the mass of SMBHs for a 
given Eddington accretion ratio, (2) a good correlation between
mass of SMBHs and mass of the host galaxies, exists at least
in the local universe (e.g., Gebhardt et al. 2000; 
Ferrarese \& Merritt 2000; Marconi \& Hunt 2003), and (3) more massive 
galaxies tend to have higher metallicity gas and stars due to
their deeper gravitational potential (e.g., Pagel \& Edmunds 1981;
Arimoto \& Yoshii 1987; Matteucci \& Tornamb\`{e} 1987;
Tremonti et al. 2004). The results presented in
this paper indicate the existence of close relation between 
mass of SMBHs and host galaxies, and that the galaxy mass-metallicity 
relation holds also at high redshift, up to $z \sim 4.5$.
We should mention that the independence of
broad emission-line flux ratios from redshift may be due to some
selection effects. For instance, quasars with low metallicity
may be dust-enshrouded in their young phase and thus very 
difficult to detect (see, e.g., Kawakatu et al. 2003).
Hard X-ray deep and wide surveys are required to examine this
possibility.

However, by comparing the measured flux ratios with the results
of the photoionization model calculations with fixed
weighting functions ($\beta$ and $\Gamma$ in \S3), the inferred
gas metallicity is apparently different depending on the 
adopted flux ratio. For instance, the observed 
ratios of N{\sc v}/C{\sc iv},
(Si{\sc iv}+O{\sc iv})/C{\sc iv}, Al{\sc iii}/C{\sc iv} and
N{\sc v}/He{\sc ii} suggest gas metallicities of 
$Z/Z_\odot > 2$, while the ratios of C{\sc ii}/C{\sc iv},
(O{\sc iii}]+Al{\sc ii})/C{\sc iv}, Si{\sc iii}]/C{\sc iv} and
C{\sc iii}]/C{\sc iv} suggest $Z/Z_\odot \la 2$ (Figure 29).
The ratio of He{\sc ii}/C{\sc iv} is also not reproduced in most
cases. This is consistent with earlier works that the estimates 
of the BLR metallicity using only the N{\sc v}/C{\sc iv} 
and/or N{\sc v}/He{\sc ii} flux ratios alone might be quite 
uncertain (see Hamann et al. 2002).
As for the ratio of Si{\sc ii}/C{\sc iv}, the observed 
value deviates completely from the range of model predictions.
The deviation of Si{\sc ii}/C{\sc iv} may be due to
the contamination of other emission lines into the Si{\sc ii}.
Indeed, we find that the flux of S{\sc ii}$\lambda$1256 becomes
high under some physical conditions. In Figure 35, we show
the predicted flux ratio of S{\sc ii}/Si{\sc ii} for
$Z/Z_\odot$ = 1.0 and 5.0, and $U = 10^{-2.5}$ and $10^{-3.5}$
as a function of gas metallicity, adopting the ionizing continuum
SED with a large UV thermal bump. The S{\sc ii}/Si{\sc ii} ratio
exceeds 0.2 and reaches $\sim$1 for models with high
gas densities. Apparently the contribution of the S{\sc ii}
flux makes the deviation of the measured Si{\sc ii}/C{\sc iv} 
ratio from the theoretically predicted range.
Apart from the Si{\sc ii} deviation issue, what causes the
discrepancies among the inferred metallicity from different 
line ratios?
SED effects are an implausible explanation of this
problem, because we have already
seen in Figure 29 that the predicted line flux ratios do not vary 
so significantly by changing the adopted SED.
Since the SEDs adopted in this paper (Figure 27) are thought to 
be extreme, opposite cases for the 
actual ionizing continuum of quasars,
the SED effects on the flux ratios should be smaller than those
presented in Figure 29.

One possibility which can cause the discrepancies of the
inferred metallicities from various emission-line flux ratios 
is that the assumption on the weighting functions $\Gamma$
and $\beta$ is an oversimplification.
To examine this possibility quantitatively,
we investigate the dependence of the theoretical predictions 
of some emission-line flux ratios on the adopted $\Gamma$ 
and $\beta$ parameters, as shown in Figure 36. 
It is clear that various predicted emission-line
flux ratios are highly dependent on the adopted values of
$\Gamma$ and $\beta$. Since the theoretical flux ratios 
presented in Figure 29 
are predicted by adopting $\Gamma = -1$ and $\beta = -1$
(as in many other studies), other assumptions on
$\beta$ and $\Gamma$ result in other predictions and thus
the metallicity inferred by each one of the observed 
emission-line flux ratios would change accordingly.
In order to derive a more accurate metallicity for each
composite spectrum, a better approach is to vary the indices 
of $\Gamma$ and $\beta$ to fit as many as emission lines as 
possible through the model predictions with a certain 
metallicity. Therefore we performed a fit of all available 
flux ratios for each composite spectrum by varying gas 
metallicity, SED, $\Gamma$ and $\beta$. We first associated 
errors on the line fluxes with the following recipe: (i) we 
set the minimum relative error to 5\%\ and (ii) in any case, 
absolute errors cannot be less than 1\% of the absolute flux 
of C{\sc iv}$\lambda$1549. The best fit $\beta$ and $\Gamma$ are 
obtained by minimizing $\chi^2$  computed using the logarithm 
of model and observed line fluxes. This allows us to give more 
weight to the points where the ratio between model and 
observed values is smaller.  We then perform the optimization 
on $\beta$ and $\Gamma$ for each spectral composite, using 
all continua and abundance sets.
In Tables 12--16 we show, for each spectral composite, the 
model with the lowest $\chi^2$ and the corresponding 
abundances set, as well as the $\beta$ and $\Gamma$ values. 
For comparison, we also show 
the best models with the classical $\beta=-1$ and $\Gamma=-1$.
The models with optimized $\beta$ and $\Gamma$ provide a 
much better description of the observed data but in most 
cases the inferred metallicities are similar: 
$Z/Z_\odot = 5.0$ for most cases, and $Z/Z_\odot = 2.0$ or 
10.0 for some other cases. In a few cases the metallicity 
obtained by varying $\beta$ and $\Gamma$ are lower than 
inferred by the models with fixed $\beta=-1$ and $\Gamma=-1$.
Our results may be partly affected by the lack
of the resolution of metallicity in our model calculation,
but they suggest that the typical metallicity of the gas in 
the BLRs is $\sim 5 Z_\odot$, or at least a super-solar value. 
In all cases the optimized $\beta$ values are lower, but close to, 
--1. $\Gamma$ is generally in the range $-2.0 < \Gamma < -1.5$, 
and always $\Gamma < -1$. It is interesting that the 
dispersions of the best-fit $\beta$ and $\Gamma$ is very small. 
The averaged values are $\beta = -1.08$ and $\Gamma = -1.52$,
and their RMS's are 0.05 and 0.13, respectively.
This result suggests that the commonly adopted values
of $\beta = -1$ and $\Gamma = -1$ are not the best choice.
The best-fit values of $\beta$ and $\Gamma$ may imply
some specific physical properties for the BLR, although we 
do not discuss this issue further in this paper.
Best-fit models and observations are compared 
graphically in Figure 37. 
N{\sc iv}]/C{\sc iv} is not shown because its weighting
factor in the fitting process is very low
(the errors are very large with respect to the
N{\sc iv}]/C{\sc iv} flux ratios; Table 12--16) and thus the 
fitting results are nearly meaningless for this flux ratio.
For the same reason, the results for C{\sc ii}/C{\sc iv}
are not good. Apart from these two flux ratios, the fitting 
results appear in better agreement with the observations 
when allowing $\beta$ and $\Gamma$ to be free.
Interestingly, large N{\sc v}/C{\sc iv} ratios can be rather
easily explained when the weighting functions are varied.
This suggests that the BLR metallicity cannot be determined
uniquely by using just N{\sc v}/C{\sc iv} (or
N{\sc v}/He{\sc ii}).

In order to illustrate our results on the metallicity trends 
in a graphical way, for the reader's convenience, Figure 38 
shows the metallicity, averaged in luminosity, as a function 
of redshift. Here we use the metallicities derived by the fit 
with varying $\beta$ and $\Gamma$ given in Tables 12--16. 
To avoid biases when calculating the average metallicity, we 
have only used the luminosity bins for which a metallicity 
determination is available at {\it all redshifts}. This 
limits the range of usable luminosities to 
$-25.5 > M_B > -28.5$, where the averaged metallicity are 
calculated. The errorbars are the estimated errors on the 
mean obtained by combining the uncertainty in the metallicity 
determination for each luminosity bin. The resulting plot 
shows what was already clear from Tables 12--16 and from our 
earlier discussion, i.e., there is no significant evolution 
of the metallicity as a function of redshift. Figure 39 shows 
the complementary diagram, i.e., the metallicity, averaged in 
redshift, as a function of luminosity. Again, to avoid biases 
when calculating the averaged metallicity, we have only used 
the redshift bins for which a metallicity determination is 
available at {\it all luminosities}, which limits the range 
of usable redshifts to $2<z<3$. The resulting diagram shows 
that the averaged metallicity increases significantly with 
absolute magnitude, as already inferred from the individual 
results in Tables 12--16.

Another possibility which may cause the discrepancies among 
the gas metallicities inferred from each emission-line flux 
ratio is the elemental abundance ratios.
In our model, the relative elemental abundances 
are assumed to scale proportionally to solar, except for
nitrogen which is assumed to scale as the square of other 
metal abundances. However these assumptions are an
oversimplification. In more realistic metallicity evolutionary 
scenarios abundances never scale linearly with the global 
metallicity (e.g., Pipino \& Matteucci 2004).
The inclusion of more realistic abundance pattern in our 
photoionization models will be presented in a forthcoming paper.

Our analysis on the composite spectra shows that there is no
apparent dependence of emission-line flux ratios on redshift
up to $z \sim 4.5$, which is consistent with the results of 
Dietrich et al. (2002). This suggests that the chemical 
composition of the gas clouds in BLRs does not change 
significantly up to $z \sim 4.5$.
Although the $\alpha$ elements, such as oxygen, can be 
enriched on a very short timescale ($\ll$ 1 Gyr) owing to 
type II supernovae, the enrichment of carbon and silicon
require longer timescales ($\sim$0.5--1 Gyr) since they are 
produced mainly by the low-mass or intermediate-mass evolved 
stars. Therefore, if the elemental abundance ratios in BLRs 
remain constant up to high redshift, this gives a strong 
constraint on the first epoch of active star-formation
in quasar host galaxies. In particular, constant elemental
abundance ratios up to $z=4.5$ suggest that the main 
star-formation epoch in the host galaxies occurred at
$z \ga 7$, when minimum timescale to enrich C and Si is
taken into account. However this kind of discussion 
requires a detailed theoretical predictions of the metal 
enrichment history based on galaxy chemical evolutionary 
models. Theoretical studies on the BLR evolution coupled 
with galaxy evolutionary models are thus crucial to 
understand the quasar formation and evolution.
Here we mention that possible variations of the 
emission-line spectra beyond $z=4$ might be present in 
Figure 23, though the statistical significance is not high 
enough. Further spectroscopic observations of  
quasars in this redshift range or even at higher redshifts, for 
sizeable samples of quasars, would be
highly insightful to examine the quasar evolutionary scenarios.

Similarly to some emission-line flux ratios, the velocity 
shift of HILs relative to LILs also correlates with the quasar 
luminosity and is independent of quasar redshift, as presented 
in Figure 26. The velocity difference between HILs and LILs 
has been analyzed for a long time to investigate various 
kinematic/geometrical models for BLRs (e.g., Gaskell 1982; 
Wilkes 1984, 1986; Espey et al. 1989; Corbin 1990; 
Vanden Berk et al. 2001). The correlation between this velocity 
difference and the quasar luminosity has been reported also by 
other studies (e.g., Corbin 1990; Richards et al. 2002b). 
Our results confirm those previous works
and reveal their independence of redshift.

Finally we discuss the nature of the "1600${\rm \AA}$
bump". This unidentified emission feature has been noted in 
earlier studies, e.g., Wilkes (1984) and Boyle (1990). Our 
analysis on the composite spectra clarifies that the 
1600${\rm \AA}$ bump is universally seen in spectra of 
quasars. Laor et al. (1994) clearly presented this emission 
feature in some low-redshift quasars. The 1600${\rm \AA}$ 
bump of the sample of Laor et al. (1994) is characterized by 
FWHM $\sim$ 12000--24000 km s$^{-1}$ and 
$\Delta v_{1600} \sim 1700-8200$ km s$^{-1}$ if it is one of 
C{\sc iv}$\lambda$1549 components. They also found a very 
redshifted broad component for Ly$\alpha$ and 
O{\sc vi}$\lambda$1034 (see Table 4 of Laor et al. 1994), 
which appears to support the interpretation that the 
1600${\rm \AA}$ bump is one of C{\sc iv}$\lambda$1549 
components. If this is the case, a slight negative 
correlation between the flux ratio of the 1600${\rm \AA}$ 
bump to C{\sc iv}$\lambda$1549 and the quasar luminosity 
(Figure 25) may be due to some luminosity dependence of the 
structure of the C{\sc iv}$\lambda$1549-emitting region in 
the BLR.
The 1600${\rm \AA}$ bump may be, otherwise, a blueshifted
component of the He{\sc ii}$\lambda$1640 emission. This 
interpretation is inferred by the negative correlation of 
the flux ratio of the 1600${\rm \AA}$ bump to 
C{\sc iv}$\lambda$1549 with the quasar luminosity, because 
the He{\sc ii}/C{\sc iv} ratio also shows the similar 
negative correlation with the quasar luminosity (Figure 25). 
However similar blueshifted spectral profile should appear 
for the emission lines with a similar ionization degree 
such as C{\sc iv}$\lambda$1549, which is not the case for 
our composite quasar spectra. One possibility might be the 
presence of outflowing very dense gas clouds with a low 
ionization parameter. As shown in Figure 28, gas clouds 
with low $\Phi$ and high $n$ radiate He{\sc ii}$\lambda$1640 
emission, but do not radiate the C{\sc iv}$\lambda$1549 
emission. This idea could be tested by examining velocity 
profiles of the other transition of He{\sc ii}. Although the 
He{\sc ii} Fowler lines ($\lambda$4686, $\lambda$3203, ...) may be 
difficult to investigate due to their blending with the strong 
Fe{\sc ii} multiplet emission, the He{\sc ii} Pickering lines, 
especially He{\sc ii}$\lambda$10124 may be useful to perform 
this test.
Alternatively, the 1600${\rm \AA}$ bump may be caused by the
UV Fe{\sc ii} multiplet emission. Sometimes at 
$\sim 1600{\rm \AA}$, the Fe{\sc ii} feature is seen in 
emission in quasars (e.g., Marziani et al. 1996; 
Vestergaard \& Wilkes 2001; Vestergaard \& Peterson 2005) or in 
absorption in low-ionization BALs (e.g., Hall et al. 2002).
As for the sample of Laor et al. (1994), the quasars with a 
stronger UV Fe{\sc ii} multiplet emission appear to show also 
stronger 1600${\rm \AA}$ bump, which may support the 
interpretation that the 1600${\rm \AA}$ bump is also a
part of the UV Fe{\sc ii} multiplet emission.
The only problem with the Fe{\sc ii} scenario is the
interpretation of the anti-correlation of 
(1600${\rm \AA}$ bump)/C{\sc iv} with luminosity, which is not
seen in other low-ionization lines.

\section{Summary}

In order to investigate the properties of BLR gas clouds as a
function of quasar luminosity and redshift, we made 
composite spectra of the SDSS DR2 quasars in the ranges of
$2.0 \leq z \leq 4.5$ and $-24.5 \geq M_B \geq -29.5$ for 
each luminosity and redshift bin with $\Delta M_B = 1.0$ mag 
and $\Delta z = 0.5$.
By analyzing these composite spectra, we obtained the
following results.
\begin{itemize}
\item The emission lines in the composite spectra are better 
      fitted with power-law profiles than with double Gaussian 
      or modified Lorentzian profiles. Such power-law profile 
      fitting method appears also to be more appropriate to
      measure broad emission-line fluxes than the method of using
      a local continuum level for each emission lines and
      then directly integrating the line flux.
\item The flux ratios of N{\sc v}/C{\sc iv}, 
      (Si{\sc iv}+O{\sc iv})/C{\sc iv}, Al{\sc iii}/C{\sc iv},
      Si{\sc iii}]/C{\sc iv} and 
      N{\sc v}/He{\sc ii} show statistically significant
      positive correlations with absolute $B$ magnitude.
\item Most of the examined flux ratios show no statistically 
      significant correlation with redshift.
\item Recession velocity differences between HILs and LILs, as well
      as emission-line widths, also show strong correlations
      with the quasar luminosity, while being independent of redshift.     
\end{itemize}
To interpret these findings, we performed extensive photoionization
model calculations. By comparing the results of our calculations
with the observational data, we obtained the following results.
\begin{itemize}
\item A natural interpretation of the dependence of the flux ratios
      on quasar luminosity is that more luminous quasars are
      characterized by more metal-rich gas in their BLR.
\item The typical metallicity of the gas in BLRs is estimated to be
      $\sim 5 Z_\odot$, or at least super-solar.
\end{itemize}
The absence of a significant metallicity variation up to $z \sim 4.5$
implies that the active star-formation epoch of quasar host galaxies
occurred at $z \ga 7$. To examine this issue further,
observations of rest-frame UV spectra of quasars at
$z \ga 4.5$ through near-infrared spectroscopy are crucially necessary.


\begin{acknowledgements}
   Funding for the creation and distribution of the SDSS Archive has 
   been provided by the Alfred P. Sloan Foundation, the Participating 
   Institutions, the National Aeronautics and Space Administration, 
   the National Science Foundation, the U.S. Department of Energy, 
   the Japanese Monbukagakusho, and the Max Planck Society. 
   The SDSS Web site is http://www.sdss.org/.
   The SDSS is managed by the Astrophysical Research Consortium (ARC) 
   for the Participating Institutions. The Participating Institutions 
   are The University of Chicago, Fermilab, the Institute for Advanced 
   Study, the Japan Participation Group, The Johns Hopkins University, 
   the Korean Scientist Group, Los Alamos National Laboratory, 
   the Max-Planck-Institute for Astronomy (MPIA), 
   the Max-Planck-Institute for Astrophysics (MPA), New Mexico State 
   University, University of Pittsburgh, Princeton University, 
   the United States Naval Observatory, and the University of Washington.
   We thank Gary Ferland for providing his excellent photoionization 
   code $Cloudy$ to the public.
   We also acknowledge the anonymous referee and M. Vestergaard
   for their useful comments.
   The numerical calculations in this work were performed partly
   with computer facilities in Astronomical Institute, Tohoku University.
   TN acknowledges financial support from the Japan Society for the
   Promotion of Science (JSPS) through JSPS Research Fellowships for 
   Young Scientists. RM acknowledges financial support from MIUR
   under grant PRIN-03-02-23.
\end{acknowledgements}


\clearpage


\begin{figure*}
\centering
\includegraphics[width=13.5cm]{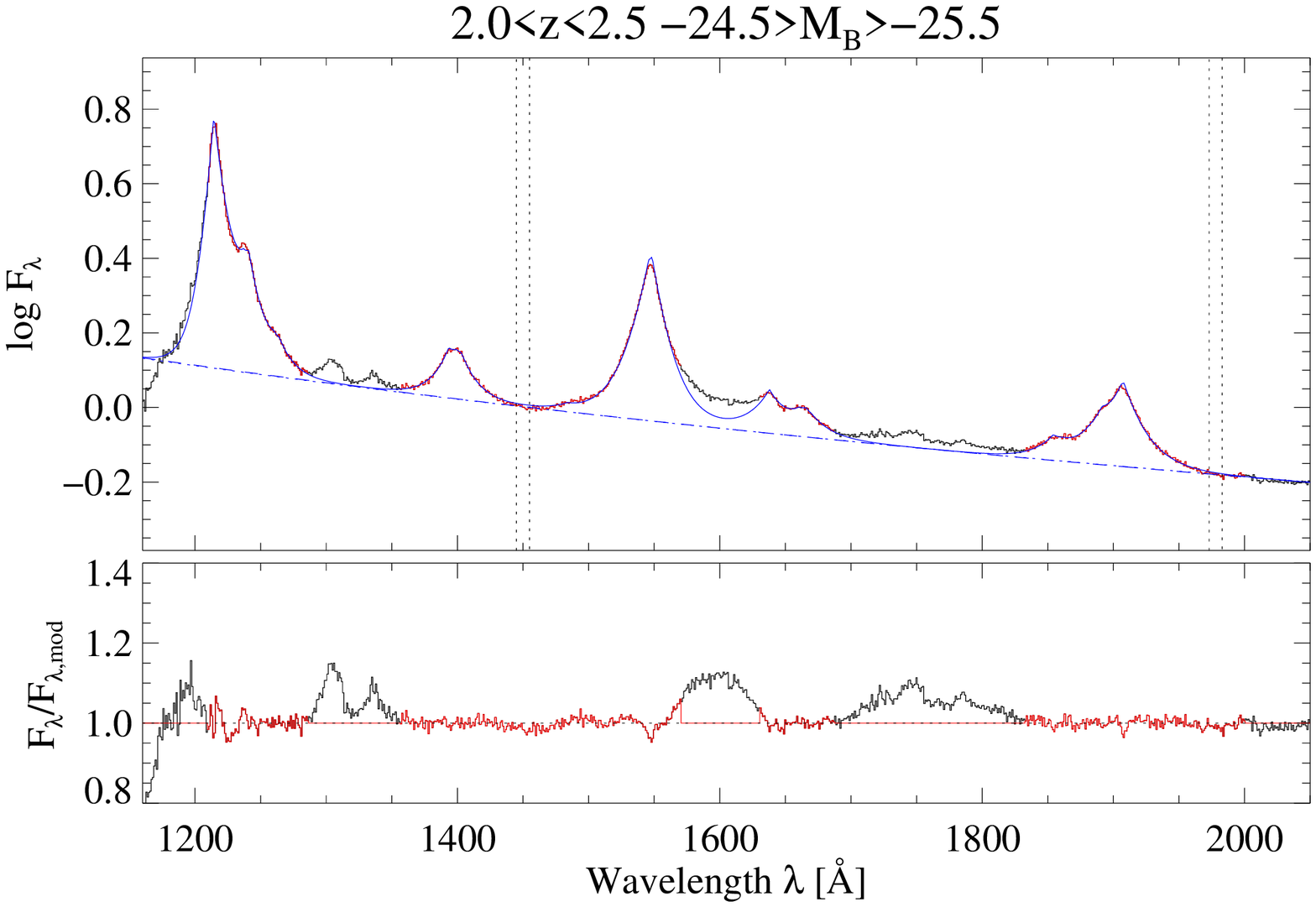}
\caption{
($Upper$) Composite spectrum of quasars with 
$-24.5 > M_B \geq -25.5$ and $2.0 \leq z < 2.5$, and
the fitting result (blue line). For the composite spectrum,
the wavelength parts which are used to the fitting are 
shown in red while the wavelength parts which are excluded
from the fitting process are shown in black. The spectral
regions denoted by vertical dotted lines are used to determine
the initial guess of the continuum level and slope
in the fitting process. The initial guess and the fitted
power-law continuum are denoted by blue dotted and dashed lines,
respectively.
($Lower$) Residual of the model fitting. Again the red parts of
the data denote the wavelength regions used in the model
fitting, and the black parts denote the excluded spectral
regions in the model fitting.
The excesses drawn by a blue, green, and yellow lines identify
O{\sc i}+Si{\sc ii} $\lambda$1305, C{\sc ii}$\lambda$1335, and 
1600${\rm \AA}$ bump, respectively.
}
\label{fig01}
\end{figure*}

\begin{figure*}
\centering
\includegraphics[width=13.5cm]{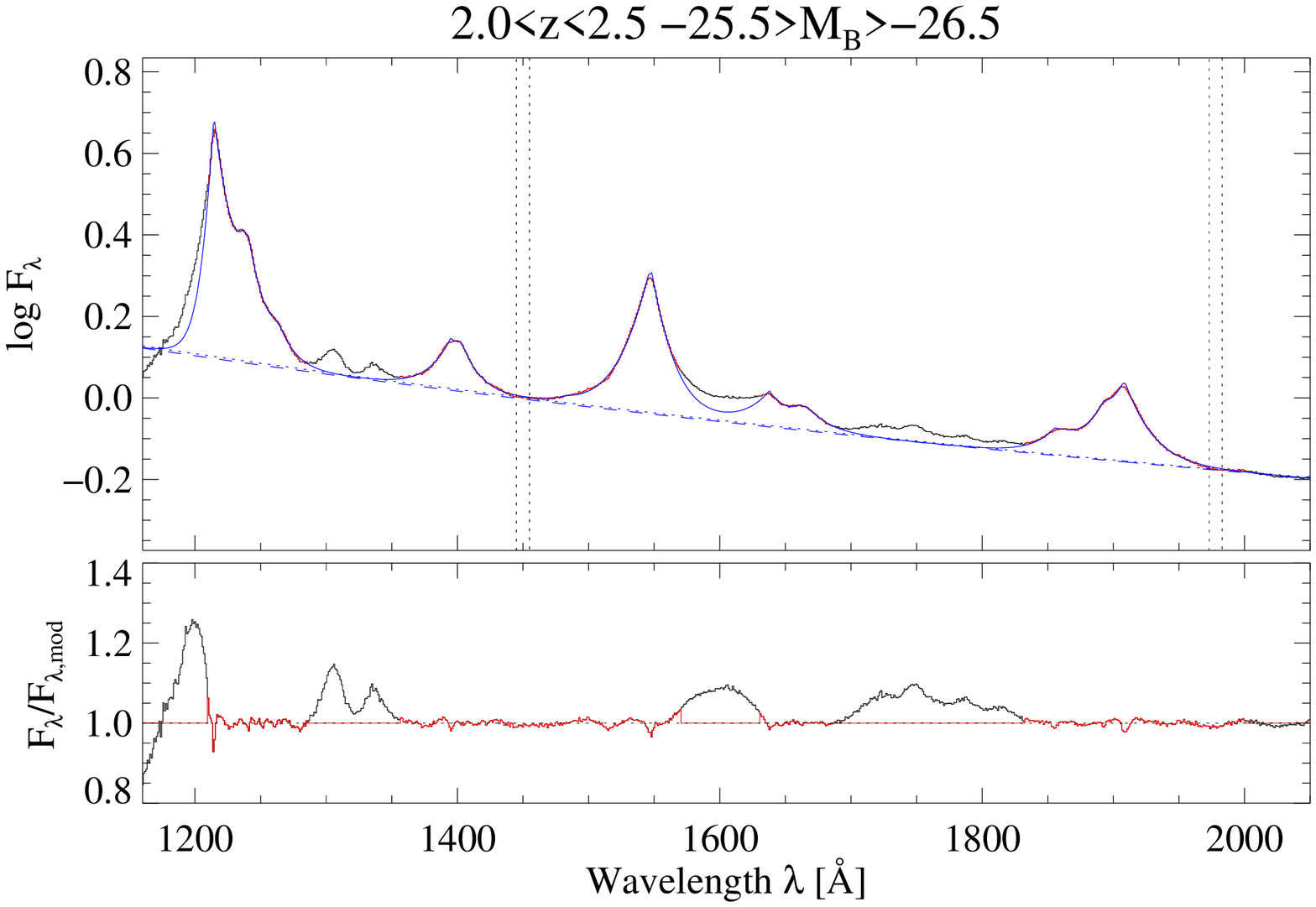}
\caption{
Same as Figure 1 but for the composite spectrum of
quasars with $-25.5 > M_B \geq -26.5$ and $2.0 \leq z < 2.5$.
}
\label{fig02}
\end{figure*}

\begin{figure*}
\centering
\includegraphics[width=13.5cm]{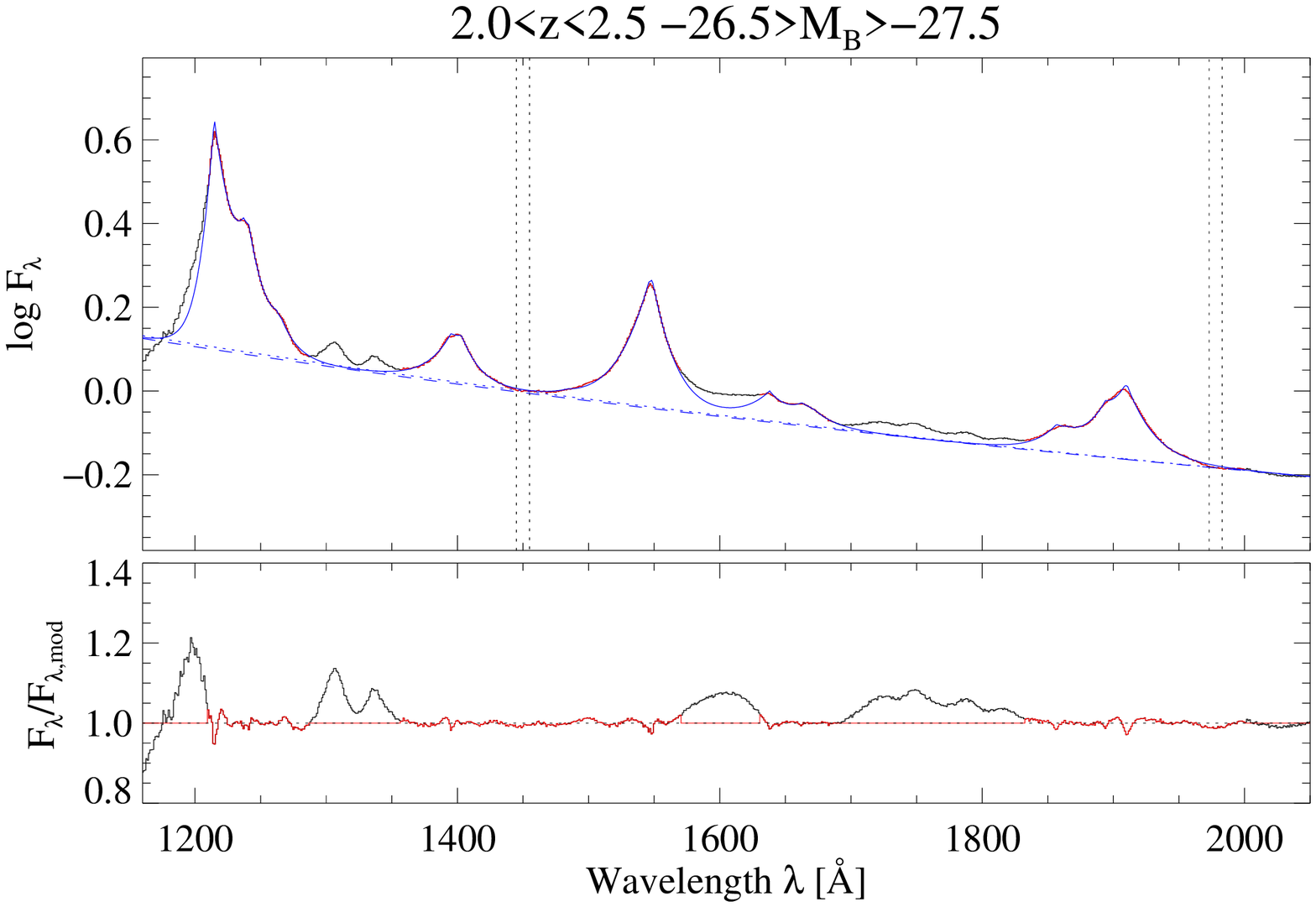}
\caption{
Same as Figure 1 but for the composite spectrum of
quasars with $-26.5 > M_B \geq -27.5$ and $2.0 \leq z < 2.5$.
}
\label{fig03}
\end{figure*}

\begin{figure*}
\centering
\includegraphics[width=13.5cm]{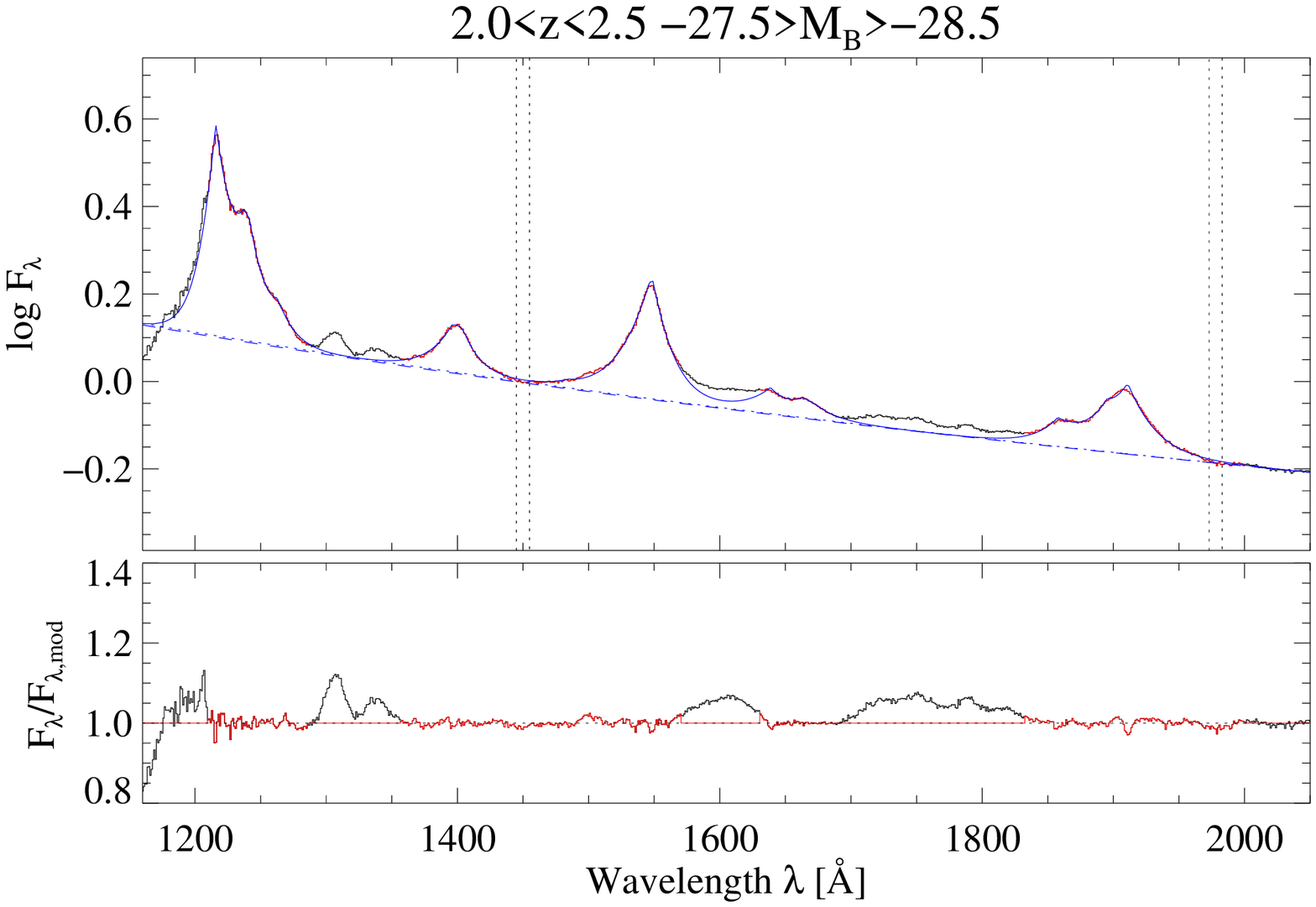}
\caption{
Same as Figure 1 but for the composite spectrum of
quasars with $-27.5 > M_B \geq -28.5$ and $2.0 \leq z < 2.5$.
}
\label{fig04}
\end{figure*}

\begin{figure*}
\centering
\includegraphics[width=13.5cm]{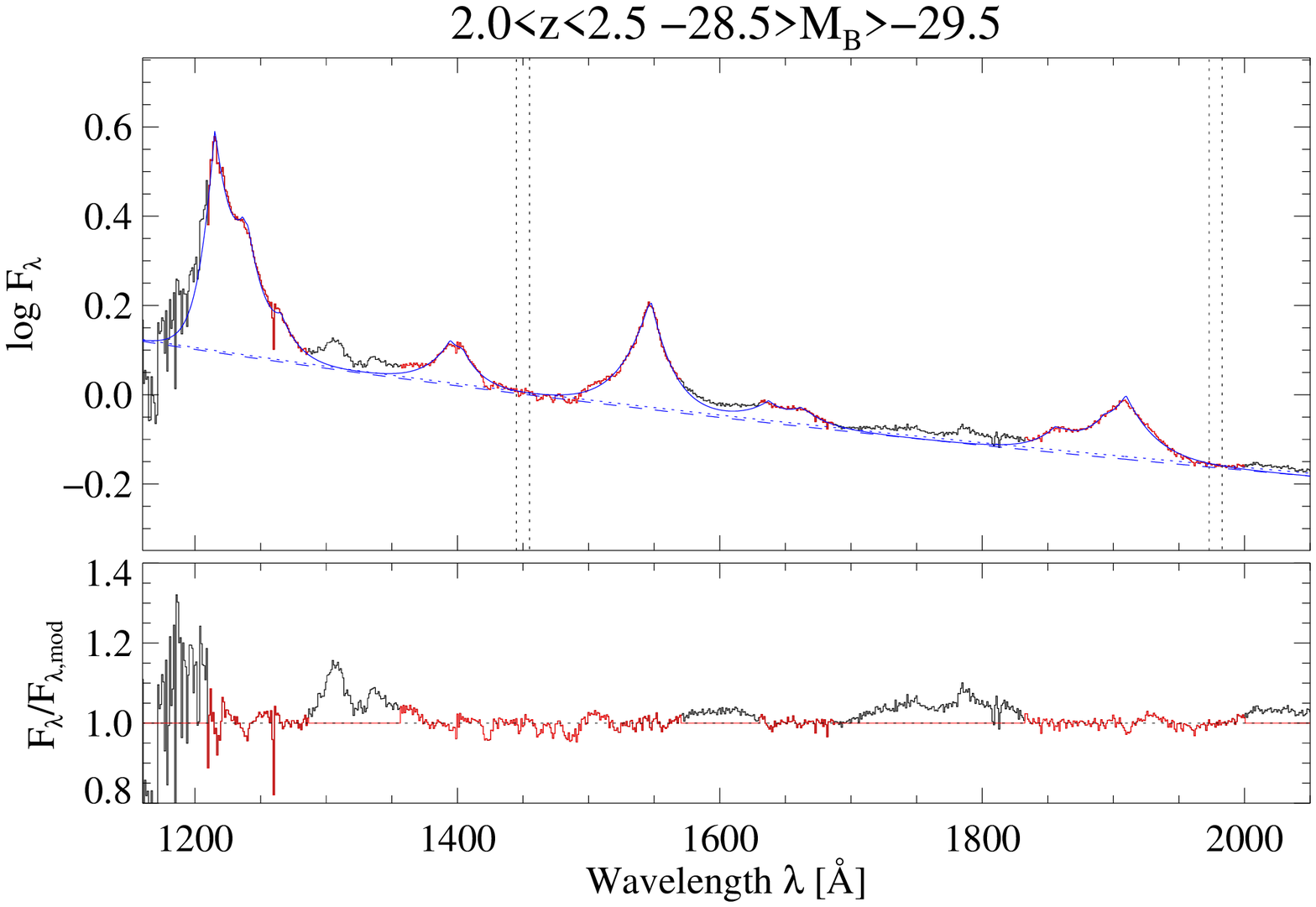}
\caption{
Same as Figure 1 but for the composite spectrum of
quasars with $-28.5 > M_B \geq -29.5$ and $2.0 \leq z < 2.5$.
}
\label{fig05}
\end{figure*}

\begin{figure*}
\centering
\includegraphics[width=13.5cm]{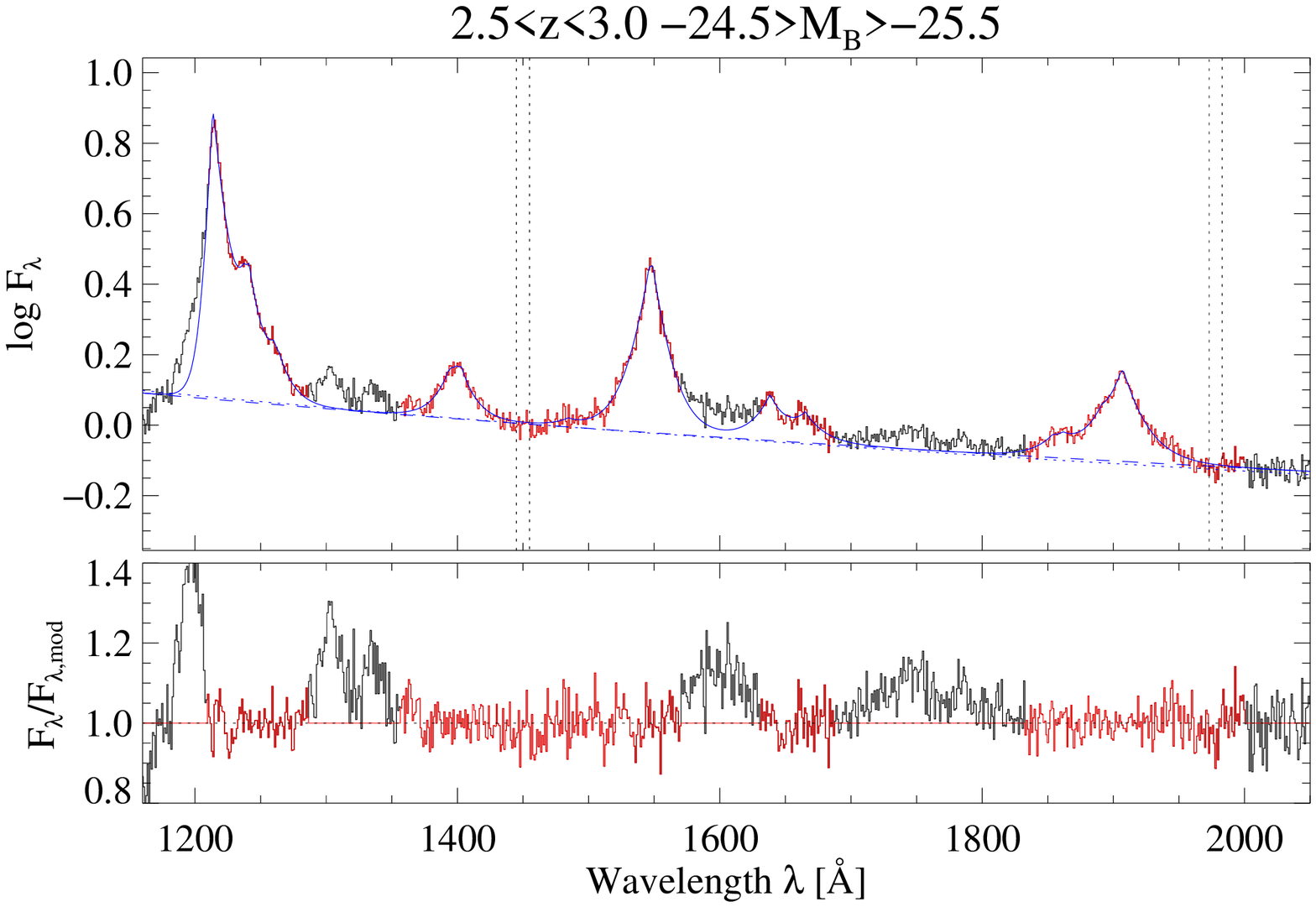}
\caption{
Same as Figure 1 but for the composite spectrum of
quasars with $-24.5 > M_B \geq -25.5$ and $2.5 \leq z < 3.0$.
}
\label{fig06}
\end{figure*}

\begin{figure*}
\centering
\includegraphics[width=13.5cm]{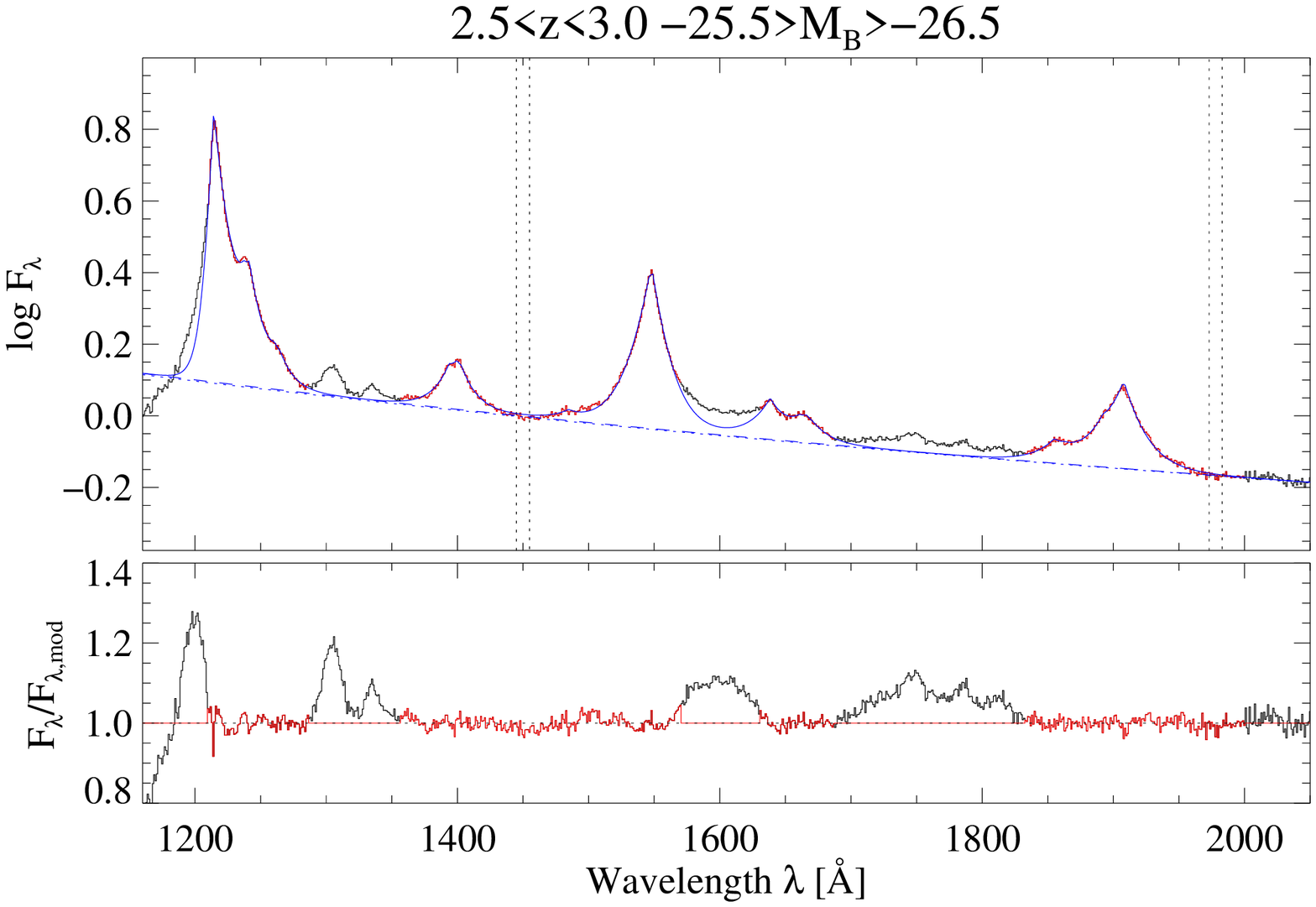}
\caption{
Same as Figure 1 but for the composite spectrum of
quasars with $-25.5 > M_B \geq -26.5$ and $2.5 \leq z < 3.0$.
}
\label{fig07}
\end{figure*}

\begin{figure*}
\centering
\includegraphics[width=13.5cm]{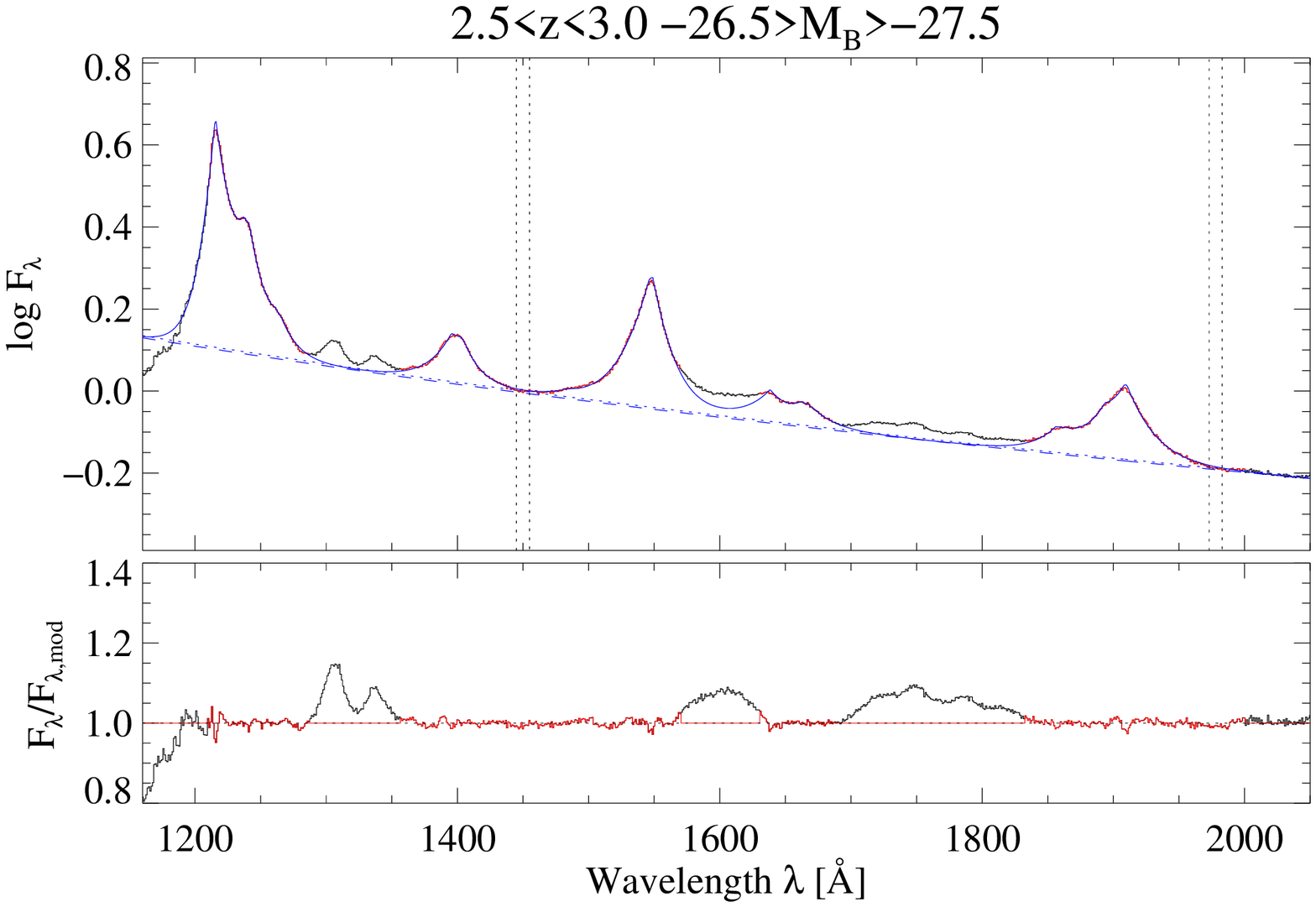}
\caption{
Same as Figure 1 but for the composite spectrum of
quasars with $-26.5 > M_B \geq -27.5$ and $2.5 \leq z < 3.0$.
}
\label{fig08}
\end{figure*}

\begin{figure*}
\centering
\includegraphics[width=13.5cm]{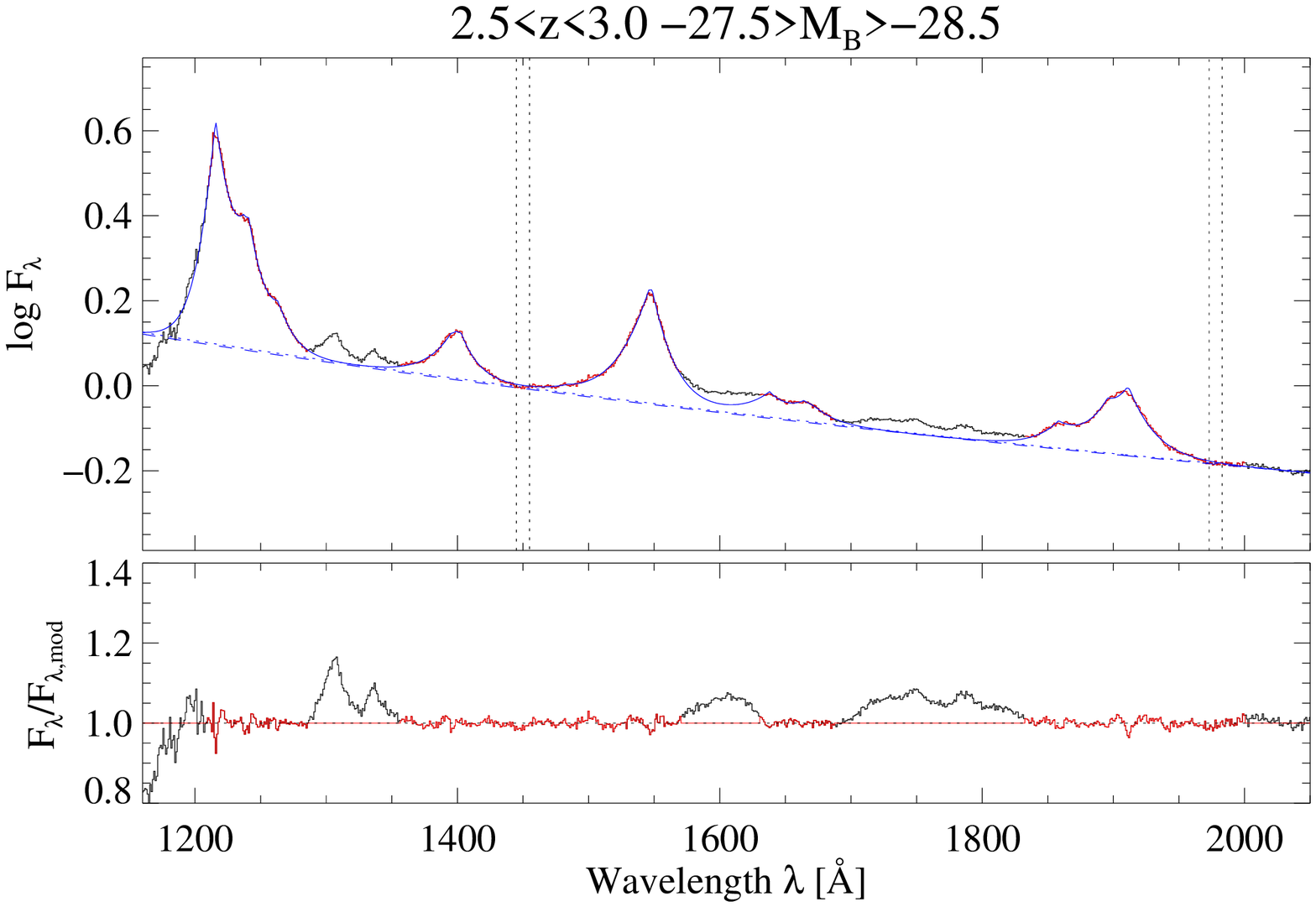}
\caption{
Same as Figure 1 but for the composite spectrum of
quasars with $-27.5 > M_B \geq -28.5$ and $2.5 \leq z < 3.0$.
}
\label{fig09}
\end{figure*}

\begin{figure*}
\center
\includegraphics[width=13.5cm]{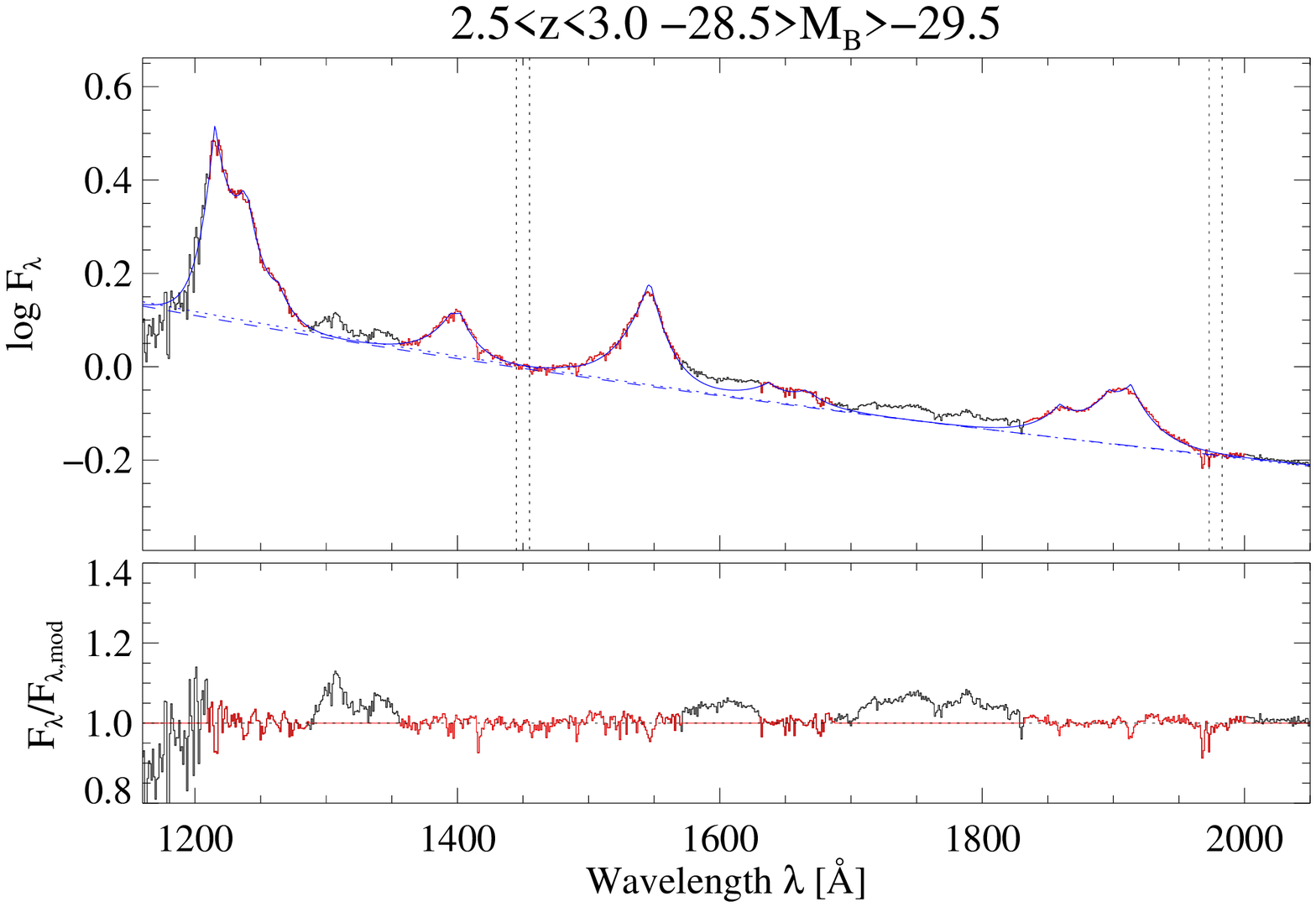}
\caption{
Same as Figure 1 but for the composite spectrum of
quasars with $-28.5 > M_B \geq -29.5$ and $2.5 \leq z < 3.0$.
}
\label{fig10}
\end{figure*}

\begin{figure*}
\centering
\includegraphics[width=13.5cm]{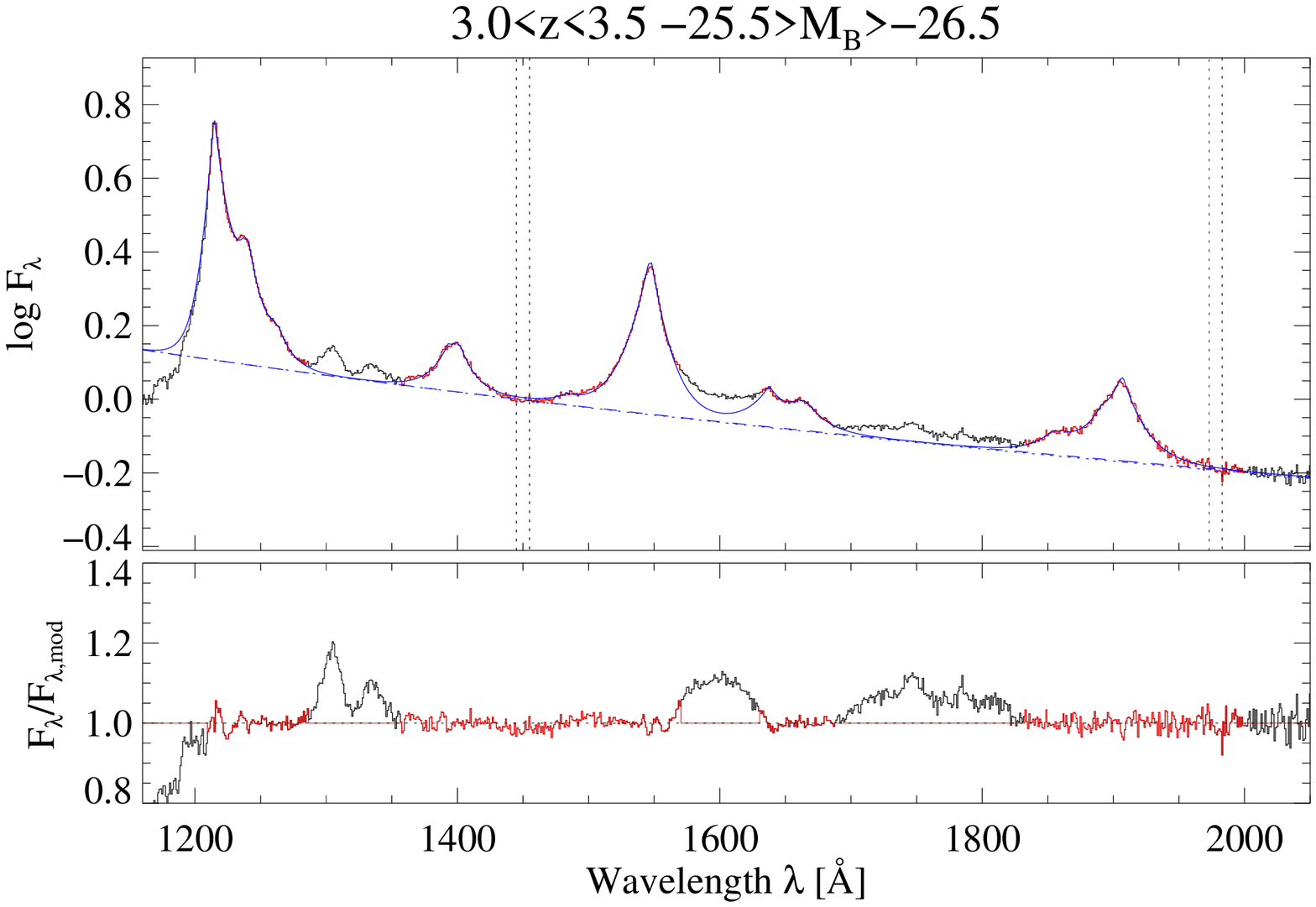}
\caption{
Same as Figure 1 but for the composite spectrum of
quasars with $-25.5 > M_B \geq -26.5$ and $3.0 \leq z < 3.5$.
}
\label{fig11}
\end{figure*}

\begin{figure*}
\centering
\includegraphics[width=13.5cm]{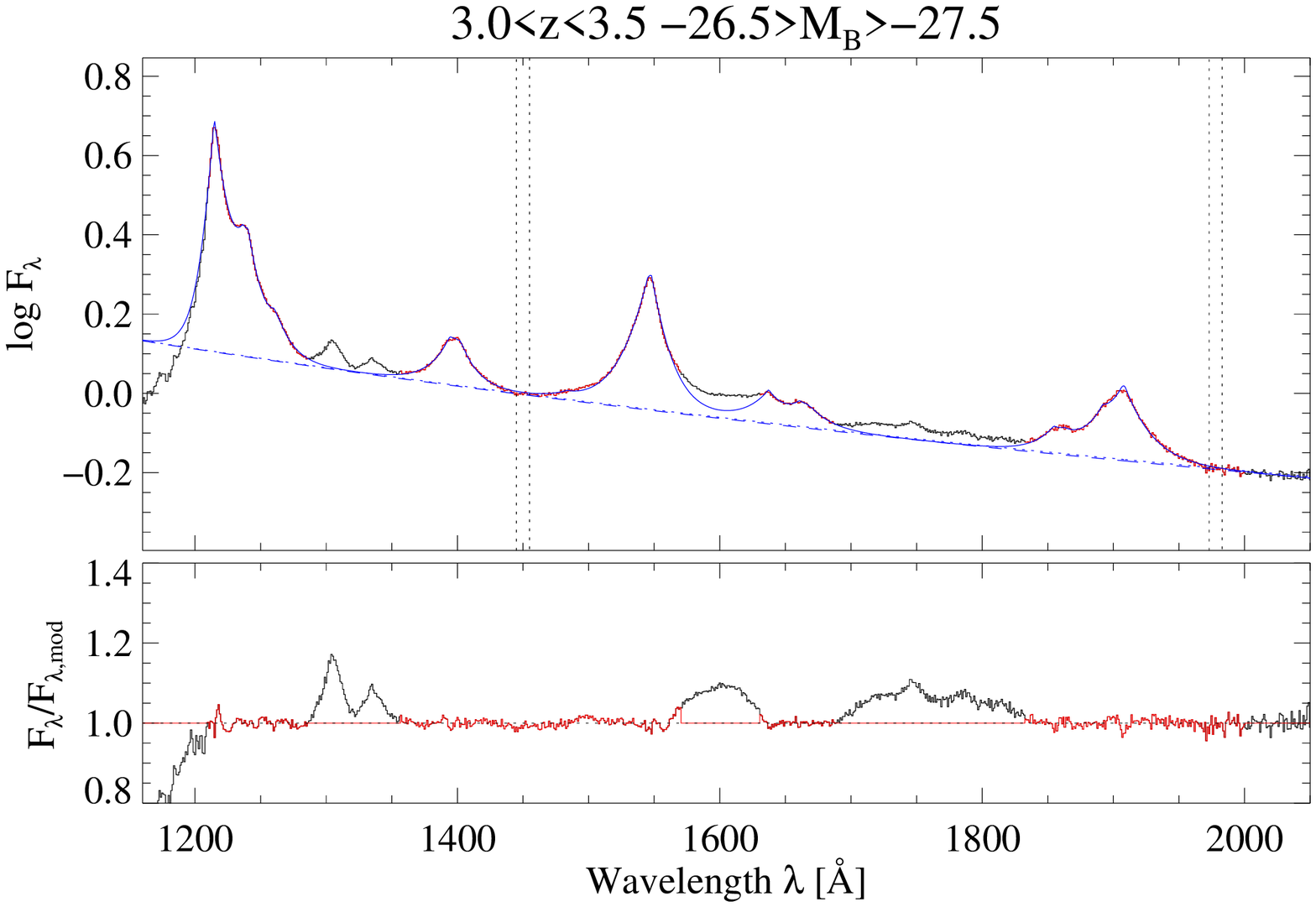}
\caption{
Same as Figure 1 but for the composite spectrum of
quasars with $-26.5 > M_B \geq -27.5$ and $3.0 \leq z < 3.5$.
}
\label{fig12}
\end{figure*}

\begin{figure*}
\centering
\includegraphics[width=13.5cm]{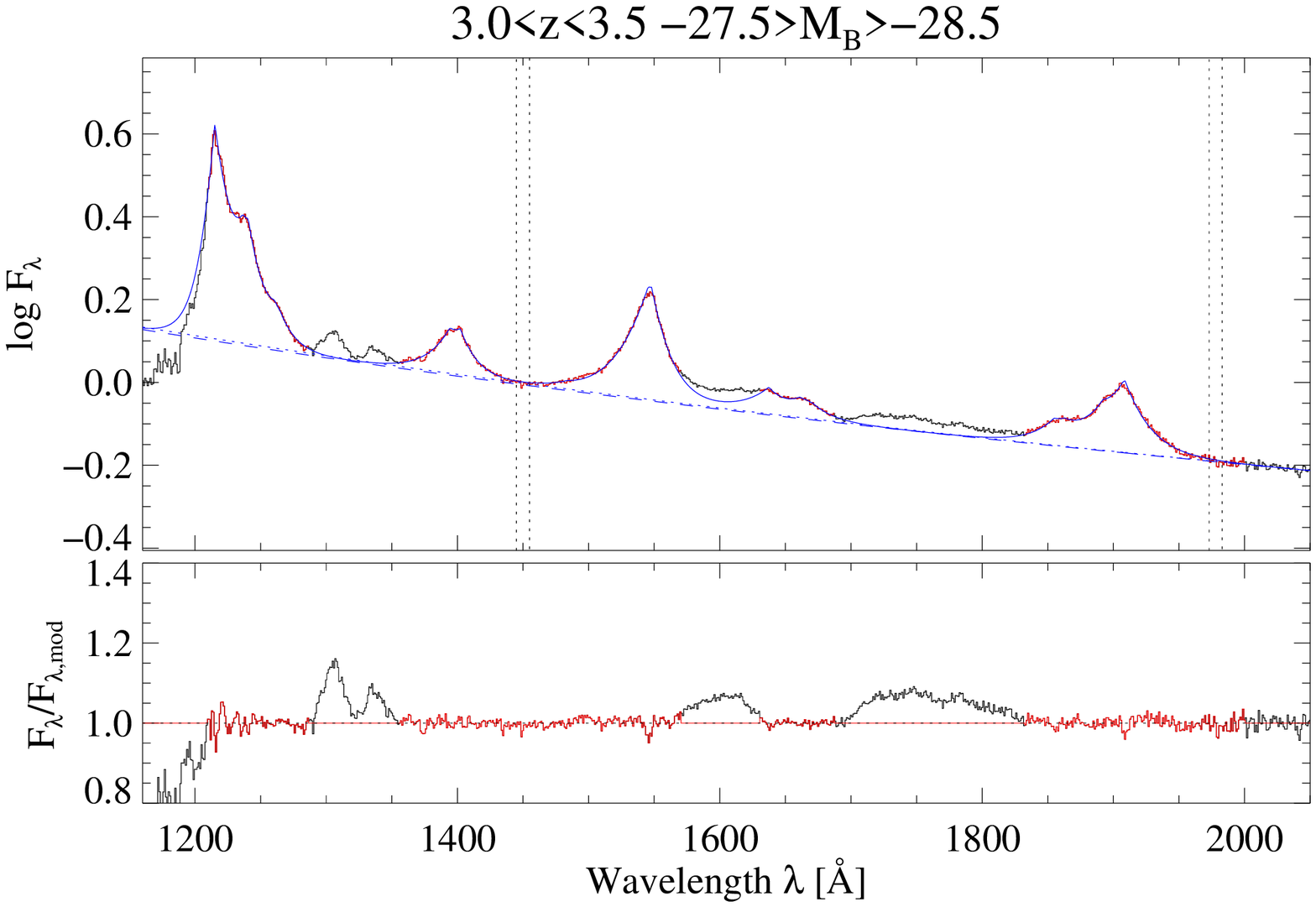}
\caption{
Same as Figure 1 but for the composite spectrum of
quasars with $-27.5 > M_B \geq -28.5$ and $3.0 \leq z < 3.5$.
}
\label{fig13}
\end{figure*}

\begin{figure*}
\centering
\includegraphics[width=13.5cm]{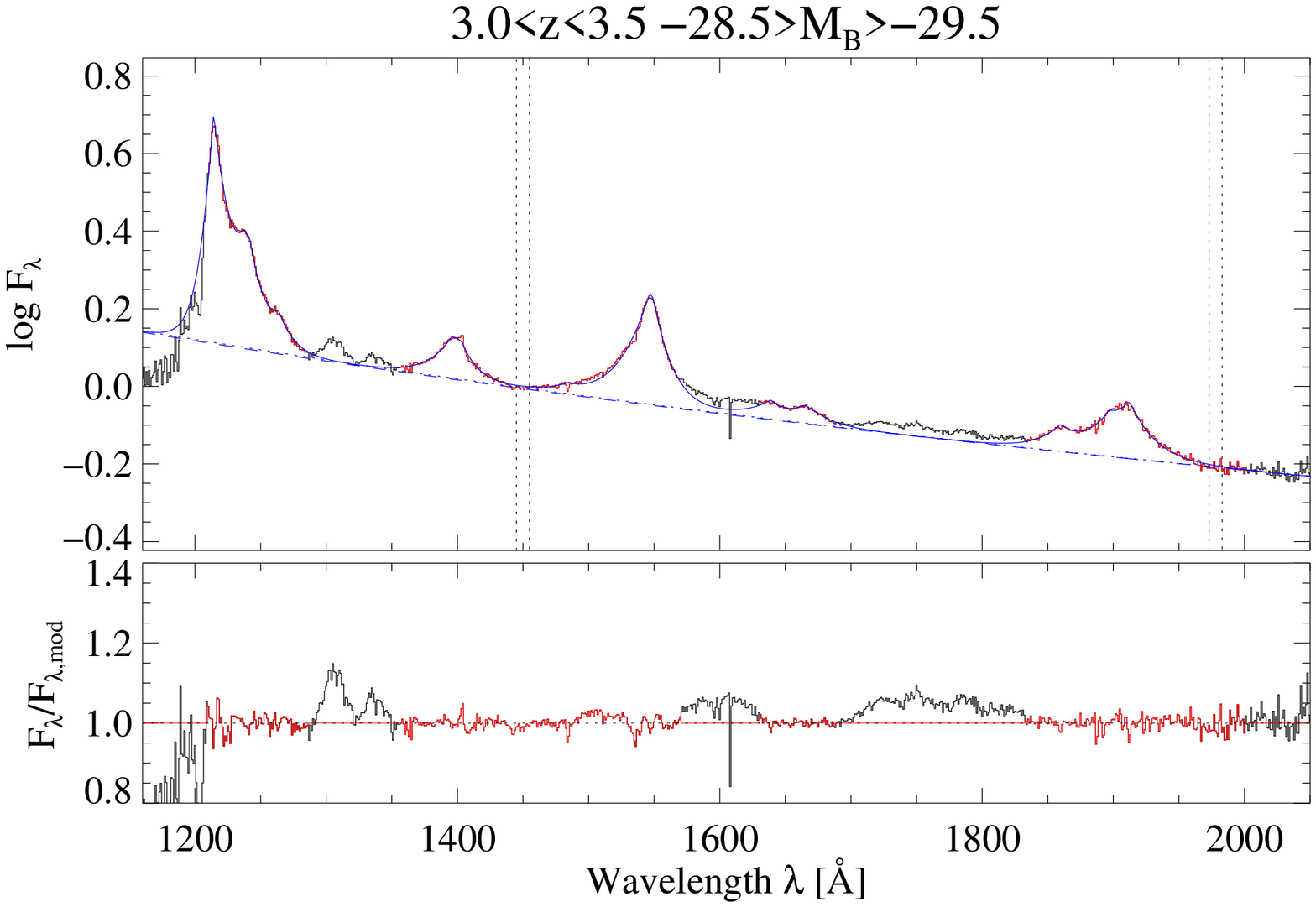}
\caption{
Same as Figure 1 but for the composite spectrum of
quasars with $-28.5 > M_B \geq -29.5$ and $3.0 \leq z < 3.5$.
}
\label{fig14}
\end{figure*}

\begin{figure*}
\centering
\includegraphics[width=13.5cm]{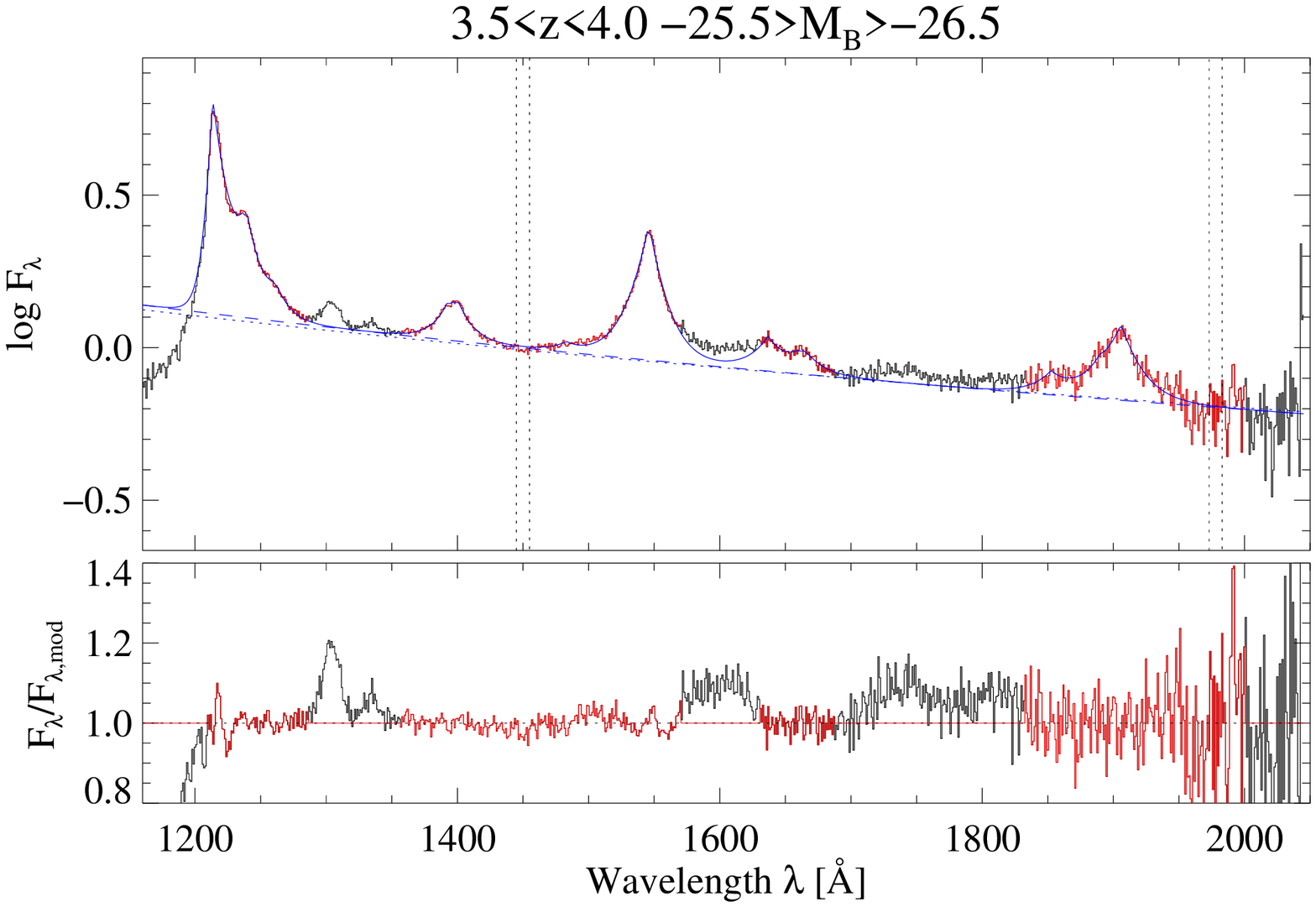}
\caption{
Same as Figure 1 but for the composite spectrum of
quasars with $-25.5 > M_B \geq -26.5$ and $3.5 \leq z < 4.0$.
}
\label{fig15}
\end{figure*}

\begin{figure*}
\centering
\includegraphics[width=13.5cm]{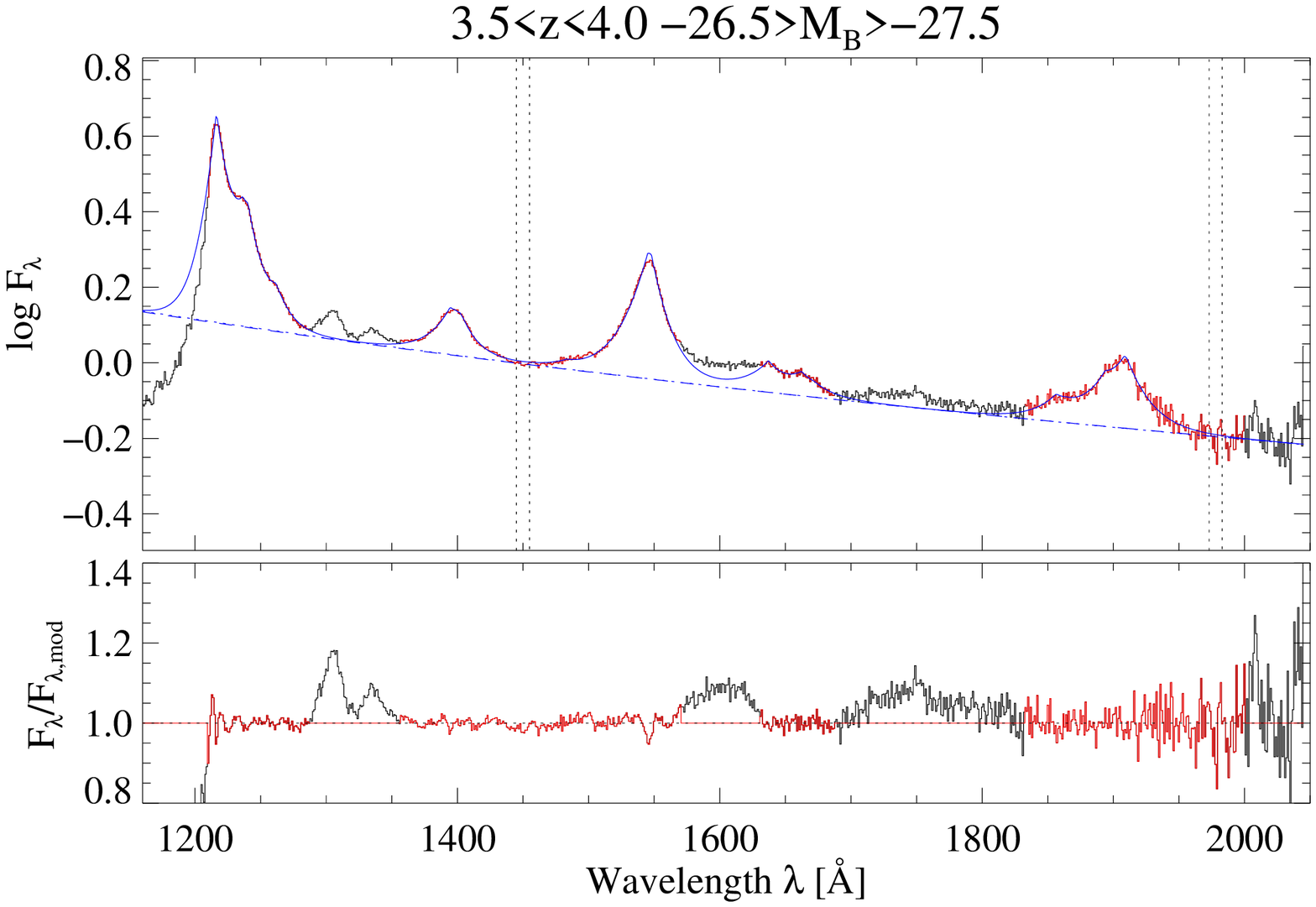}
\caption{
Same as Figure 1 but for the composite spectrum of
quasars with $-26.5 > M_B \geq -27.5$ and $3.5 \leq z < 4.0$.
}
\label{fig16}
\end{figure*}

\clearpage

\begin{figure*}
\centering
\includegraphics[width=13.5cm]{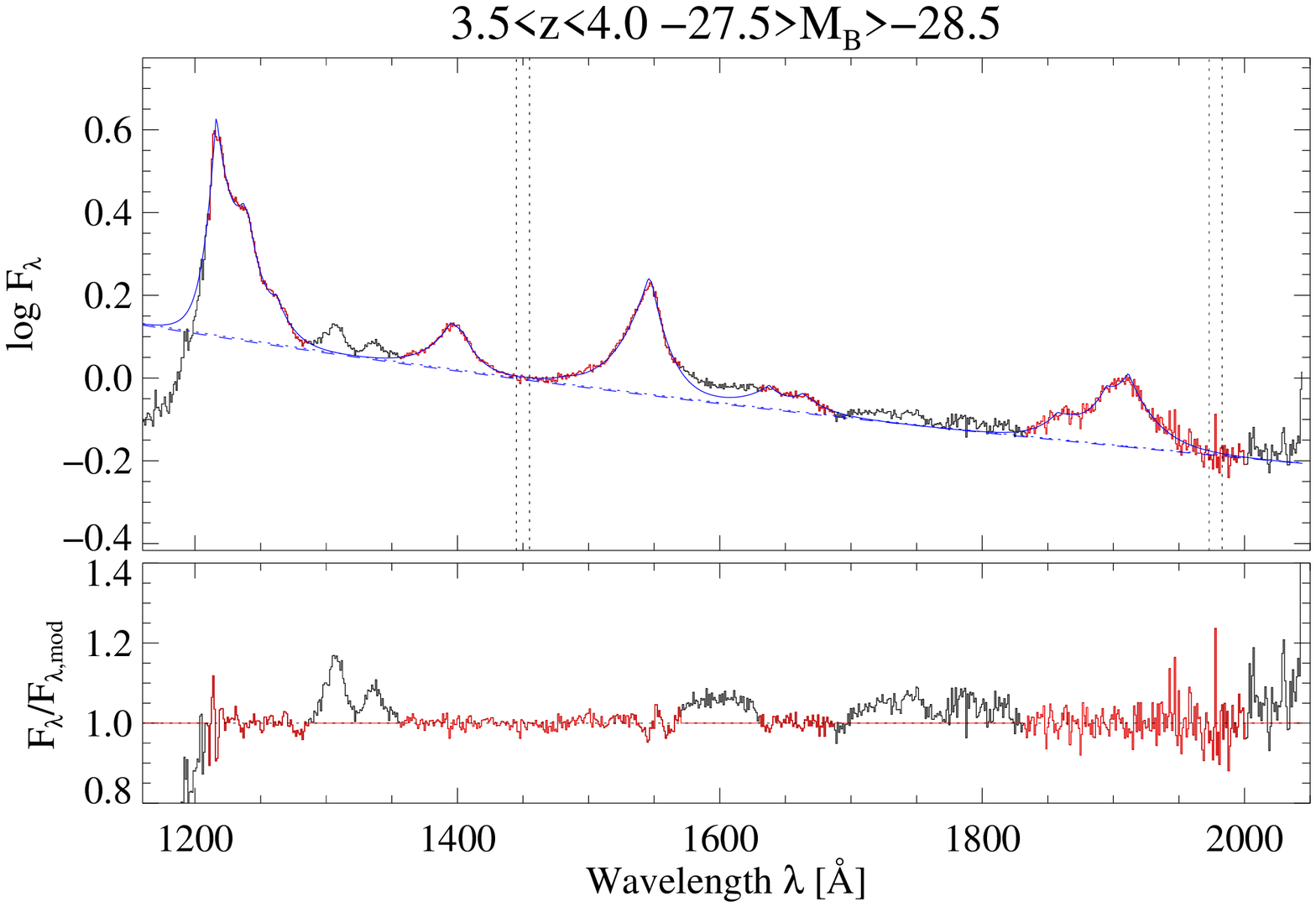}
\caption{
Same as Figure 1 but for the composite spectrum of
quasars with $-27.5 > M_B \geq -28.5$ and $3.5 \leq z < 4.0$.
}
\label{fig17}
\end{figure*}

\begin{figure*}
\centering
\includegraphics[width=13.5cm]{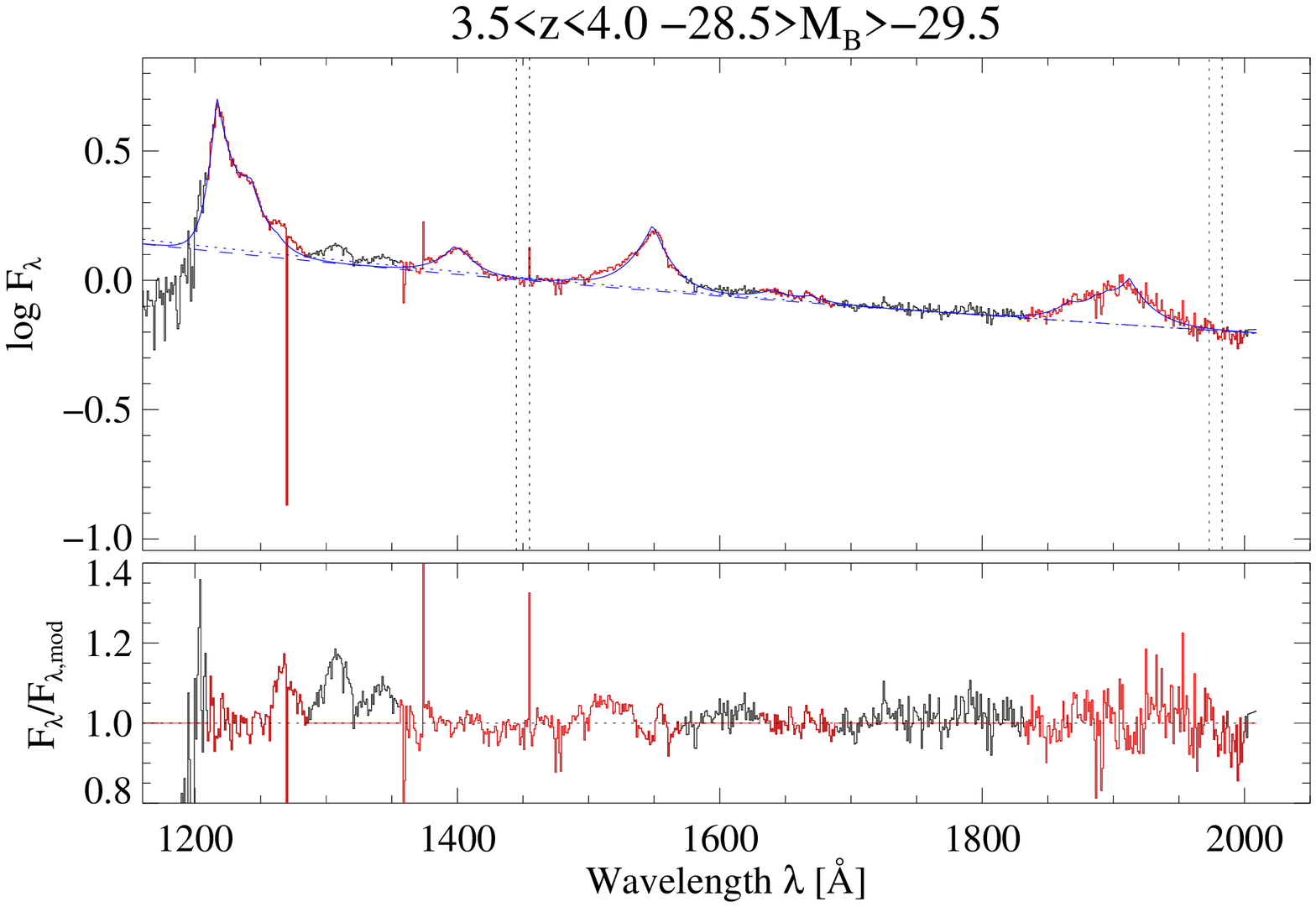}
\caption{
Same as Figure 1 but for the composite spectrum of
quasars with $-28.5 > M_B \geq -29.5$ and $3.5 \leq z < 4.0$.
}
\label{fig18}
\end{figure*}

\begin{figure*}
\centering
\includegraphics[width=13.5cm]{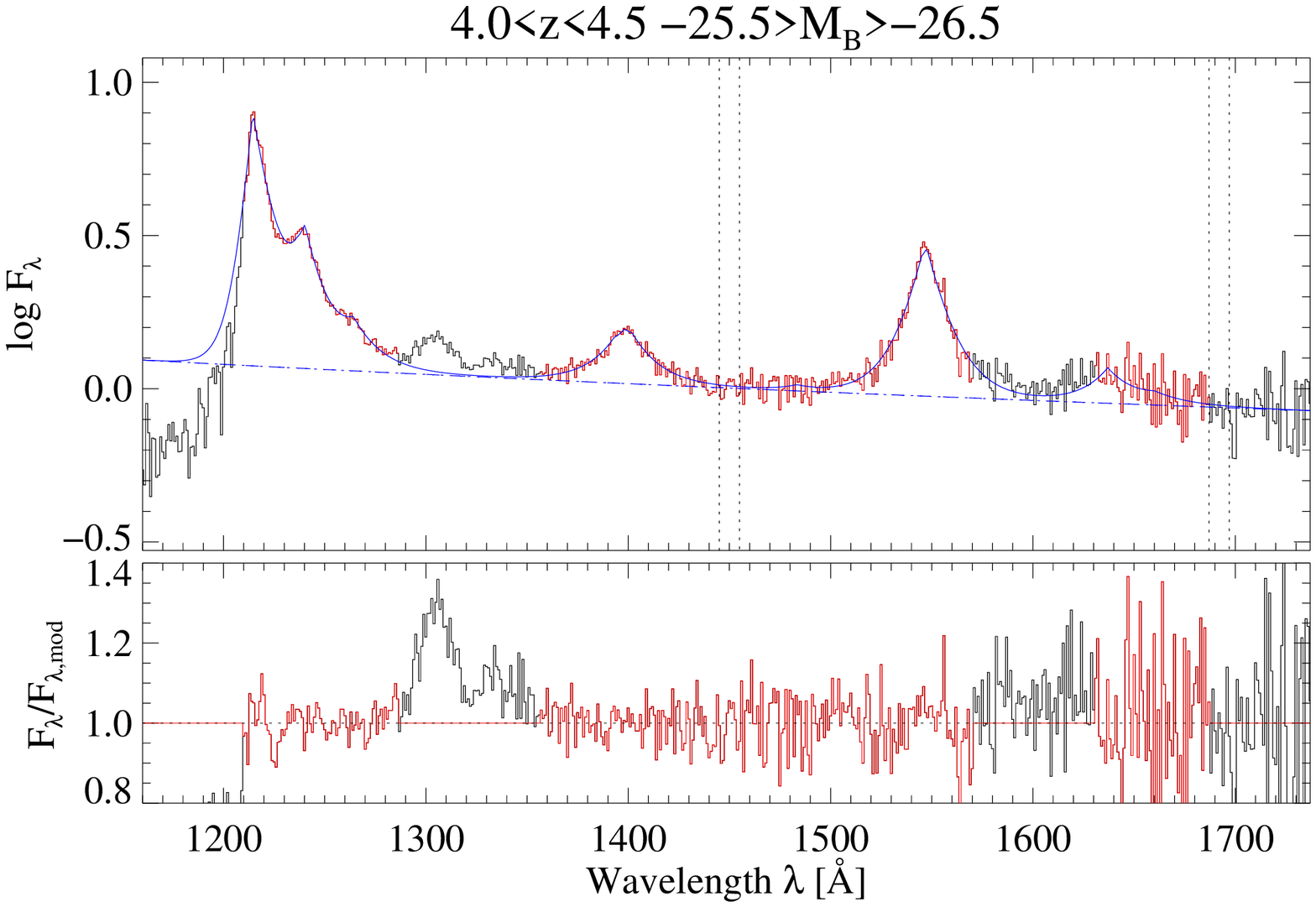}
\caption{
Same as Figure 1 but for the composite spectrum of
quasars with $-25.5 > M_B \geq -26.5$ and $4.0 \leq z < 4.5$.
}
\label{fig19}
\end{figure*}

\begin{figure*}
\centering
\includegraphics[width=13.5cm]{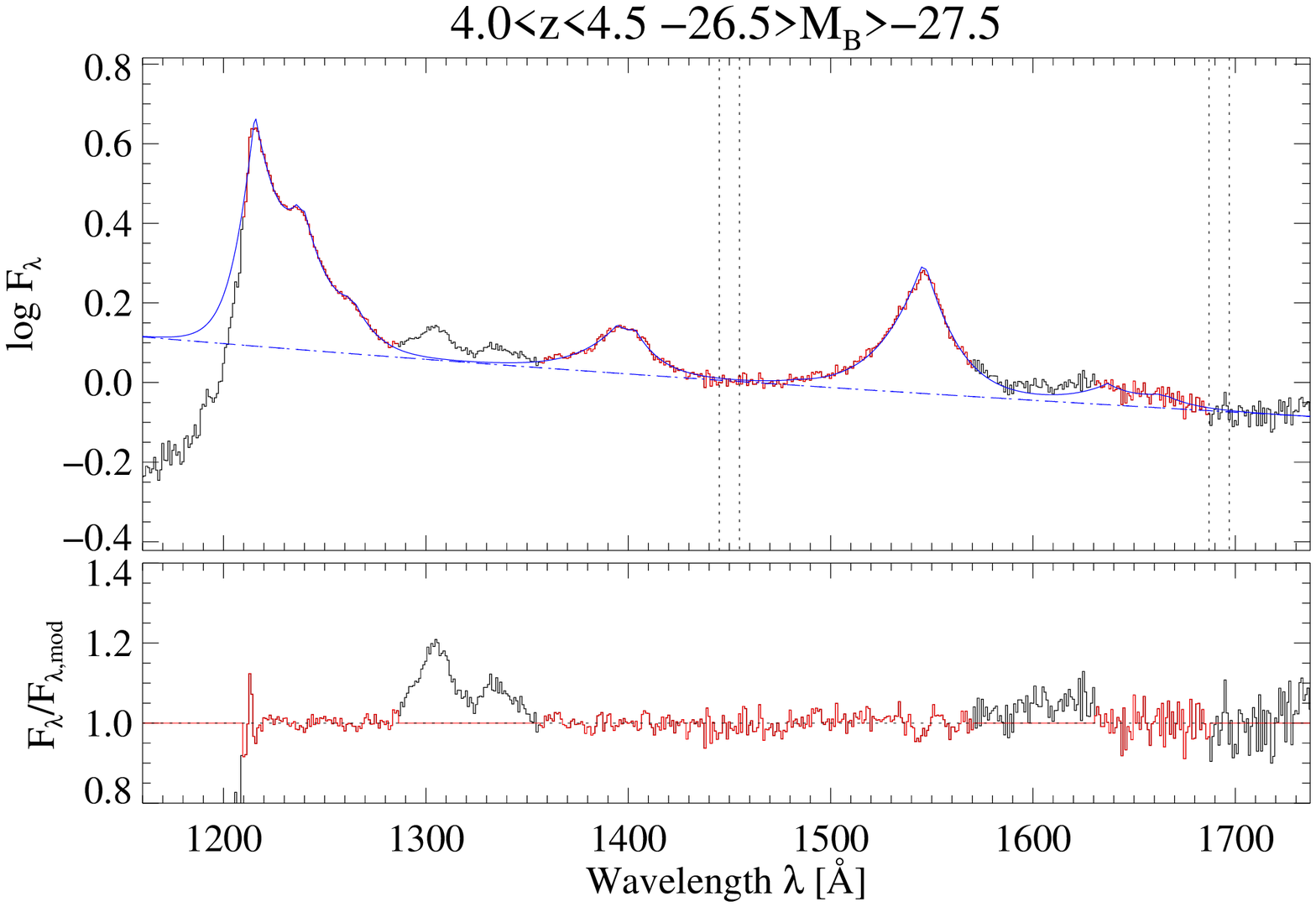}
\caption{
Same as Figure 1 but for the composite spectrum of
quasars with $-26.5 > M_B \geq -27.5$ and $4.0 \leq z < 4.5$.
}
\label{fig20}
\end{figure*}

\begin{figure*}
\centering
\includegraphics[width=13.5cm]{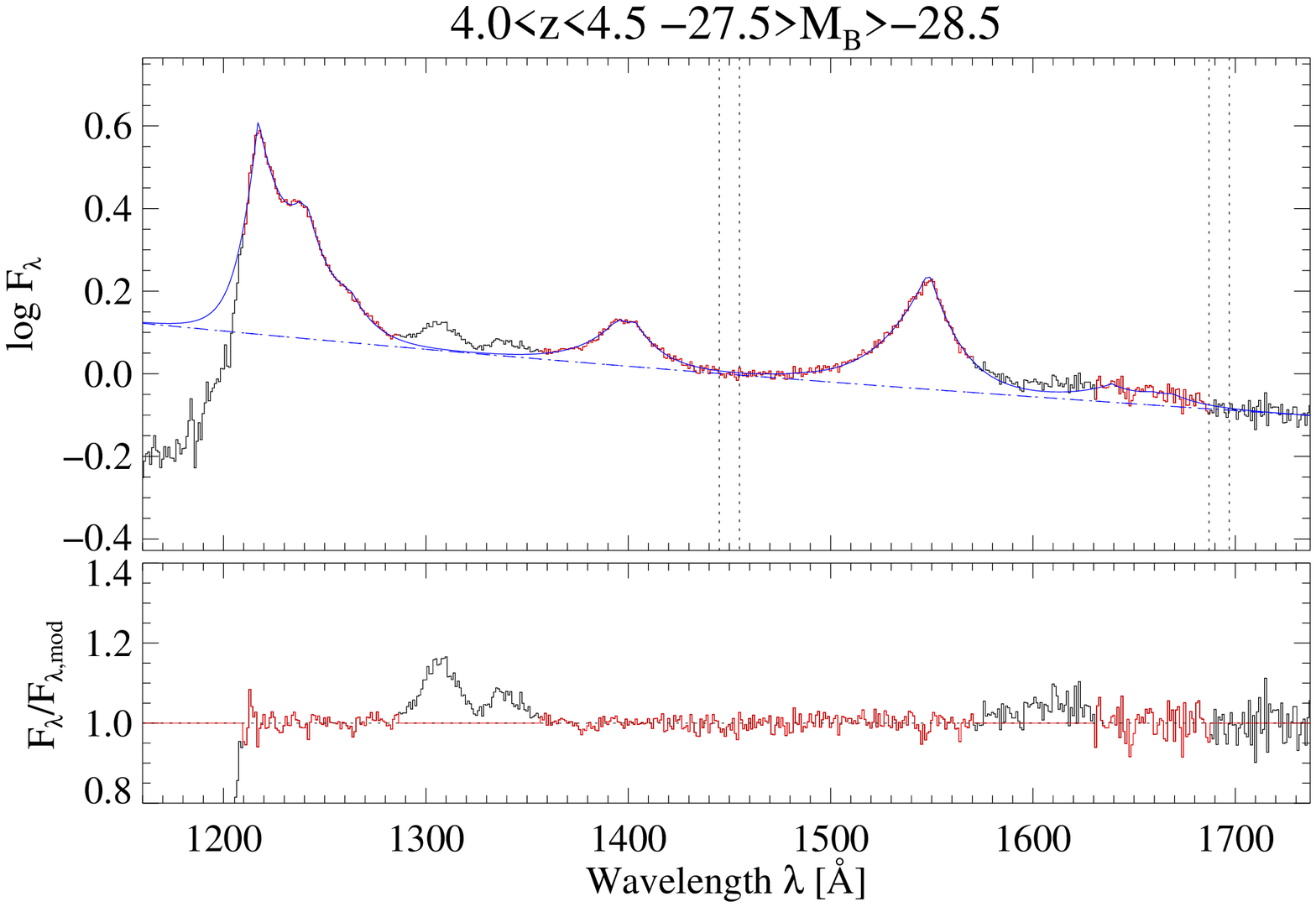}
\caption{
Same as Figure 1 but for the composite spectrum of
quasars with $-27.5 > M_B \geq -28.5$ and $4.0 \leq z < 4.5$.
}
\label{fig21}
\end{figure*}

\clearpage

\begin{figure*}
\centering
\vspace{2cm}
\rotatebox{-90}{\includegraphics[width=13.5cm]{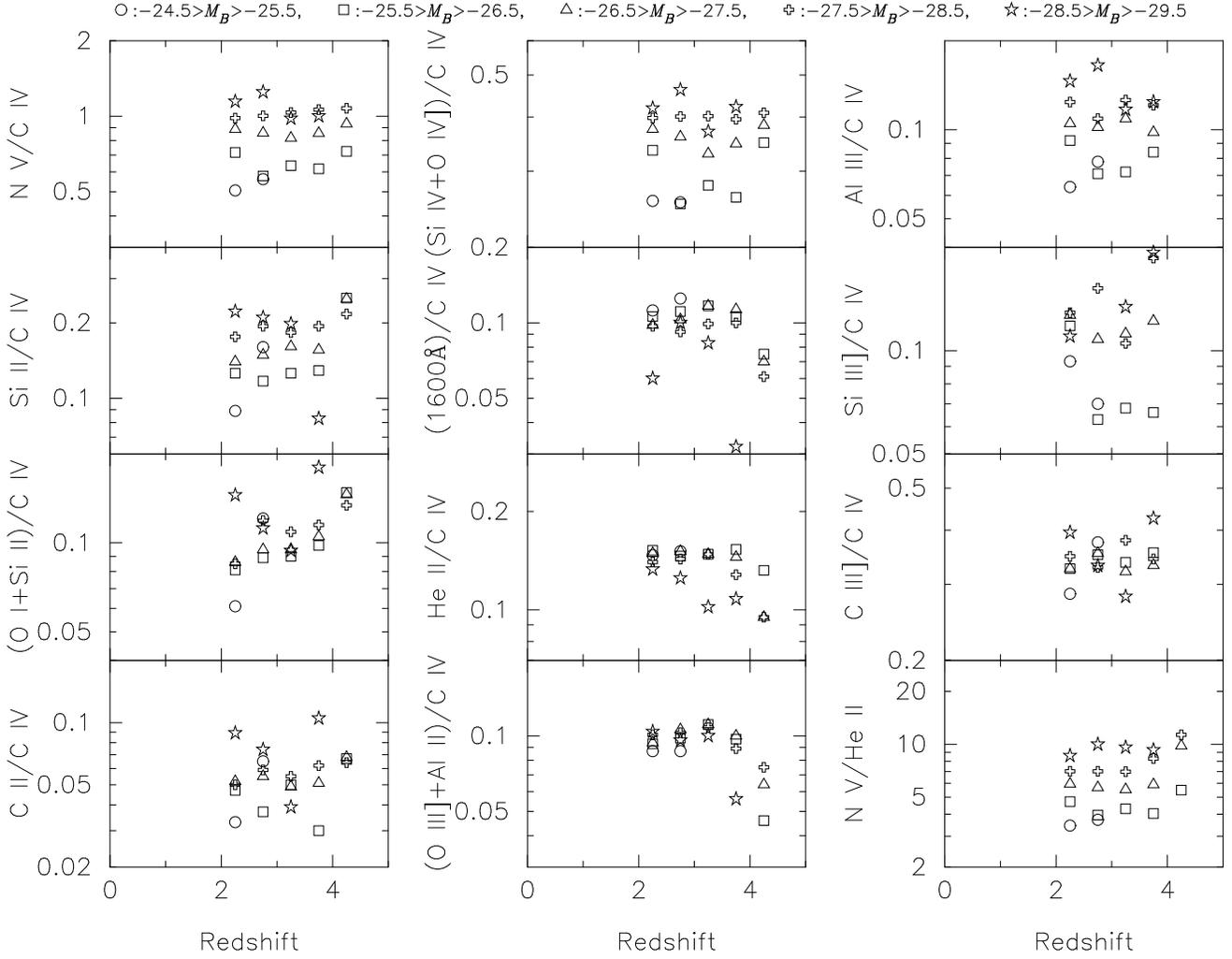}}
\caption{
Measured flux ratios as a function of redshift. Open circles,
squares, triangles, pluses, and stars denote the composite
spectra for $-24.5 > M_B \geq -25.5$, $-25.5 > M_B \geq -26.5$,
$-26.5 > M_B \geq -27.5$, $-27.5 > M_B \geq -28.5$, and
$-28.5 > M_B \geq -29.5$, respectively.
}
\label{fig22}
\end{figure*}

\clearpage

\begin{figure*}
\centering
\vspace{2cm}
\rotatebox{-90}{\includegraphics[width=13.5cm]{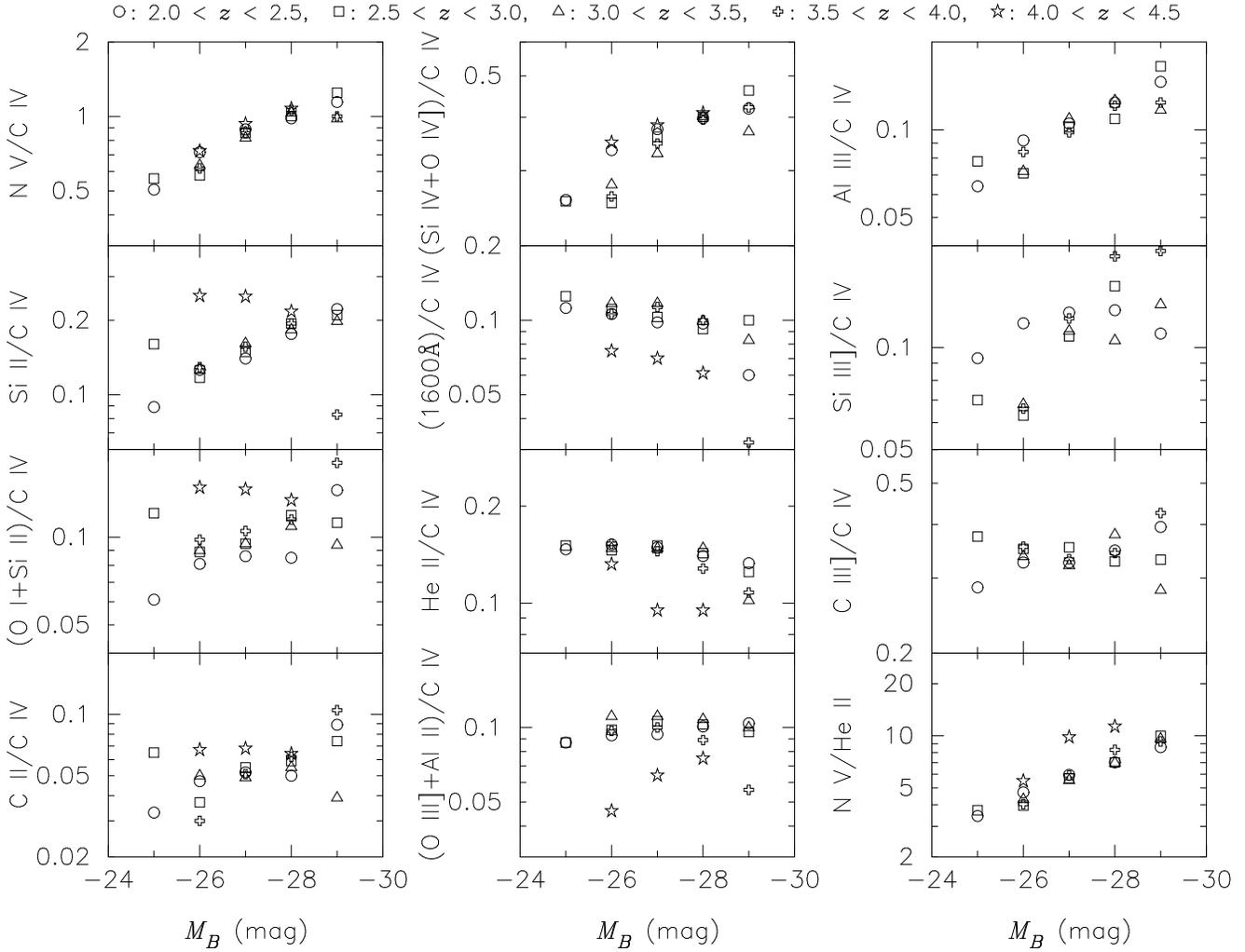}}
\caption{
Measured flux ratios as a function of absolute $B$ magnitude.
Open circles, squares, triangles, pluses, and stars denote the 
composite spectra for $2.0 \leq z < 2.5$, $2.5 \leq z < 3.0$,
$3.0 \leq z < 3.5$, $3.5 \leq z < 4.0$, and $4.0 \leq z < 4.5$,
respectively.
}
\label{fig23}
\end{figure*}

\clearpage

\begin{figure*}
\centering
\vspace{2cm}
\rotatebox{-90}{\includegraphics[width=15.5cm]{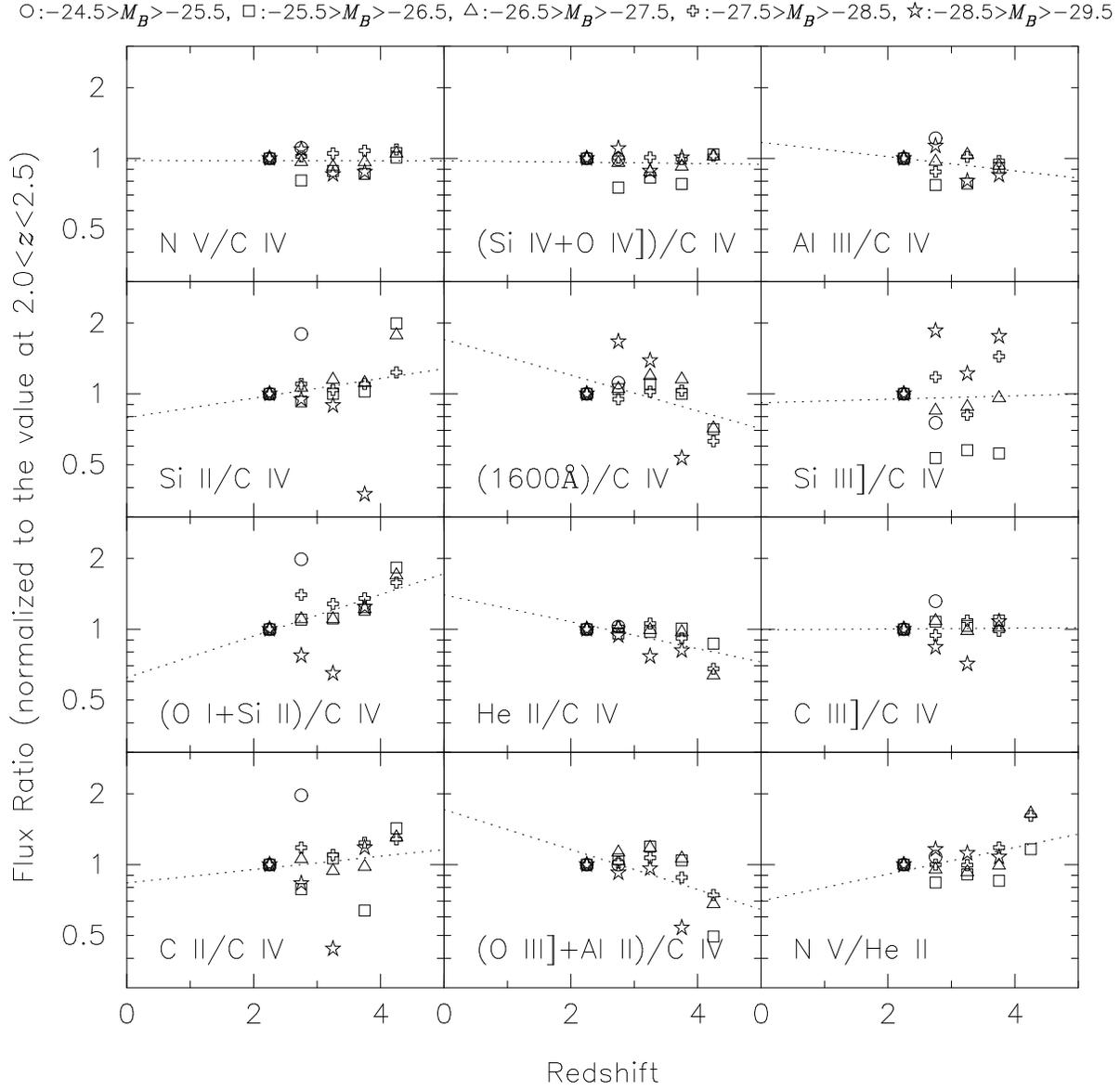}}
\caption{
Flux ratios normalized by the value measured on
composite spectra at $2.0 \leq z < 2.5$ for the individual
luminosity bins, as a function of redshift. Symbols are the
same as those in Figure 22. Results of the linear fitting are
shown by dotted lines.
}
\label{fig24}
\end{figure*}

\clearpage

\begin{figure*}
\centering
\vspace{2cm}
\rotatebox{-90}{\includegraphics[width=15.5cm]{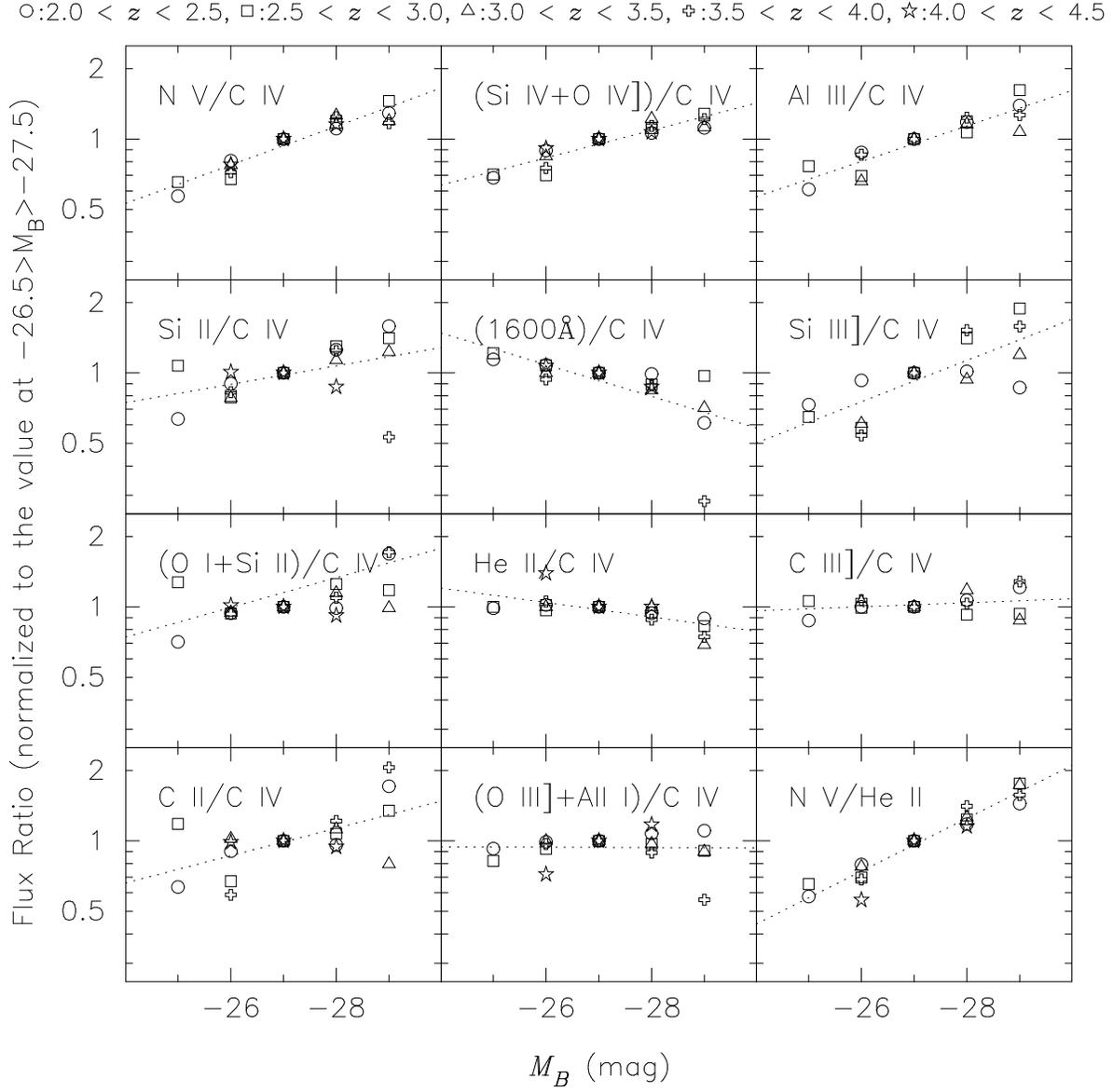}}
\caption{
Flux ratios normalized by the value measured on
composite spectra at $-26.5 > M_B \geq -27.5$ for the
individual redshift bins, as a function of absolute $B$ 
magnitude. Symbols are the same as those in Figure 23. 
Results of the linear fitting are
shown by dotted lines.
}
\label{fig25}
\end{figure*}

\clearpage

\begin{figure*}
\centering
\vspace{2cm}
\rotatebox{-90}{\includegraphics[width=14.5cm]{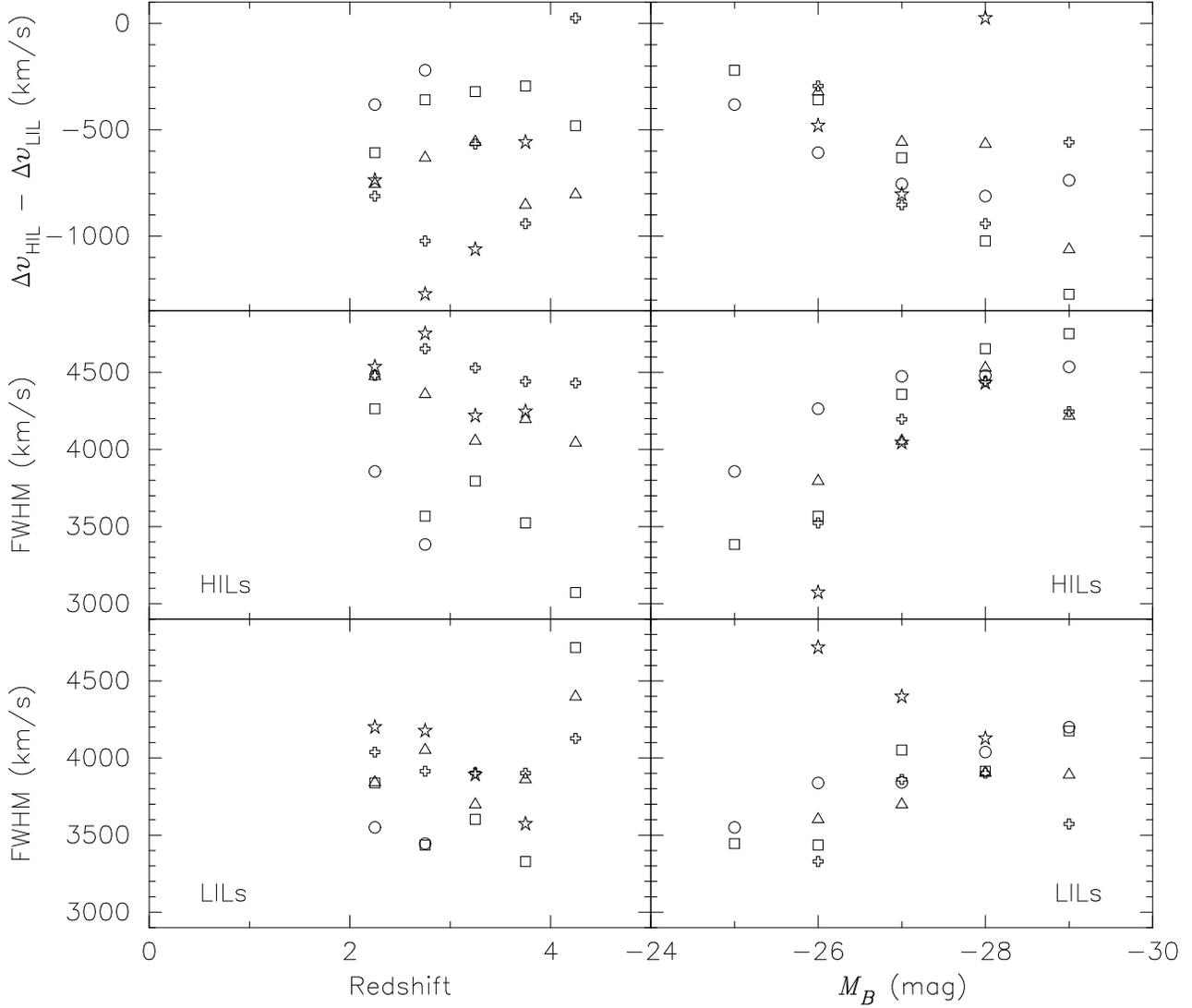}}
\caption{
Velocity shift of high-ionization lines (HILs) in relative to 
that of low-ionization lines (LILs)
(upper panels), FWHM of HILs (middle panels) and LILs
(lower panels), as functions
of redshift (left panels) and absolute $B$ magnitude (right panels).
In the left panel, open circles, squares, triangles, pluses, and 
stars denote the data for $-24.5 > M_B \geq -25.5$, 
$-25.5 > M_B \geq -26.5$, $-26.5 > M_B \geq -27.5$, 
$-27.5 > M_B \geq -28.5$, and $-28.5 > M_B \geq -29.5$, 
respectively. In the right panel, open circles, squares, 
triangles, pluses, and stars denote the data for 
$2.0 \leq z < 2.5$, $2.5 \leq z < 3.0$,
$3.0 \leq z < 3.5$, $3.5 \leq z < 4.0$, and $4.0 \leq z < 4.5$,
respectively. 
}
\label{fig26}
\end{figure*}

\clearpage

\begin{figure*}
\centering
\rotatebox{-90}{\includegraphics[width=6.0cm]{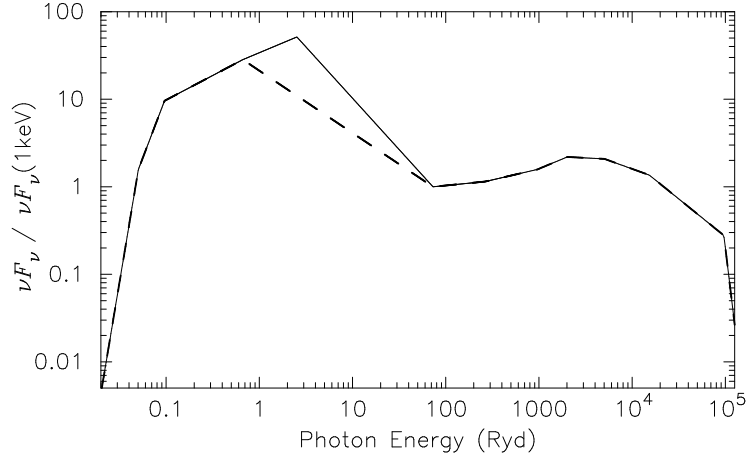}}
\caption{
Adopted SEDs for our photoionization model calculations.
Thin solid line denotes a SED with a strong UV bump and
thick dashed line denotes a SED with a weak UV bump.
These SEDs are normalized to the flux at 1 keV ($\simeq$73 Ryd).
}
\label{fig27}
\end{figure*}

\begin{figure*}
\centering
\rotatebox{-90}{\includegraphics[width=13.5cm]{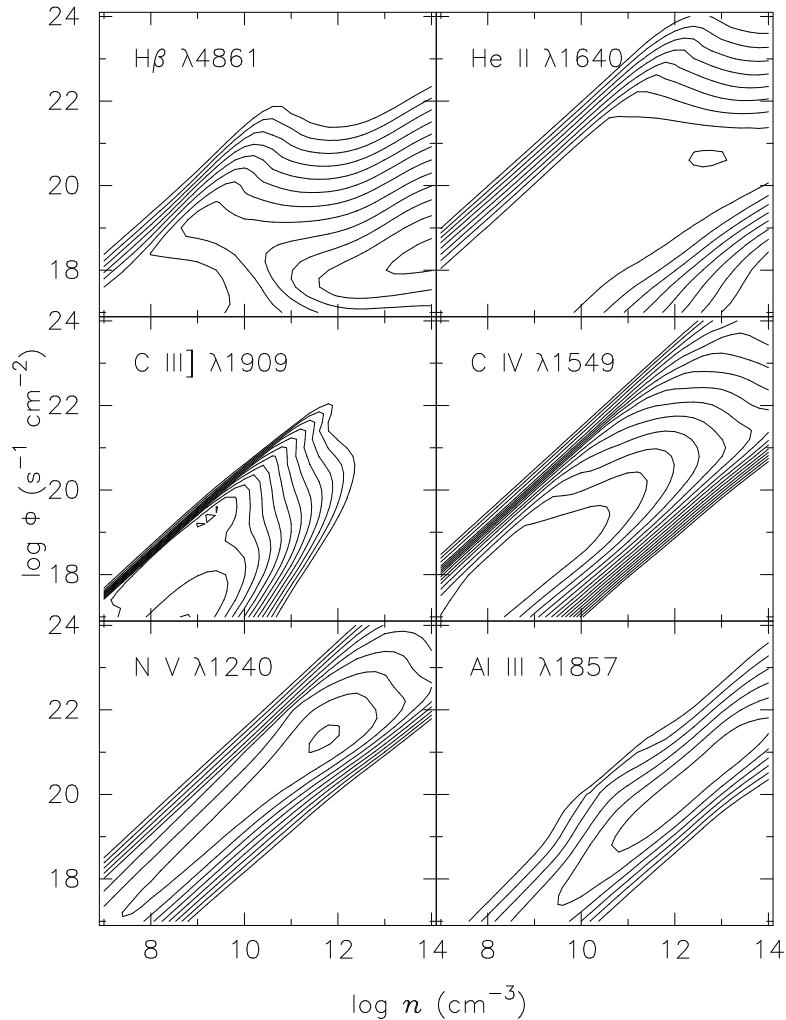}}
\caption{
Results of model calculations for 
H$\beta$ $\lambda$4861,
He{\sc ii}$\lambda$1640, C{\sc iii}]$\lambda$1909,
C{\sc iv}$\lambda$1640, N{\sc v}$\lambda$1240 and 
Al{\sc iii}$\lambda$1857. The equivalent widths referenced to
the incident continuum at 1215${\rm \AA}$ are displayed as 
contours. The smallest decade contour corresponds to 1${\rm \AA}$,
and the contour step size is 0.2 dex in logscale. Only models with 
an input SED with a large UV thermal bump and a metallicity 
of $Z/Z_\odot = 1.0$ are shown as examples.
}
\label{fig28}
\end{figure*}

\clearpage

\begin{figure*}
\centering
\rotatebox{-90}{\includegraphics[width=10cm]{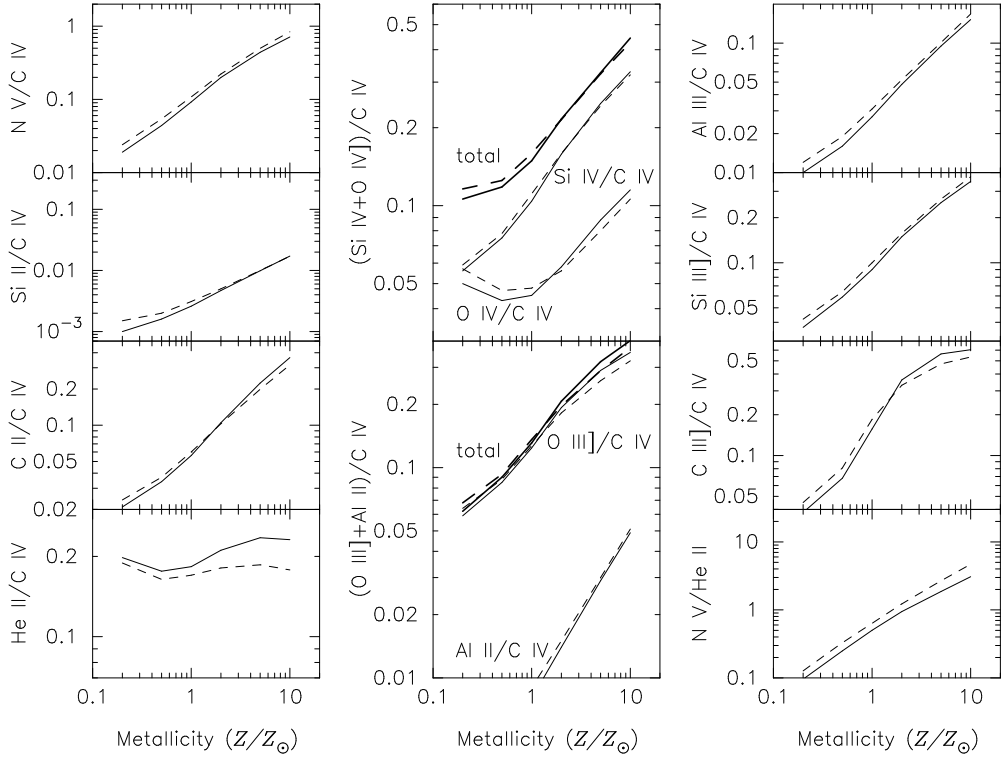}}
\caption{
Theoretical emission-line flux ratios calculated by the LOC
photoionization model, as a function of gas metallicity. 
Solid and dashed lines denote the
models with a large and a small UV thermal bump.
For blended lines, in addition to the individual emission
lines, their total flux are drawn with
thick lines.
}
\label{fig29}
\end{figure*}

\begin{figure*}
\centering
\rotatebox{90}{\includegraphics[width=9cm]{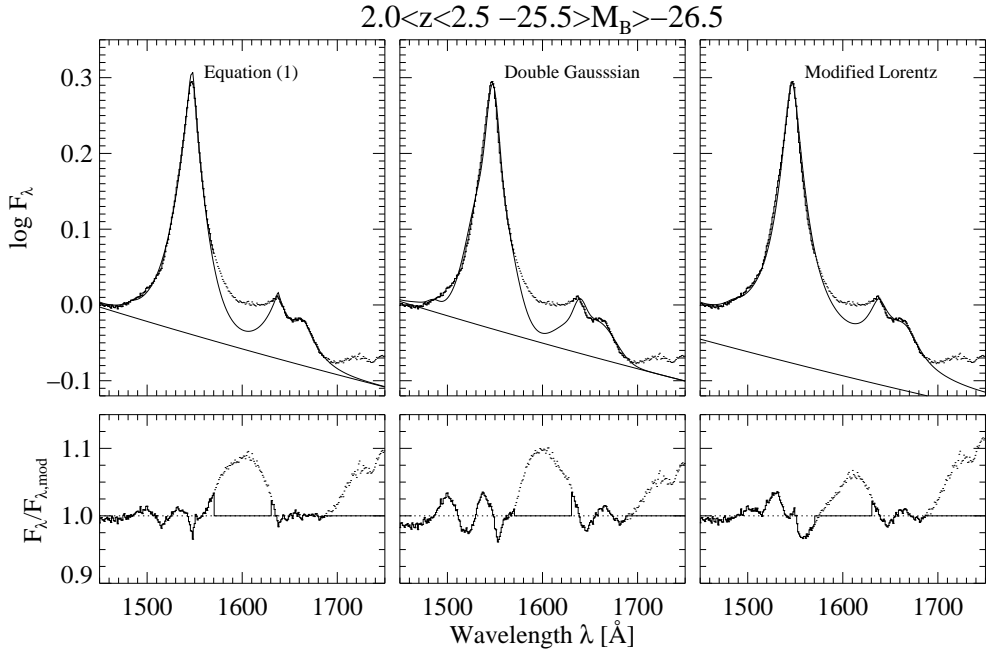}}
\caption{
Spectral fitting by adopting equation (1) ($left$),
double Gaussian ($middle$) and modified Lorentzian ($right$),
for emission lines. The composite spectrum and the model fit
are shown in upper panels while the residual spectrum is shown
in lower panels. The presented composite spectrum is
for quasars at $2.0 \leq z < 2.5$ and $-26.5 > M_B \geq -27.5$.
}
\label{fig30}
\end{figure*}

\clearpage

\begin{figure*}
\centering
\rotatebox{90}{\includegraphics[width=10cm]{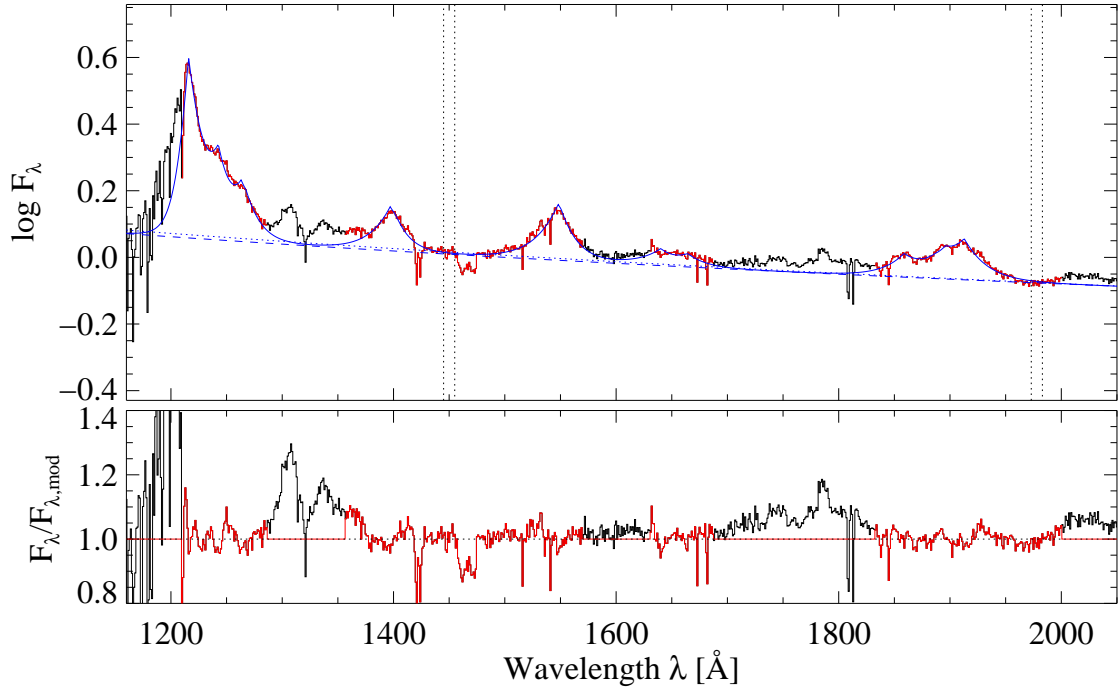}}
\caption{
Same as Figure 1 but for the individual spectrum of 
SDSS J085417.6+532735 ($z=2.42$, $M_B = -28.6$), not for 
quasar composite spectrum.
}
\label{fig31}
\end{figure*}

\begin{figure*}
\centering
\rotatebox{90}{\includegraphics[width=10cm]{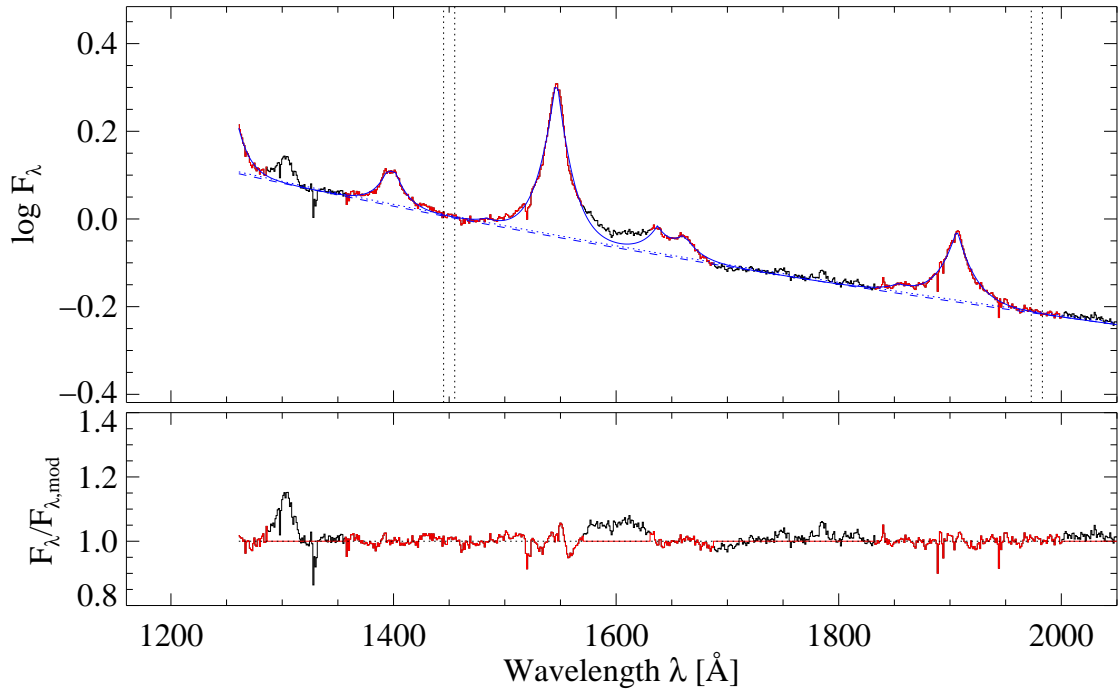}}
\caption{
Same as Figure 31 but for 
SDSS J080342.0+302254 ($z=2.03$, $M_B = -28.9$).
}
\label{fig32}
\end{figure*}

\begin{figure*}
\centering
\rotatebox{90}{\includegraphics[width=10cm]{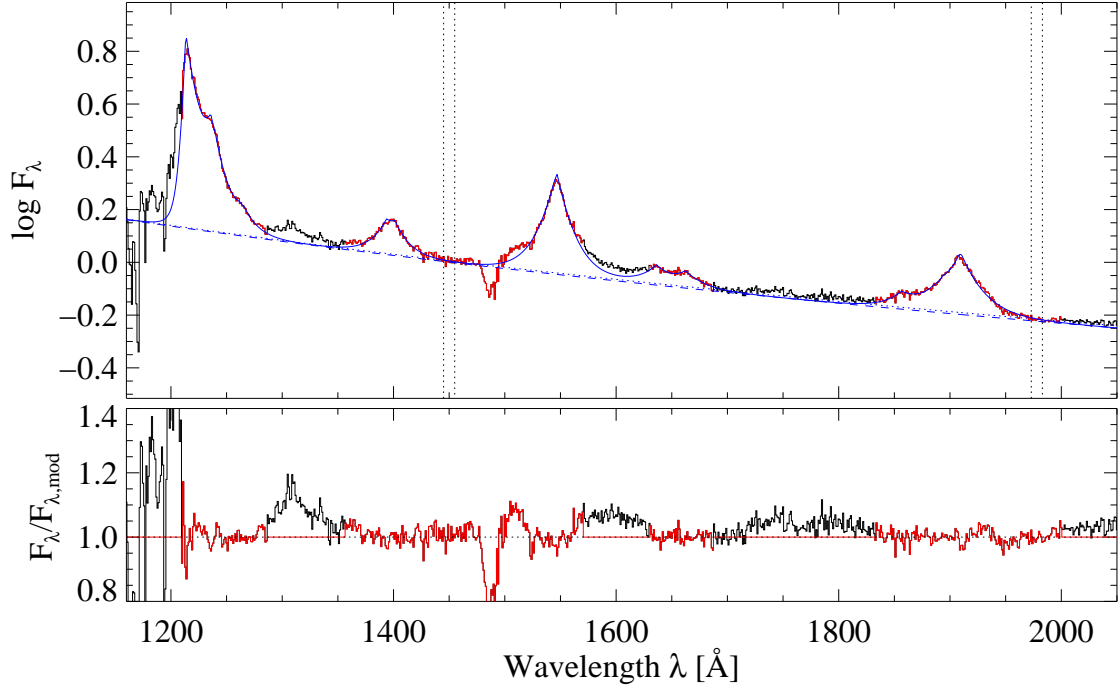}}
\caption{
Same as Figure 31 but for
SDSS J154359.4+535903 ($z=2.37$, $M_B = -28.5$).
}
\label{fig33}
\end{figure*}

\begin{figure*}
\centering
\vspace{0.5cm}
\rotatebox{-90}{\includegraphics[width=7.0cm]{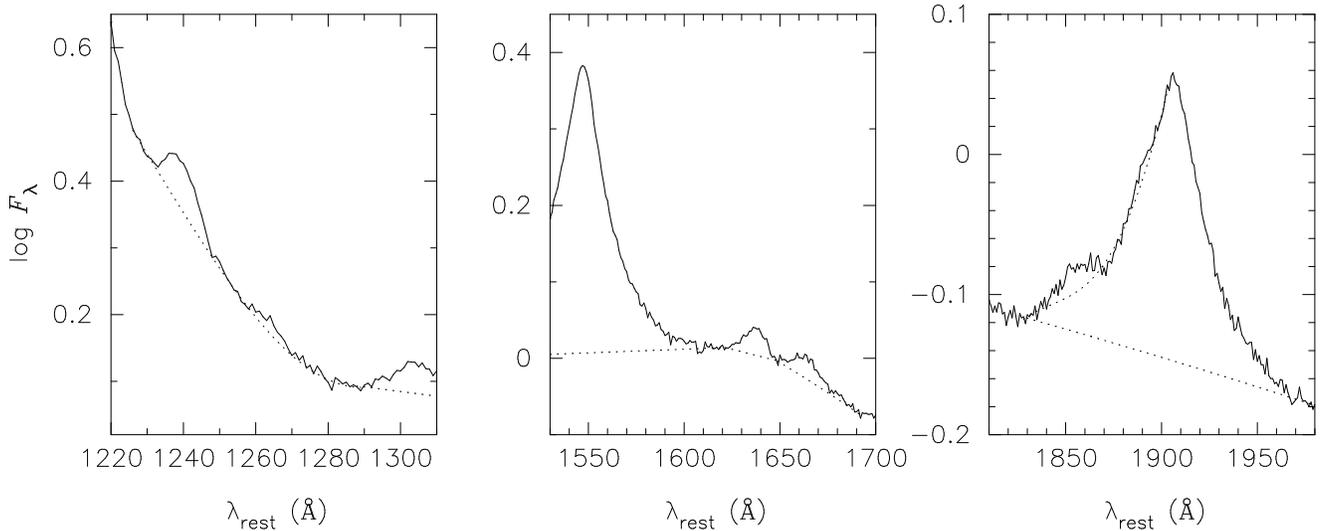}}
\caption{
Examples of the estimated local continuum (dotted lines) for
a few lines. The composite spectrum used here is
for quasars at $2.0 \leq z < 2.5$ and $-26.5 > M_B \geq -27.5$.
}
\label{fig34}
\end{figure*}

\clearpage

\begin{figure*}
\centering
\rotatebox{-90}{\includegraphics[width=8.5cm]{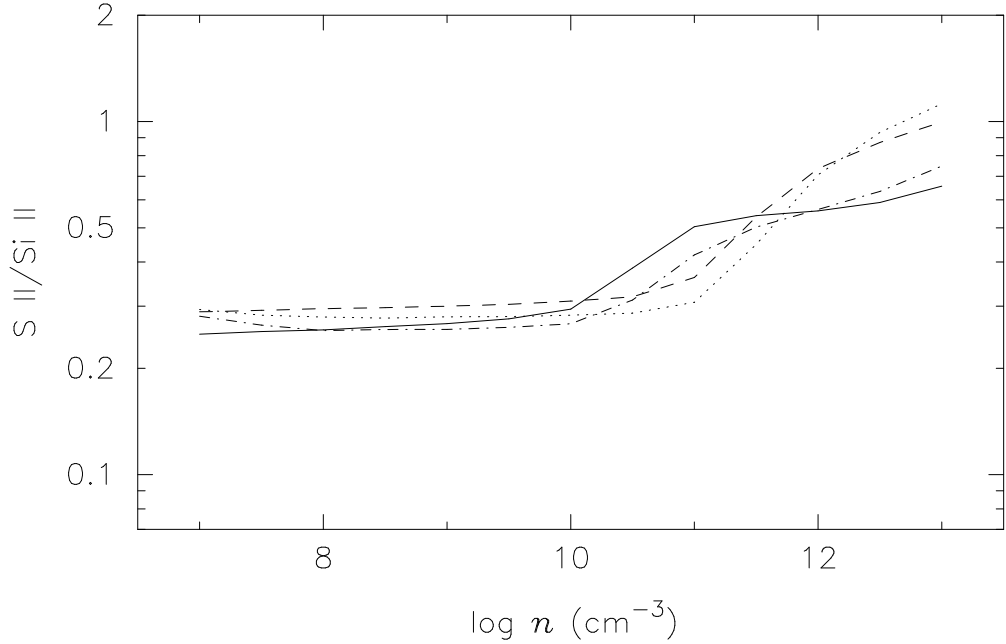}}
\caption{
Predicted S{\sc ii}$\lambda$1257/Si{\sc ii}$\lambda$1263 flux ratio
as a function of gas density. Solid, dashed, dot-dashed and dotted
lines denote the models with ($Z/Z_\odot$, log $U$) = (1.0, --2.5),
(1.0, --3.5), (5.0, --2.5), and (5.0, --3.5), respectively.
}
\label{fig35}
\end{figure*}

\begin{figure*}
\centering
\vspace{1cm}
\rotatebox{90}{\includegraphics[width=9.0cm]{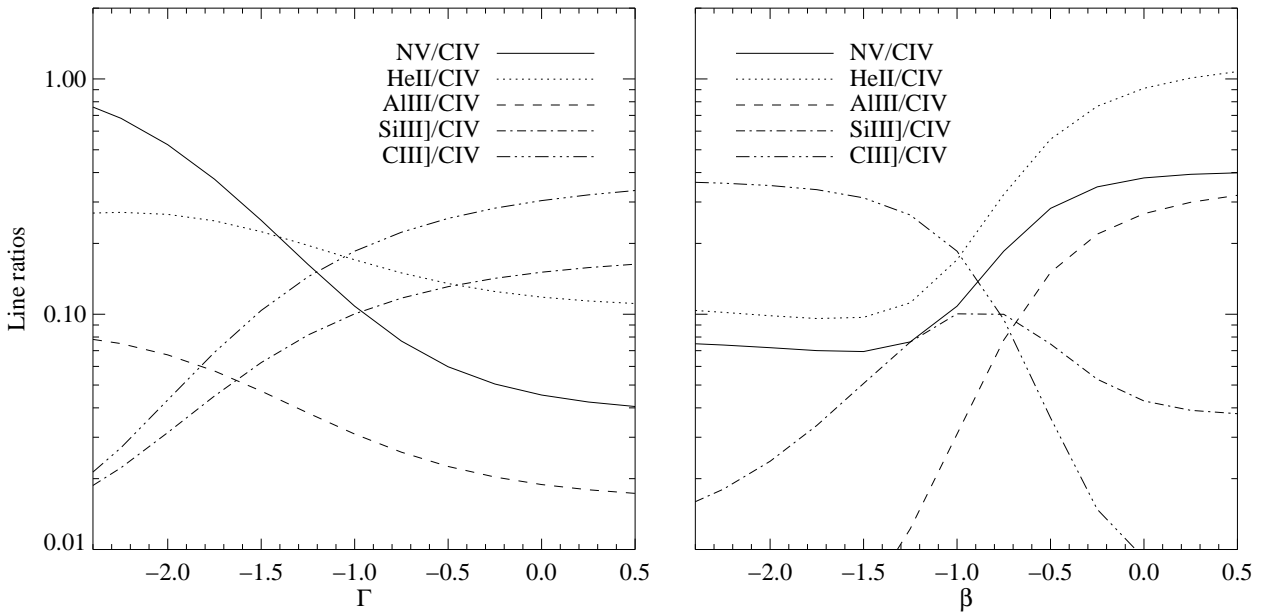}}
\caption{
Integrated theoretical flux ratios as a function of $\Gamma$ 
($left$) and $\beta$ ($right$). Only the models with 
$Z/Z_\odot$ = 1.0 and with a small UV bump SED are shown.
Predicted flux ratios of N{\sc v}/C{\sc iv}, He{\sc ii}/C{\sc iv},
Al{\sc iii}/C{\sc iv}, Si{\sc iii}]/C{\sc iv} and C{\sc iii}]/C{\sc iv}
are denoted by solid, dotted, dashed, dot-dashed, and three-dot-dashed
lines, respectively.
}
\label{fig36}
\end{figure*}

\clearpage

\begin{figure*}
\centering
\vspace{1cm}
\rotatebox{-90}{\includegraphics[width=14.5cm]{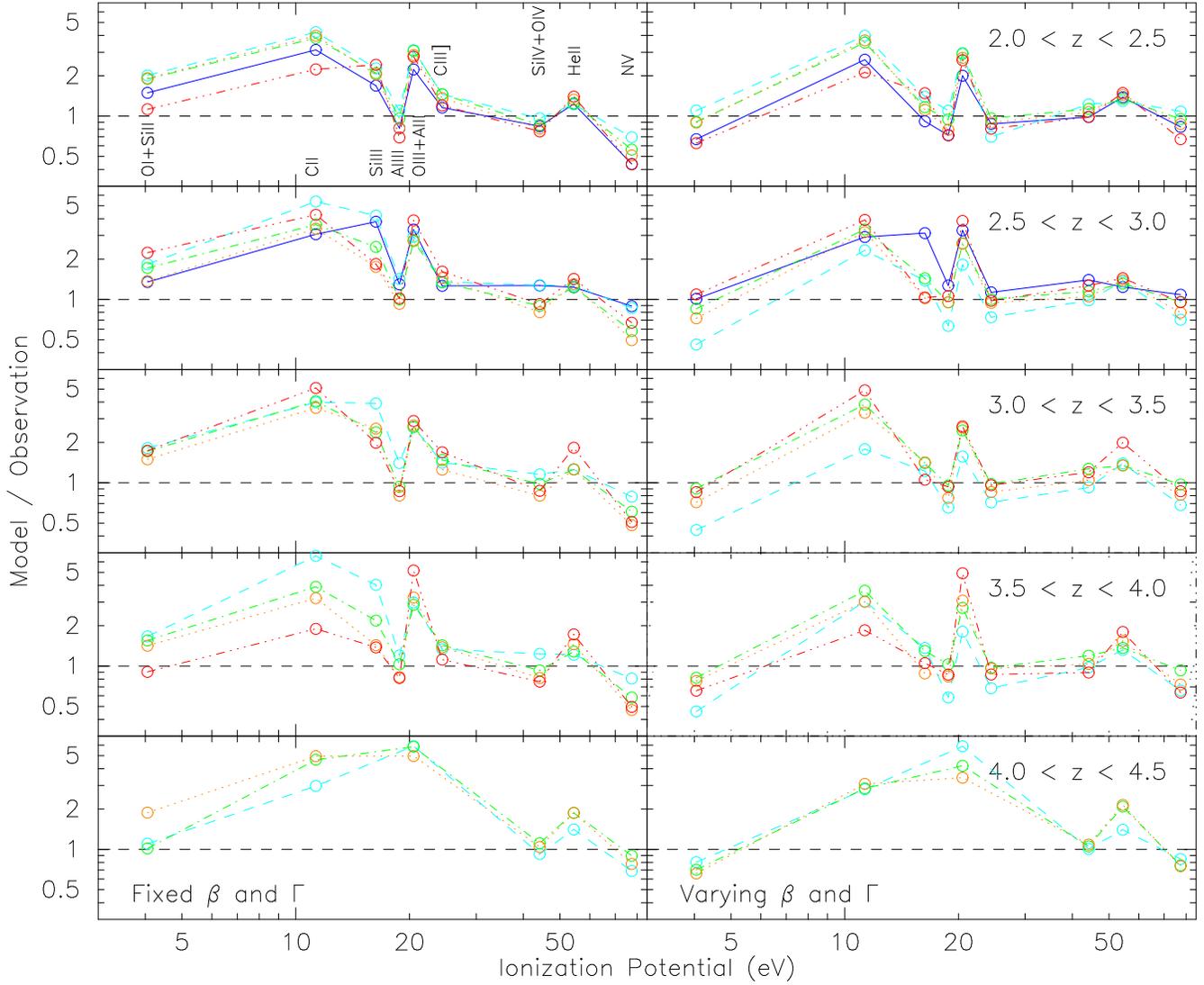}}
\caption{
Ratios of predicted to observed emission-line fluxes normalized
to the C{\sc iv} flux, as a function of the ionization potential
of corresponding ions. Model predictions with fixed $\beta$
and $\Gamma$ and those with varying $\beta$ and $\Gamma$ are
presented in the left and right panels, respectively.
Results for composite spectra of $2.0 \leq z < 2.5$,
$2.5 \leq z < 3.0$, $3.0 \leq z < 3.5$, $3.5 \leq z < 4.0$
and $4.0 \leq z < 4.5$ quasars are shown at upper to lower panels.
Results for composite spectra of $-24.5 > M_B \geq -25.5$,
$-25.5 > M_B \geq -26.5$, $-26.5 > M_B \geq -27.5$,
$-27.5 > M_B \geq -28.5$ and $-28.5 > M_B \geq -29.5$ are
denoted by solid blue lines, dashed light-blue lines, 
dash-dotted green lines, dotted orange lines, and 
dash-dot-dot-dotted red lines, respectively.
Dashed horizontal lines denote the unity of ratios of
predicted to observed fluxes.
For ionization parameters of emission-line pairs,
the average values are adopted just for the presentation; 
i.e., 4.05 eV for O{\sc i}+Si{\sc ii}, 20.5 eV for 
O{\sc iii}+Al{\sc ii}, and 44.2 eV for Si{\sc iv}+O{\sc iv}.
}
\label{fig37}
\end{figure*}

\clearpage

\begin{figure*}
\centering
\includegraphics[width=16cm]{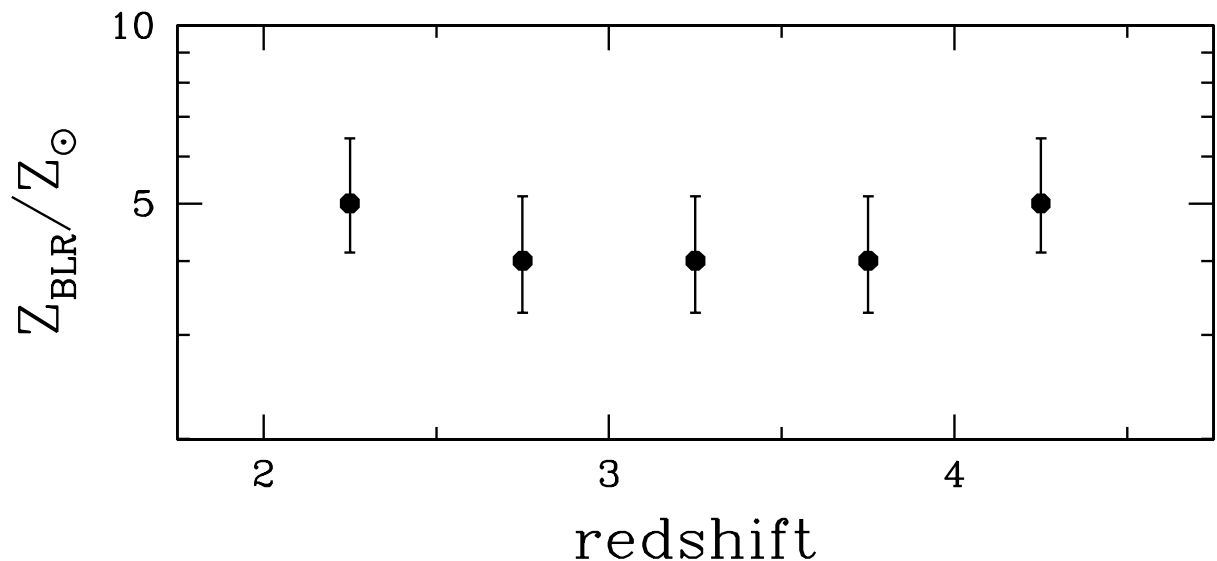}
\caption{
Estimated metallicities from our composite spectra, averaged
in the luminosity range $-25.5 > M_B > -28.5$, as a function of
redshift. The estimation of the metallicity given in this figure
is derived from the fit with the varying $\beta$ and $\Gamma$,
which are presented in Tables 12--16.
}
\label{fig38}
\end{figure*}

\begin{figure*}
\centering
\includegraphics[width=16.5cm]{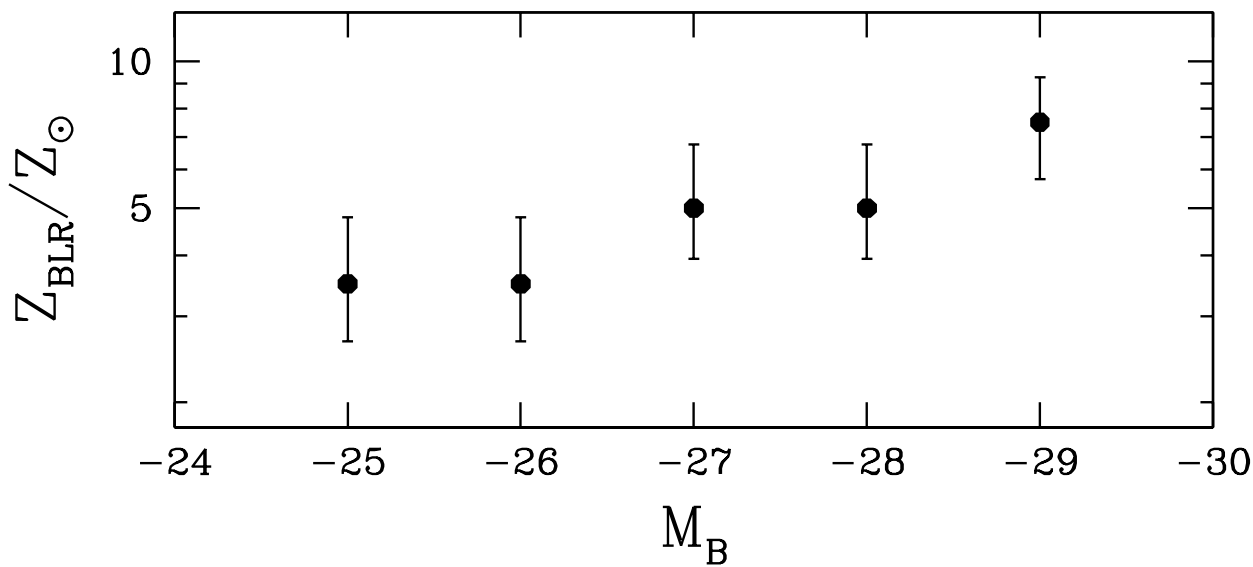}
\caption{
Estimated metallicities from our composite spectra, averaged
in the redshift range $2.0 \leq z < 3.0$, as a function of
luminosity. The estimation of the metallicity given in this figure
is derived from the fit with the varying $\beta$ and $\Gamma$,
which are presented in Tables 12 and 13.
}
\label{fig39}
\end{figure*}


\begin{table*}
\centering
\caption{Number distribution of our quasar sample
         among the redshift and the absolute magnitude bins}
\label{table:01}
\begin{tabular}{c c c c c c r}
\hline\hline
 &
 $2.0 \leq z < 2.5$ &
 $2.5 \leq z < 3.0$ &
 $3.0 \leq z < 3.5$ &
 $3.5 \leq z < 4.0$ &
 $4.0 \leq z < 4.5$ &
 total \\
\hline 
$-24.5 > M_B \geq -25.5$ &  643 &  50 &   1 &   0 &   0 &  694 \\
$-25.5 > M_B \geq -26.5$ & 1497 & 284 & 332 & 153 &  25 & 2291 \\
$-26.5 > M_B \geq -27.5$ &  917 & 385 & 323 & 222 & 120 & 1967 \\
$-27.5 > M_B \geq -28.5$ &  105 &  71 &  76 &  53 &  45 &  350 \\
$-28.5 > M_B \geq -29.5$ &    5 &  11 &  16 &   5 &   3 &   40 \\
$-29.5 > M_B \geq -30.5$ &    0 &   1 &   1 &   0 &   0 &    2 \\
total                    & 3167 & 802 & 749 & 433 & 193 & 5344 \\
\hline
\end{tabular}
\end{table*}


\begin{table*}
\centering
\caption{Ionization potentials of ions}
\label{table:02}
\begin{tabular}{l c c c}
\hline\hline
 Ion &
 Ionization Potential &
 Ionization Potential &
 Classification \\
 &
 Lower (eV) &
 Upper (eV) &
 \\
\hline
O{\sc i}    &  0.0 & 13.6 & LIL \\
Al{\sc ii}  &  5.9 & 18.8 & LIL \\
Si{\sc ii}  &  8.1 & 16.3 & LIL \\
C{\sc ii}   & 11.3 & 24.4 & LIL \\
Si{\sc iii} & 16.3 & 33.5 & LIL \\
Al{\sc iii} & 18.8 & 28.4 & LIL \\
C{\sc iii}  & 24.4 & 47.9 & LIL \\
He{\sc ii}  & 24.6 & 54.4 & HIL$^{\mathrm{a}}$ \\
Si{\sc iv}  & 33.5 & 45.1 & LIL \\
O{\sc iii}  & 35.1 & 54.9 & LIL \\
N{\sc iv}   & 47.4 & 77.4 & HIL \\
C{\sc iv}   & 47.9 & 64.5 & HIL \\
O{\sc iv}   & 54.9 & 77.4 & HIL \\
N{\sc v}    & 77.4 & 97.9 & HIL \\
\hline
\end{tabular}
\begin{list}{}{}
\item[$^{\mathrm{a}}$]  
         Classified as a HIL because 
         He{\sc ii}$\lambda$1640 is a recombination line
         and thus the $upper$ ionization potential is
         important rather than the $lower$ ionization
         potential.
\end{list}
\end{table*}


\begin{table*}
\centering
\caption{Measured line fluxes; $2.0 \leq z < 2.5$}
\label{table:03}
\begin{tabular}{l c c c c c}
\hline\hline
 Line &
 $-24.5 \! > \! M_B \! \geq \! -25.5$ &
 $-25.5 \! > \! M_B \! \geq \! -26.5$ &
 $-26.5 \! > \! M_B \! \geq \! -27.5$ &
 $-27.5 \! > \! M_B \! \geq \! -28.5$ &
 $-28.5 \! > \! M_B \! \geq \! -29.5$ \\
\hline
N{\sc v}$\lambda$1240 &
  0.506 $\pm$ 0.004 &
  0.717 $\pm$ 0.007 &
  0.887 $\pm$ 0.010 &
  0.983 $\pm$ 0.012 &
  1.144 $\pm$ 0.016 \\
Si{\sc ii}$\lambda$1263 &
  0.089 $\pm$ 0.001 &
  0.126 $\pm$ 0.002 &
  0.140 $\pm$ 0.002 &
  0.176 $\pm$ 0.003 &
  0.222 $\pm$ 0.003 \\
O{\sc i} $\!\!+\!\!$ Si{\sc ii}$\lambda$1305 &
  0.061 $\pm$ 0.002 &
  0.081 $\pm$ 0.002 &
  0.086 $\pm$ 0.002 &
  0.085 $\pm$ 0.002 &
  0.145 $\pm$ 0.004 \\
C{\sc ii}$\lambda$1335 &
  0.033 $\pm$ 0.002 &
  0.047 $\pm$ 0.001 &
  0.052 $\pm$ 0.001 &
  0.050 $\pm$ 0.002 &
  0.089 $\pm$ 0.004 \\
Si{\sc iv}$\lambda$1397 &
  0.111 $\pm$ 0.002 &
  0.027 $\pm$ 0.003 &
  0.001 $\pm$ 0.016 &
  0.033 $\pm$ 0.003 &
  ---               \\
O{\sc iv}]$\lambda$1402 &
  0.145 $\pm$ 0.008 &
  0.308 $\pm$ 0.008 &
  0.374 $\pm$ 0.019 &
  0.365 $\pm$ 0.010 &
  0.419 $\pm$ 0.008 \\
N{\sc iv}]$\lambda$1486 &
  0.017 $\pm$ 0.000 &
  0.013 $\pm$ 0.000 &
  0.015 $\pm$ 0.000 &
  0.020 $\pm$ 0.000 &
  ---               \\
C{\sc iv}$\lambda$1549 &
  1.000 $\pm$ 0.007 &
  1.000 $\pm$ 0.008 &
  1.000 $\pm$ 0.010 &
  1.000 $\pm$ 0.010 &
  1.000 $\pm$ 0.012 \\
1600{\rm \AA} bump &
  0.112 $\pm$ 0.002 &
  0.106 $\pm$ 0.002 &
  0.098 $\pm$ 0.002 &
  0.097 $\pm$ 0.003 &
  0.060 $\pm$ 0.005 \\
He{\sc ii}$\lambda$1640 &
  0.147 $\pm$ 0.001 &
  0.152 $\pm$ 0.001 &
  0.149 $\pm$ 0.002 &
  0.140 $\pm$ 0.001 &
  0.133 $\pm$ 0.002 \\
O{\sc iii}]$\lambda$1663 &
  0.077 $\pm$ 0.001 &
  0.083 $\pm$ 0.002 &
  0.072 $\pm$ 0.002 &
  0.083 $\pm$ 0.003 &
  0.104 $\pm$ 0.002 \\
Al{\sc ii}$\lambda$1671 &
  0.010 $\pm$ 0.001 &
  0.010 $\pm$ 0.001 &
  0.022 $\pm$ 0.001 &
  0.018 $\pm$ 0.001 &
  ---               \\
Al{\sc iii}$\lambda$1857 &
  0.064 $\pm$ 0.001 &
  0.092 $\pm$ 0.001 &
  0.105 $\pm$ 0.001 &
  0.124 $\pm$ 0.002 &
  0.146 $\pm$ 0.002 \\
Si{\sc iii}]$\lambda$1892 &
  0.093 $\pm$ 0.001 &
  0.118 $\pm$ 0.001 &
  0.127 $\pm$ 0.001 &
  0.129 $\pm$ 0.002 &
  0.110 $\pm$ 0.002 \\
C{\sc iii}]$\lambda$1909 &
  0.285 $\pm$ 0.005 &
  0.326 $\pm$ 0.008 &
  0.326 $\pm$ 0.008 &
  0.348 $\pm$ 0.010 &
  0.395 $\pm$ 0.012 \\
\hline
\end{tabular}
\end{table*}


\begin{table*}
\centering
\caption{Measured line fluxes; $2.5 \leq z < 3.0$}
\label{table:04}
\begin{tabular}{l c c c c c}
\hline\hline
 Line &
 $-24.5 \! > \! M_B \! \geq \! -25.5$ &
 $-25.5 \! > \! M_B \! \geq \! -26.5$ &
 $-26.5 \! > \! M_B \! \geq \! -27.5$ &
 $-27.5 \! > \! M_B \! \geq \! -28.5$ &
 $-28.5 \! > \! M_B \! \geq \! -29.5$ \\
\hline
N{\sc v}$\lambda$1240 &
  0.561 $\pm$ 0.004 &
  0.578 $\pm$ 0.005 &
  0.858 $\pm$ 0.010 & 
  1.004 $\pm$ 0.013 &
  1.248 $\pm$ 0.018 \\
Si{\sc ii}$\lambda$1263 &
  0.160 $\pm$ 0.001 &
  0.117 $\pm$ 0.001 &
  0.149 $\pm$ 0.002 &
  0.194 $\pm$ 0.003 &
  0.210 $\pm$ 0.003 \\
O{\sc i} $\!\!+\!\!$ Si{\sc ii}$\lambda$1305 &
  0.121 $\pm$ 0.006 &
  0.089 $\pm$ 0.002 &
  0.095 $\pm$ 0.002 &
  0.119 $\pm$ 0.002 & 
  0.112 $\pm$ 0.005 \\
C{\sc ii}$\lambda$1335 &
  0.065 $\pm$ 0.006 &
  0.037 $\pm$ 0.002 &
  0.055 $\pm$ 0.002 &
  0.059 $\pm$ 0.002 &
  0.074 $\pm$ 0.004 \\
Si{\sc iv}$\lambda$1397 &
  0.052 $\pm$ 0.002 &
  0.160 $\pm$ 0.002 &
  ---               &
  ---               &
  ---               \\ 
O{\sc iv}]$\lambda$1402 &
  0.202 $\pm$ 0.003 &
  0.092 $\pm$ 0.002 &
  0.360 $\pm$ 0.008 &
  0.401 $\pm$ 0.007 &
  0.462 $\pm$ 0.027 \\
N{\sc iv}]$\lambda$1486 &
  0.021 $\pm$ 0.000 &
  0.029 $\pm$ 0.000 &
  0.022 $\pm$ 0.000 &
  0.019 $\pm$ 0.000 &
  0.012 $\pm$ 0.001 \\
C{\sc iv}$\lambda$1549 &
  1.000 $\pm$ 0.006 &
  1.000 $\pm$ 0.007 &
  1.000 $\pm$ 0.010 &
  1.000 $\pm$ 0.011 &
  1.000 $\pm$ 0.012 \\
1600{\rm \AA} bump &
  0.125 $\pm$ 0.008 &
  0.111 $\pm$ 0.003 &
  0.103 $\pm$ 0.002 &
  0.092 $\pm$ 0.003 &
  0.100 $\pm$ 0.006 \\
He{\sc ii}$\lambda$1640 &
  0.151 $\pm$ 0.001 &
  0.146 $\pm$ 0.001 &
  0.151 $\pm$ 0.002 &
  0.143 $\pm$ 0.002 &
  0.125 $\pm$ 0.002 \\
O{\sc iii}]$\lambda$1663 &
  0.087 $\pm$ 0.001 &
  0.083 $\pm$ 0.002 &
  0.093 $\pm$ 0.003 &
  0.086 $\pm$ 0.003 &
  0.096 $\pm$ 0.002 \\
Al{\sc ii}$\lambda$1671 &
  ---               &
  0.015 $\pm$ 0.001 &
  0.013 $\pm$ 0.001 &
  0.017 $\pm$ 0.001 &
  ---               \\
Al{\sc iii}$\lambda$1857 &
  0.078 $\pm$ 0.001 &
  0.071 $\pm$ 0.001 &
  0.102 $\pm$ 0.001 &
  0.109 $\pm$ 0.002 &
  0.165 $\pm$ 0.003 \\
Si{\sc iii}]$\lambda$1892 &
  0.070 $\pm$ 0.001 &
  0.063 $\pm$ 0.001 &
  0.108 $\pm$ 0.001 &
  0.152 $\pm$ 0.002 &
  0.204 $\pm$ 0.003 \\
C{\sc iii}]$\lambda$1909 &
  0.375 $\pm$ 0.006 &
  0.351 $\pm$ 0.007 &
  0.354 $\pm$ 0.010 &
  0.328 $\pm$ 0.008 &
  0.331 $\pm$ 0.004 \\
\hline
\end{tabular}
\end{table*}


\begin{table*}
\centering
\caption{Measured line fluxes; $3.0 \leq z < 3.5$}
\label{table:05}
\begin{tabular}{l c c c c}
\hline\hline
 Line &
 $-25.5 \! > \! M_B \! \geq \! -26.5$ &
 $-26.5 \! > \! M_B \! \geq \! -27.5$ &
 $-27.5 \! > \! M_B \! \geq \! -28.5$ &
 $-28.5 \! > \! M_B \! \geq \! -29.5$ \\
\hline
N{\sc v}$\lambda$1240 &
  0.635 $\pm$ 0.005 &
  0.819 $\pm$ 0.009 &
  1.034 $\pm$ 0.013 &
  0.979 $\pm$ 0.011 \\
Si{\sc ii}$\lambda$1263 &
  0.126 $\pm$ 0.002 &
  0.161 $\pm$ 0.002 &
  0.183 $\pm$ 0.003 &
  0.198 $\pm$ 0.003 \\
O{\sc i} $\!\!+\!\!$ Si{\sc ii}$\lambda$1305 &
  0.090 $\pm$ 0.002 &
  0.095 $\pm$ 0.002 &
  0.109 $\pm$ 0.003 &
  0.094 $\pm$ 0.004 \\
C{\sc ii}$\lambda$1335 &
  0.050 $\pm$ 0.002 &
  0.049 $\pm$ 0.002 &
  0.055 $\pm$ 0.003 &
  0.039 $\pm$ 0.004 \\
Si{\sc iv}$\lambda$1397 &
  0.102 $\pm$ 0.001 &
  0.043 $\pm$ 0.003 &
  0.075 $\pm$ 0.003 &
  ---               \\
O{\sc iv}]$\lambda$1402 &
  0.176 $\pm$ 0.004 &
  0.286 $\pm$ 0.007 &
  0.327 $\pm$ 0.005 &
  0.370 $\pm$ 0.006 \\
N{\sc iv}]$\lambda$1486 &
  0.030 $\pm$ 0.000 &
  0.023 $\pm$ 0.000 &
  0.019 $\pm$ 0.000 &
  0.045 $\pm$ 0.001 \\
C{\sc iv}$\lambda$1549 &
  1.000 $\pm$ 0.007 &
  1.000 $\pm$ 0.010 &
  1.000 $\pm$ 0.011 &
  1.000 $\pm$ 0.010 \\
1600{\rm \AA} bump &
  0.117 $\pm$ 0.003 &
  0.117 $\pm$ 0.003 &
  0.099 $\pm$ 0.004 &
  0.083 $\pm$ 0.005 \\
He{\sc ii}$\lambda$1640 &
  0.148 $\pm$ 0.001 &
  0.148 $\pm$ 0.001 &
  0.148 $\pm$ 0.002 &
  0.102 $\pm$ 0.001 \\
O{\sc iii}]$\lambda$1663 &
  0.084 $\pm$ 0.002 &
  0.092 $\pm$ 0.002 &
  0.078 $\pm$ 0.003 &
  0.088 $\pm$ 0.003 \\
Al{\sc ii}$\lambda$1671 &
  0.027 $\pm$ 0.001 &
  0.019 $\pm$ 0.001 &
  0.030 $\pm$ 0.001 &
  0.012 $\pm$ 0.001 \\
Al{\sc iii}$\lambda$1857 &
  0.072 $\pm$ 0.001 &
  0.109 $\pm$ 0.001 &
  0.126 $\pm$ 0.002 &
  0.117 $\pm$ 0.001 \\
Si{\sc iii}]$\lambda$1892 &
  0.068 $\pm$ 0.001 &
  0.112 $\pm$ 0.001 &
  0.105 $\pm$ 0.001 &
  0.134 $\pm$ 0.001 \\
C{\sc iii}]$\lambda$1909 &
  0.337 $\pm$ 0.007 &
  0.321 $\pm$ 0.007 &
  0.379 $\pm$ 0.010 &
  0.281 $\pm$ 0.007 \\
\hline
\end{tabular}
\end{table*}


\begin{table*}
\centering
\caption{Measured line fluxes; $3.5 \leq z < 4.0$}
\label{table:06}
\begin{tabular}{l c c c c}
\hline\hline
 Line &
 $-25.5 \! > \! M_B \! \geq \! -26.5$ &
 $-26.5 \! > \! M_B \! \geq \! -27.5$ &
 $-27.5 \! > \! M_B \! \geq \! -28.5$ &
 $-28.5 \! > \! M_B \! \geq \! -29.5$ \\
\hline
N{\sc v}$\lambda$1240 &
  0.617 $\pm$ 0.005 &
  0.856 $\pm$ 0.010 &
  1.062 $\pm$ 0.012 &
  1.001 $\pm$ 0.015 \\
Si{\sc ii}$\lambda$1263 &
  0.129 $\pm$ 0.001 &
  0.156 $\pm$ 0.002 &
  0.194 $\pm$ 0.002 &
  0.083 $\pm$ 0.001 \\
O{\sc i} $\!\!+\!\!$ Si{\sc ii}$\lambda$1305 &
  0.098 $\pm$ 0.006 &
  0.105 $\pm$ 0.004 &
  0.115 $\pm$ 0.005 &
  0.180 $\pm$ 0.013 \\
C{\sc ii}$\lambda$1335 &
  0.030 $\pm$ 0.006 &
  0.051 $\pm$ 0.004 &
  0.062 $\pm$ 0.005 &
  0.105 $\pm$ 0.013 \\
Si{\sc iv}$\lambda$1397 &
  0.196 $\pm$ 0.002 &
  ---               &
  ---               &
  0.024 $\pm$ 0.016 \\
O{\sc iv}]$\lambda$1402 &
  0.065 $\pm$ 0.007 &
  0.347 $\pm$ 0.008 &
  0.395 $\pm$ 0.009 &
  0.398 $\pm$ 0.017 \\
N{\sc iv}]$\lambda$1486 &
  0.038 $\pm$ 0.000 &
  0.025 $\pm$ 0.000 &
  0.016 $\pm$ 0.000 &
  0.031 $\pm$ 0.001 \\
C{\sc iv}$\lambda$1549 &
  1.000 $\pm$ 0.006 &
  1.000 $\pm$ 0.010 &
  1.000 $\pm$ 0.010 &
  1.000 $\pm$ 0.013 \\
1600{\rm \AA} bump &
  0.106 $\pm$ 0.008 &
  0.113 $\pm$ 0.005 &
  0.100 $\pm$ 0.007 &
  0.032 $\pm$ 0.016 \\
He{\sc ii}$\lambda$1640 &
  0.153 $\pm$ 0.001 &
  0.145 $\pm$ 0.002 &
  0.128 $\pm$ 0.001 &
  0.108 $\pm$ 0.001 \\
O{\sc iii}]$\lambda$1663 &
  0.081 $\pm$ 0.001 &
  0.083 $\pm$ 0.001 &
  0.089 $\pm$ 0.002 &
  0.056 $\pm$ 0.001 \\
Al{\sc ii}$\lambda$1671 &
  0.016 $\pm$ 0.001 &
  0.017 $\pm$ 0.001 &
  ---               &
  ---               \\
Al{\sc iii}$\lambda$1857 &
  0.084 $\pm$ 0.001 &
  0.098 $\pm$ 0.001 &
  0.121 $\pm$ 0.002 &
  0.124 $\pm$ 0.002 \\
Si{\sc iii}]$\lambda$1892 &
  0.066 $\pm$ 0.001 &
  0.122 $\pm$ 0.001 &
  0.186 $\pm$ 0.002 &
  0.193 $\pm$ 0.003 \\
C{\sc iii}]$\lambda$1909 &
  0.355 $\pm$ 0.006 &
  0.332 $\pm$ 0.008 &
  0.343 $\pm$ 0.007 &
  0.426 $\pm$ 0.010 \\
\hline
\end{tabular}
\end{table*}


\begin{table*}
\centering
\caption{Measured line fluxes; $4.0 \leq z < 4.5$}
\label{table:07}
\begin{tabular}{l c c c}
\hline\hline
 Line &
 $-25.5 \! > \! M_B \! \geq \! -26.5$ &
 $-26.5 \! > \! M_B \! \geq \! -27.5$ &
 $-27.5 \! > \! M_B \! \geq \! -28.5$ \\
\hline
N{\sc v}$\lambda$1240 &
  0.724 $\pm$ 0.004 &
  0.933 $\pm$ 0.010 &
  1.075 $\pm$ 0.013 \\
Si{\sc ii}$\lambda$1263 &
  0.251 $\pm$ 0.003 &
  0.249 $\pm$ 0.004 &
  0.217 $\pm$ 0.005 \\
O{\sc i} $\!\!+\!\!$ Si{\sc ii}$\lambda$1305 &
  0.148 $\pm$ 0.012 &
  0.146 $\pm$ 0.005 &
  0.134 $\pm$ 0.005 \\
C{\sc ii}$\lambda$1335 &
  0.067 $\pm$ 0.012 &
  0.068 $\pm$ 0.005 &
  0.064 $\pm$ 0.005 \\
Si{\sc iv}$\lambda$1397 &
  0.205 $\pm$ 0.003 &
  0.379 $\pm$ 0.009 &
  0.055 $\pm$ 0.006 \\
O{\sc iv}]$\lambda$1402 &
  0.144 $\pm$ 0.007 &
  0.004 $\pm$ 0.015 &
  0.354 $\pm$ 0.010 \\
N{\sc iv}]$\lambda$1486 &
  0.019 $\pm$ 0.000 &
  0.010 $\pm$ 0.000 &
  0.007 $\pm$ 0.000 \\
C{\sc iv}$\lambda$1549 &
  1.000 $\pm$ 0.004 &
  1.000 $\pm$ 0.009 &
  1.000 $\pm$ 0.011 \\
1600{\rm \AA} bump &
  0.075 $\pm$ 0.015 &
  0.070 $\pm$ 0.007 &
  0.061 $\pm$ 0.007 \\
He{\sc ii}$\lambda$1640 &
  0.132 $\pm$ 0.001 &
  0.095 $\pm$ 0.001 &
  0.095 $\pm$ 0.001 \\
O{\sc iii}]$\lambda$1663 &
  0.048 $\pm$ 0.002 &
  0.064 $\pm$ 0.002 &
  0.029 $\pm$ 0.001 \\
Al{\sc ii}$\lambda$1671 &
  0.000 $\pm$ 0.001 &
  ---               &
  0.046 $\pm$ 0.001 \\
\hline
\end{tabular}
\end{table*}


\begin{table*}
\centering
\caption{Measured velocity shift and line width of 
         HILs and LILs}
\label{table:08}
\begin{tabular}{c c c c c c c}
\hline\hline
 Redshift &
 Magnitude &
 Line &
 Velocity Shift &
 \multicolumn{2}{c}{Profile Parameter} &
 FWHM$^{\mathrm{a}}$ \\
  &
  &
  &
 (km/s) &
 $\alpha$ &
 $\beta$ &
 (km/s) \\
\hline
$2.0 \! \leq \! z \! < \! 2.5$
  & $-24.5 \! > \! M_B \! \geq \! -25.5$ &
      HIL & --464.9  $\pm$ 1.2  & 
              117.30 $\pm$ 0.02 &
               99.37 $\pm$ 0.02 & 3858 \\
  & & LIL &  --83.4  $\pm$ 1.9  &
              114.49 $\pm$ 0.05 &
              119.60 $\pm$ 0.07 & 3550 \\
  & $-25.5 \! > \! M_B \! \geq \! -26.5$ &
      HIL & --495.6  $\pm$ 1.4  &
              112.90 $\pm$ 0.02 &
               85.50 $\pm$ 0.02 & 4264 \\
  & & LIL &  +111.5  $\pm$ 2.5  &
              106.01 $\pm$ 0.05 &
              110.53 $\pm$ 0.07 & 3838 \\
  & $-26.5 \! > \! M_B \! \geq \! -27.5$ &
      HIL & --446.5  $\pm$ 1.7  &
              110.31 $\pm$ 0.02 &
               79.98 $\pm$ 0.03 & 4474 \\
  & & LIL &  +308.1  $\pm$ 2.5  &
              106.49 $\pm$ 0.07 &
              109.75 $\pm$ 0.09 & 3842 \\
  & $-27.5 \! > \! M_B \! \geq \! -28.5$ &
      HIL & --327.0  $\pm$ 1.8  &
              114.32 $\pm$ 0.02 &
               77.77 $\pm$ 0.02 & 4480 \\
  & & LIL &  +484.8  $\pm$ 2.8  &
              101.64 $\pm$ 0.07 &
              104.12 $\pm$ 0.08 & 4038 \\
  & $-28.5 \! > \! M_B \! \geq \! -29.5$ &
      HIL & --583.7  $\pm$ 2.0  &
              106.74 $\pm$ 0.02 &
               80.08 $\pm$ 0.03 & 4535 \\
  & & LIL &  +153.3  $\pm$ 2.4  &
               93.17 $\pm$ 0.05 &
              105.50 $\pm$ 0.05 & 4199 \\
$2.5 \! \leq \! z \! < \! 3.0$
  & $-24.5 \! > \! M_B \! \geq \! -25.5$ &
      HIL & --390.6  $\pm$ 1.0  &
              128.57 $\pm$ 0.02 &
              117.31 $\pm$ 0.03 & 3384 \\
  & & LIL & --170.7  $\pm$ 1.5  &
              115.80 $\pm$ 0.04 &
              125.72 $\pm$ 0.06 & 3445 \\
  & $-25.5 \! > \! M_B \! \geq \! -26.5$ &
      HIL & --354.7  $\pm$ 1.2  &
              127.48 $\pm$ 0.02 &
              106.94 $\pm$ 0.03 & 3568 \\
  & & LIL &    +4.6  $\pm$ 2.0  &
              125.05 $\pm$ 0.06 & 
              116.90 $\pm$ 0.05 & 3436 \\
  & $-26.5 \! > \! M_B \! \geq \! -27.5$ &
      HIL & --371.4  $\pm$ 1.8  & 
              112.18 $\pm$ 0.01 &
               82.69 $\pm$ 0.02 & 4358 \\
  & & LIL &  +259.9  $\pm$ 2.9  &
              101.31 $\pm$ 0.05 &
              103.77 $\pm$ 0.09 & 4051 \\
  & $-27.5 \! > \! M_B \! \geq \! -28.5$ &
      HIL & --451.1  $\pm$ 1.9  &
              109.93 $\pm$ 0.02 &
               74.99 $\pm$ 0.03 & 4653 \\
  & & LIL &  +571.5  $\pm$ 2.4  &
              109.82 $\pm$ 0.07 &
              102.59 $\pm$ 0.07 & 3914 \\
  & $-28.5 \! > \! M_B \! \geq \! -29.5$ &
      HIL & --535.5  $\pm$ 2.2  &
              108.21 $\pm$ 0.03 &
               73.18 $\pm$ 0.03 & 4751 \\
  & & LIL &  +737.3  $\pm$ 1.7  &
              104.08 $\pm$ 0.05 &
               95.22 $\pm$ 0.05 & 4175 \\
$3.0 \! \leq \! z \! < \! 3.5$
  & $-25.5 \! > \! M_B \! \geq \! -26.5$ &
      HIL & --507.0  $\pm$ 1.2  &
              122.96 $\pm$ 0.02 &
               98.43 $\pm$ 0.03 & 3795 \\
  & & LIL & --186.5  $\pm$ 2.1  &
              119.09 $\pm$ 0.05 &
              111.70 $\pm$ 0.06 & 3602 \\
  & $-26.5 \! > \! M_B \! \geq \! -27.5$ &
      HIL & --576.9  $\pm$ 1.6  &
              119.72 $\pm$ 0.02 &
               89.27 $\pm$ 0.02 & 4056 \\
  & & LIL &  --20.1  $\pm$ 2.2  &
              106.02 $\pm$ 0.05 &
              119.36 $\pm$ 0.06 & 3699 \\
  & $-27.5 \! > \! M_B \! \geq \! -28.5$ &
      HIL & --549.8  $\pm$ 1.8  &
              118.15 $\pm$ 0.03 &
               74.78 $\pm$ 0.02 & 4528 \\
  & & LIL &   +16.9  $\pm$ 2.7  &
              110.84 $\pm$ 0.05 &
              102.22 $\pm$ 0.06 & 3904 \\
  & $-28.5 \! > \! M_B \! \geq \! -29.5$ &
      HIL & --298.9  $\pm$ 1.7  &
              126.01 $\pm$ 0.02 &
               80.60 $\pm$ 0.03 & 4218 \\
  & & LIL &  +763.3  $\pm$ 2.1  &
              110.34 $\pm$ 0.06 &
              103.28 $\pm$ 0.06 & 3891 \\
$3.5 \! \leq \! z \! < \! 4.0$
  & $-25.5 \! > \! M_B \! \geq \! -26.5$ &
      HIL & --644.7  $\pm$ 1.1  &
              126.18 $\pm$ 0.02 &
              110.47 $\pm$ 0.02 & 3524 \\
  & & LIL & --350.6  $\pm$ 1.5  &
              124.67 $\pm$ 0.04 &
              124.82 $\pm$ 0.07 & 3329 \\
  & $-26.5 \! > \! M_B \! \geq \! -27.5$ &
      HIL & --588.7  $\pm$ 1.8  &
              116.88 $\pm$ 0.02 &
               85.67 $\pm$ 0.02 & 4196 \\
  & & LIL &  +263.9  $\pm$ 2.2  &
              111.61 $\pm$ 0.07 &
              103.73 $\pm$ 0.06 & 3861 \\
  & $-27.5 \! > \! M_B \! \geq \! -28.5$ &
      HIL & --507.8  $\pm$ 1.7  &
              120.30 $\pm$ 0.02 &
               76.28 $\pm$ 0.02 & 4441 \\
  & & LIL &  +433.2  $\pm$ 1.6  &
               99.75 $\pm$ 0.04 &
              114.15 $\pm$ 0.07 & 3902 \\
  & $-28.5 \! > \! M_B \! \geq \! -29.5$ &
      HIL &   --5.3  $\pm$ 2.3  &
              109.58 $\pm$ 0.02 &
               88.25 $\pm$ 0.03 & 4245 \\
  & & LIL &  +553.2  $\pm$ 1.7  &
              116.68 $\pm$ 0.05 &
              115.84 $\pm$ 0.07 & 3572 \\
$4.0 \! \leq \! z \! < \! 4.5$
  & $-25.5 \! > \! M_B \! \geq \! -26.5$ &
      HIL & --625.7  $\pm$ 0.7  &
              133.03 $\pm$ 0.02 &
              137.22 $\pm$ 0.03 & 3073 \\
  & & LIL & --145.0  $\pm$ 2.8  &
               89.42 $\pm$ 0.05 &
               86.79 $\pm$ 0.06 & 4716 \\
  & $-26.5 \! > \! M_B \! \geq \! -27.5$ &
      HIL & --634.4  $\pm$ 1.6  &
              116.43 $\pm$ 0.03 &
               91.71 $\pm$ 0.03 & 4044 \\
  & & LIL &  +169.8  $\pm$ 3.9  &
              123.05 $\pm$ 0.10 &
               76.44 $\pm$ 0.05 & 4397 \\
  & $-27.5 \! > \! M_B \! \geq \! -28.5$ &
      HIL & --262.2  $\pm$ 1.9  &
              115.83 $\pm$ 0.02 &
               78.54 $\pm$ 0.02 & 4431 \\
  & & LIL & --287.4  $\pm$ 5.8  &
              111.97 $\pm$ 0.25 &
               91.26 $\pm$ 0.14 & 4127 \\
\hline
\end{tabular}
\begin{list}{}{}
\item[$^{\mathrm{a}}$] 
   Instrumental broadening was corrected by assuming
   that the instrumental width of 150 km s$^{-1}$.
\end{list}
\end{table*}

\clearpage


\begin{table*}
\centering
\caption{Dependences of the normalized flux ratios
         on redshift and absolute $B$ magnitude}
\label{table:09}
\begin{tabular}{l c c c}
\hline\hline
 Line Ratio &
 slope &
 $r_{\rm S}$$^{\mathrm{a}}$ &
 $p(r_{\rm S})$$^{\mathrm{b}}$ \\
\hline
\multicolumn{4}{c}{vs. redshift} \\
\hline
N{\sc v}/C{\sc iv}   & 
  -0.000 $\pm$ 0.013 & +0.05 & $8.2 \times 10^{-1}$ \\
Si{\sc ii}/C{\sc iv} & 
  +0.042 $\pm$ 0.045 & +0.47 & $3.2 \times 10^{-2}$ \\
(O{\sc i}+Si{\sc ii})/C{\sc iv} &
  +0.086 $\pm$ 0.032 & +0.66 & $1.0 \times 10^{-3}$ \\
C{\sc ii}/C{\sc iv}  & 
  +0.028 $\pm$ 0.042 & +0.34 & $1.3 \times 10^{-1}$ \\
(Si{\sc iv}+O{\sc iv})/C{\sc iv}  &  
 --0.003 $\pm$ 0.014 & +0.12 & $5.9 \times 10^{-1}$ \\
(1600${\rm \AA}$ bump)/C{\sc iv}  & 
 --0.076 $\pm$ 0.031 &--0.20 & $3.8 \times 10^{-1}$ \\
He{\sc ii}/C{\sc iv}  &
 --0.057 $\pm$ 0.015 &--0.59 & $5.3 \times 10^{-3}$ \\ 
(O{\sc iii}]+Al{\sc ii})/C{\sc ii}  &  
 --0.085 $\pm$ 0.027 &--0.31 & $1.7 \times 10^{-1}$ \\
Al{\sc iii}/C{\sc iv}  & 
 --0.030 $\pm$ 0.022 &--0.37 & $1.3 \times 10^{-1}$ \\
Si{\sc iii}]/C{\sc iv}  & 
  +0.007 $\pm$ 0.066 &--0.07 & $7.9 \times 10^{-1}$ \\
C{\sc iii}]/C{\sc iv}  & 
  +0.002 $\pm$ 0.023 & +0.19 & $4.5 \times 10^{-1}$ \\
N{\sc v}/He{\sc ii}  & 
  +0.057 $\pm$ 0.020 & +0.32 & $1.6 \times 10^{-1}$ \\
\hline
\multicolumn{4}{c}{vs. absolute $B$ magnitude} \\
\hline
N{\sc v}/C{\sc iv}  &  
  +0.082 $\pm$ 0.006 & +0.95 & $2.7 \times 10^{-11}$ \\
Si{\sc ii}/C{\sc iv}  &  
  +0.039 $\pm$ 0.018 & +0.53 & $1.3 \times 10^{-2}$ \\
(O{\sc i}+Si{\sc ii})/C{\sc iv}  &  
  +0.039 $\pm$ 0.013 & +0.48 & $2.9 \times 10^{-2}$ \\
C{\sc ii}/C{\sc iv}  &  
  +0.059 $\pm$ 0.018 & +0.50 & $2.2 \times 10^{-2}$ \\
(Si{\sc iv}+O{\sc iv})/C{\sc iv}  &  
  +0.059 $\pm$ 0.006 & +0.95 & $6.4 \times 10^{-11}$ \\
(1600${\rm \AA}$ bump)/C{\sc iv}  &  
 --0.068 $\pm$ 0.018 &--0.88 & $1.4 \times 10^{-7}$ \\
He{\sc ii}/C{\sc iv}  &  
 --0.031 $\pm$ 0.008 &--0.70 & $3.9 \times 10^{-4}$ \\
(O{\sc iii}]+Al{\sc ii})/C{\sc ii}  &  
 --0.001 $\pm$ 0.012 & +0.10 & $6.6 \times 10^{-1}$ \\
Al{\sc iii}/C{\sc iv}  &  
  +0.076 $\pm$ 0.009 & +0.95 & $3.3 \times 10^{-9}$ \\
Si{\sc iii}]/C{\sc iv}  &  
  +0.088 $\pm$ 0.018 & +0.77 & $2.0 \times 10^{-4}$ \\
C{\sc iii}]/C{\sc iv}  &  
  +0.009 $\pm$ 0.008 & +0.16 & $5.3 \times 10^{-1}$ \\
N{\sc v}/He{\sc ii}  &  
  +0.113 $\pm$ 0.007 & +0.97 & $1.1 \times 10^{-13}$ \\
\hline
\end{tabular}
\begin{list}{}{}
\item[$^{\mathrm{a}}$]
   Spearman rank-order correlation coefficient.
\item[$^{\mathrm{b}}$]
   Probability of the data being consistent with the
   null hypothesis that the flux ratio is not
   correlated with redshift or absolute $B$ magnitude.
\end{list}
\end{table*}

\clearpage


\begin{table*}
\centering
\caption{Model predictions of emission-line flux ratios}
\label{table:10}
\begin{tabular}{l c c c c c c} 
\hline\hline
 Line Ratio &
 $Z/Z_\odot \! = \! 0.2$ &
 $Z/Z_\odot \! = \! 0.5$ &
 $Z/Z_\odot \! = \! 1.0$ &
 $Z/Z_\odot \! = \! 2.0$ &
 $Z/Z_\odot \! = \! 5.0$ &
 $Z/Z_\odot \! = \! 10.0$ \\
\hline
\multicolumn{7}{c}{SED with a large UV thermal bump}\\
\hline
   N{\sc v}/C{\sc iv}  &  
      0.019 & 0.044 & 0.092 & 0.199 & 0.439 & 0.712 \\
   Si{\sc ii}/C{\sc iv}  &  
      0.001 & 0.002 & 0.003 & 0.005 & 0.010 & 0.017 \\
   C{\sc ii}/C{\sc iv}  &  
      0.021 & 0.034 & 0.056 & 0.105 & 0.224 & 0.364 \\
   Si{\sc iv}/C{\sc iv}  &  
      0.056 & 0.075 & 0.104 & 0.158 & 0.248 & 0.329 \\
   O{\sc iv}]/C{\sc iv}  &  
      0.050 & 0.043 & 0.045 & 0.058 & 0.088 & 0.115 \\
   He{\sc ii}/C{\sc iv}  &  
      0.198 & 0.176 & 0.183 & 0.211 & 0.235 & 0.231 \\
   O{\sc iii}]/C{\sc iv}  &  
      0.059 & 0.085 & 0.124 & 0.193 & 0.290 & 0.353 \\
   Al{\sc ii}/C{\sc iv}  &  
      0.003 & 0.005 & 0.008 & 0.014 & 0.029 & 0.049 \\
   Al{\sc iii}/C{\sc iv}  &  
      0.010 & 0.016 & 0.027 & 0.048 & 0.095 & 0.152 \\
   Si{\sc iii}]/C{\sc iv}  &  
      0.037 & 0.059 & 0.090 & 0.149 & 0.252 & 0.349 \\
   C{\sc iii}]/C{\sc iv}  &  
      0.038 & 0.068 & 0.157 & 0.361 & 0.562 & 0.604 \\
\hline
\multicolumn{7}{c}{SED with a small UV thermal bump} \\
\hline
   N{\sc v}/C{\sc iv}  &  
      0.024 & 0.054 & 0.108 & 0.222 & 0.499 & 0.837 \\
   Si{\sc ii}/C{\sc iv}  &  
      0.002 & 0.002 & 0.003 & 0.005 & 0.010 & 0.017 \\
   C{\sc ii}/C{\sc iv}  &  
      0.024 & 0.037 & 0.060 & 0.103 & 0.199 & 0.317 \\
   Si{\sc iv}/C{\sc iv}  &  
      0.059 & 0.078 & 0.111 & 0.159 & 0.243 & 0.320 \\
   O{\sc iv}]/C{\sc iv}  &  
      0.057 & 0.047 & 0.048 & 0.056 & 0.080 & 0.106 \\
   He{\sc ii}/C{\sc iv}  &  
      0.189 & 0.164 & 0.170 & 0.181 & 0.186 & 0.178 \\
   O{\sc iii}]/C{\sc iv}  &  
      0.064 & 0.088 & 0.127 & 0.182 & 0.259 & 0.322 \\
   Al{\sc ii}/C{\sc iv}  &  
      0.004 & 0.005 & 0.009 & 0.015 & 0.030 & 0.051 \\
   Al{\sc iii}/C{\sc iv}  &  
      0.012 & 0.019 & 0.031 & 0.052 & 0.101 & 0.168 \\
   Si{\sc iii}]/C{\sc iv}  &  
      0.042 & 0.064 & 0.100 & 0.157 & 0.266 & 0.375 \\
   C{\sc iii}]/C{\sc iv}  &  
      0.045 & 0.081 & 0.185 & 0.330 & 0.474 & 0.534 \\
\hline
\end{tabular}
\end{table*}


\begin{table*}
\centering
\vspace{1cm}
\caption{Measured fluxes by adopting local continuum method}
\label{table:11}
\begin{tabular}{l c c c}
\hline\hline
 Line Ratio &
 \multicolumn{2}{c}{our composite$^{\mathrm{a}}$} &
 Vanden Berk et al. (2001) \\
  &
 local cont. &
 equation (1)$^{\mathrm{b}}$ &
 local cont. \\
\hline
N{\sc v}/C{\sc iv}     & 
  0.097 $\pm$ 0.001 & 0.506 $\pm$ 0.004 & 0.097 $\pm$ 0.008 \\ 
Si{\sc ii}/C{\sc iv}   & 
  0.012 $\pm$ 0.001 & 0.089 $\pm$ 0.001 & 0.012 $\pm$ 0.000 \\ 
He{\sc ii}/C{\sc iv}   & 
  0.027 $\pm$ 0.001 & 0.147 $\pm$ 0.001 & 0.021 $\pm$ 0.001 \\ 
Al{\sc iii}/C{\sc iv}  & 
  0.016 $\pm$ 0.001 & 0.064 $\pm$ 0.001 & 0.013 $\pm$ 0.001 \\ 
Si{\sc iii}]/C{\sc iv} & 
  0.006 $\pm$ 0.001 & 0.093 $\pm$ 0.001 & 0.006 $\pm$ 0.001 \\ 
C{\sc iii}]/C{\sc iv}  & 
  0.424 $\pm$ 0.003 & 0.285 $\pm$ 0.005 & 0.630 $\pm$ 0.002 \\ 
\hline
\end{tabular}
\begin{list}{}{}
\item[$^{\mathrm{a}}$]  
   The quasar composite spectrum for
   $2.0 \leq z < 2.5$ and 
   $-24.5 > M_B \geq -25.5$ is used.
\item[$^{\mathrm{b}}$]
   The given values are the same as those in Table 3.
\end{list}
\end{table*}

\clearpage


\begin{table}
\caption{Observed and model line ratios; $2.0 \leq z < 2.5$}
\label{table:12a}
\centering
\begin{tabular}{l c c c c}
\hline\hline
  Parameter                                &
  Obs.                                     &
  Err.$^{\mathrm{a}}$                      &
  \multicolumn{2}{c}{Models$^{\mathrm{b}}$}\\
\hline
\multicolumn{5}{c}{$-24.5 \!\! > \!\! M_B \!\! \geq \!\! -25.5$} \\
\hline
  $Z/Z_\odot$                       &       &       &   2.0 &   2.0 \\
  $\beta$                           &       &       &--1.00 &--1.17 \\
  $\Gamma$                          &       &       &--1.00 &--1.66 \\
  N{\sc v}/C{\sc iv}                & 0.506 & 0.025 & 0.222 & 0.420 \\
  (O{\sc i}+Si{\sc ii})/C{\sc iv}   & 0.061 & 0.010 & 0.091 & 0.041 \\
  C{\sc ii}/C{\sc iv}               & 0.033 & 0.010 & 0.103 & 0.087 \\
  (Si{\sc iv}+O{\sc iv}])/C{\sc iv} & 0.256 & 0.013 & 0.215 & 0.252 \\
  N{\sc iv}]/C{\sc iv}              & 0.016 & 0.010 & 0.062 & 0.066 \\
  He{\sc ii}/C{\sc iv}              & 0.147 & 0.010 & 0.181 & 0.203 \\
  (O{\sc iii}]+Al{\sc ii})/C{\sc iv}& 0.088 & 0.010 & 0.196 & 0.176 \\
  Al{\sc iii}/C{\sc iv}             & 0.064 & 0.010 & 0.052 & 0.046 \\
  Si{\sc iii}]/C{\sc iv}            & 0.093 & 0.010 & 0.157 & 0.085 \\
  C{\sc iii}]/C{\sc iv}             & 0.285 & 0.014 & 0.330 & 0.249 \\
\hline
\multicolumn{5}{c}{$-25.5 \!\! > \!\! M_B \!\! \geq \!\! -26.5$} \\
\hline
  $Z/Z_\odot$                   &       &       &   5.0 &   5.0 \\
  $\beta$                           &       &       &--1.00 &--1.08 \\
  $\Gamma$                          &       &       &--1.00 &--1.51 \\
  N{\sc v}/C{\sc iv}                & 0.717 & 0.036 & 0.499 & 0.776 \\
  (O{\sc i}+Si{\sc ii})/C{\sc iv}   & 0.081 & 0.010 & 0.163 & 0.089 \\
  C{\sc ii}/C{\sc iv}               & 0.047 & 0.010 & 0.199 & 0.187 \\
  (Si{\sc iv}+O{\sc iv}])/C{\sc iv} & 0.334 & 0.017 & 0.322 & 0.409 \\
  N{\sc iv}]/C{\sc iv}              & 0.013 & 0.010 & 0.152 & 0.148 \\
  He{\sc ii}/C{\sc iv}              & 0.152 & 0.010 & 0.186 & 0.197 \\
  (O{\sc iii}]+Al{\sc ii})/C{\sc iv}& 0.093 & 0.010 & 0.289 & 0.275 \\
  Al{\sc iii}/C{\sc iv}             & 0.092 & 0.010 & 0.101 & 0.101 \\ 
  Si{\sc iii}]/C{\sc iv}            & 0.117 & 0.010 & 0.266 & 0.163 \\
  C{\sc iii}]/C{\sc iv}             & 0.326 & 0.016 & 0.474 & 0.332 \\
\hline
\multicolumn{5}{c}{$-26.5 \!\! > \!\! M_B \!\! \geq \!\! -27.5$} \\
\hline
  $Z/Z_\odot$                       &       &       &   5.0 &   5.0 \\
  $\beta$                           &       &       &--1.00 &--1.10 \\
  $\Gamma$                          &       &       &--1.00 &--1.61 \\
  N{\sc v}/C{\sc iv}                & 0.887 & 0.044 & 0.499 & 0.843 \\
  (O{\sc i}+Si{\sc ii})/C{\sc iv}   & 0.086 & 0.010 & 0.163 & 0.078 \\
  C{\sc ii}/C{\sc iv}               & 0.052 & 0.010 & 0.199 & 0.184 \\
  (Si{\sc iv}+O{\sc iv}])/C{\sc iv} & 0.375 & 0.019 & 0.322 & 0.424 \\
  N{\sc iv}]/C{\sc iv}              & 0.015 & 0.010 & 0.152 & 0.150 \\
  He{\sc ii}/C{\sc iv}              & 0.149 & 0.010 & 0.186 & 0.200 \\
  (O{\sc iii}]+Al{\sc ii})/C{\sc iv}& 0.094 & 0.010 & 0.289 & 0.274 \\
  Al{\sc iii}/C{\sc iv}             & 0.105 & 0.010 & 0.101 & 0.099 \\
  Si{\sc iii}]/C{\sc iv}            & 0.127 & 0.010 & 0.266 & 0.148 \\
  C{\sc iii}]/C{\sc iv}             & 0.326 & 0.016 & 0.474 & 0.314 \\
\hline
\end{tabular}
\end{table}

\begin{table}
\addtocounter{table}{-1}
\caption{Observed and model line ratios; $2.0 \leq z < 2.5$ (continued)}
\label{table:12b}
\centering
\begin{tabular}{l c c c c}
\hline\hline
  Parameter                                &
  Obs.                                     &
  Err.$^{\mathrm{a}}$                      &
  \multicolumn{2}{c}{Models$^{\mathrm{b}}$}\\
\hline
\multicolumn{5}{c}{$-27.5 \!\! > \!\! M_B \!\! \geq \!\! -28.5$} \\
\hline
  $Z/Z_\odot$                   &       &       &   5.0 &   5.0 \\
  $\beta$                           &       &       &--1.00 &--1.11 \\
  $\Gamma$                          &       &       &--1.00 &--1.63 \\
  N{\sc v}/C{\sc iv}                & 0.983 & 0.049 & 0.499 & 0.858 \\
  (O{\sc i}+Si{\sc ii})/C{\sc iv}   & 0.085 & 0.010 & 0.163 & 0.076 \\
  C{\sc ii}/C{\sc iv}               & 0.050 & 0.010 & 0.199 & 0.184 \\
  (Si{\sc iv}+O{\sc iv}])/C{\sc iv} & 0.398 & 0.020 & 0.322 & 0.427 \\
  N{\sc iv}]/C{\sc iv}              & 0.020 & 0.010 & 0.152 & 0.150 \\
  He{\sc ii}/C{\sc iv}              & 0.140 & 0.010 & 0.186 & 0.201 \\
  (O{\sc iii}]+Al{\sc ii})/C{\sc iv}& 0.101 & 0.010 & 0.289 & 0.275 \\
  Al{\sc iii}/C{\sc iv}             & 0.124 & 0.010 & 0.101 & 0.098 \\
  Si{\sc iii}]/C{\sc iv}            & 0.130 & 0.010 & 0.266 & 0.145 \\
  C{\sc iii}]/C{\sc iv}             & 0.348 & 0.017 & 0.474 & 0.311 \\
\hline
\multicolumn{5}{c}{$-28.5 \!\! > \!\! M_B \!\! \geq \!\! -29.5$} \\
\hline
  $Z/Z_\odot$                   &       &       &   5.0 &   5.0 \\
  $\beta$                           &       &       &--1.00 &--1.06 \\
  $\Gamma$                          &       &       &--1.00 &--1.50 \\
  N{\sc v}/C{\sc iv}                & 1.145 & 0.057 & 0.499 & 0.770 \\
  (O{\sc i}+Si{\sc ii})/C{\sc iv}   & 0.145 & 0.010 & 0.163 & 0.091 \\
  C{\sc ii}/C{\sc iv}               & 0.089 & 0.010 & 0.199 & 0.189 \\
  (Si{\sc iv}+O{\sc iv}])/C{\sc iv} & 0.419 & 0.021 & 0.322 & 0.413 \\
  N{\sc iv}]/C{\sc iv}              & 0.000 & 0.010 & 0.152 & 0.144 \\
  He{\sc ii}/C{\sc iv}              & 0.133 & 0.010 & 0.186 & 0.198 \\
  (O{\sc iii}]+Al{\sc ii})/C{\sc iv}& 0.104 & 0.010 & 0.289 & 0.272 \\
  Al{\sc iii}/C{\sc iv}             & 0.146 & 0.010 & 0.101 & 0.105 \\
  Si{\sc iii}]/C{\sc iv}            & 0.110 & 0.010 & 0.266 & 0.163 \\
  C{\sc iii}]/C{\sc iv}             & 0.395 & 0.020 & 0.474 & 0.318 \\
\hline
\end{tabular}
\begin{list}{}{}
\item[$^{\mathrm{a}}$]
   Errors adopted for carrying out model fittings.
   Values are changed from those presented in
   Table 3; see text for details.
\item[$^{\mathrm{b}}$]
   Model predictions with fixed $\beta$ and $\Gamma$
   and those with varying $\beta$ and $\Gamma$ are
   given in the left and right columns, respectively.
\end{list}
\end{table}

\clearpage


\begin{table}
\caption{Observed and model line ratios; $2.5 \leq z < 3.0$}
\label{table:13a}
\centering
\begin{tabular}{l c c c c}
\hline\hline
  Parameter                                &
  Obs.                                     &
  Err.$^{\mathrm{a}}$                      &
  \multicolumn{2}{c}{Models$^{\mathrm{b}}$}\\
\hline
\multicolumn{5}{c}{$-24.5 \!\! > \!\! M_B \!\! \geq \!\! -25.5$} \\
\hline
  $Z/Z_\odot$                       &       &       &   5.0 &   5.0 \\
  $\beta$                           &       &       &--1.00 &--1.04 \\
  $\Gamma$                          &       &       &--1.00 &--1.23 \\
  N{\sc v}/C{\sc iv}                & 0.561 & 0.028 & 0.499 & 0.608 \\
  (O{\sc i}+Si{\sc ii})/C{\sc iv}   & 0.121 & 0.010 & 0.163 & 0.122 \\
  C{\sc ii}/C{\sc iv}               & 0.065 & 0.010 & 0.199 & 0.190 \\
  (Si{\sc iv}+O{\sc iv}])/C{\sc iv} & 0.254 & 0.013 & 0.322 & 0.355 \\
  N{\sc iv}]/C{\sc iv}              & 0.021 & 0.010 & 0.152 & 0.152 \\
  He{\sc ii}/C{\sc iv}              & 0.151 & 0.010 & 0.186 & 0.187 \\
  (O{\sc iii}]+Al{\sc ii})/C{\sc iv}& 0.087 & 0.010 & 0.289 & 0.285 \\
  Al{\sc iii}/C{\sc iv}             & 0.078 & 0.010 & 0.101 & 0.099 \\
  Si{\sc iii}]/C{\sc iv}            & 0.070 & 0.010 & 0.266 & 0.219 \\
  C{\sc iii}]/C{\sc iv}             & 0.375 & 0.019 & 0.474 & 0.423 \\
\hline
\multicolumn{5}{c}{$-25.5 \!\! > \!\! M_B \!\! \geq \!\! -26.5$} \\
\hline
  $Z/Z_\odot$                       &       &       &   5.0 &   2.0 \\
  $\beta$                           &       &       &--1.00 &--1.17 \\
  $\Gamma$                          &       &       &--1.00 &--1.64 \\
  N{\sc v}/C{\sc iv}                & 0.578 & 0.029 & 0.499 & 0.407 \\
  (O{\sc i}+Si{\sc ii})/C{\sc iv}   & 0.089 & 0.010 & 0.163 & 0.041 \\
  C{\sc ii}/C{\sc iv}               & 0.037 & 0.010 & 0.199 & 0.086 \\
  (Si{\sc iv}+O{\sc iv}])/C{\sc iv} & 0.252 & 0.013 & 0.322 & 0.248 \\
  N{\sc iv}]/C{\sc iv}              & 0.029 & 0.010 & 0.152 & 0.067 \\
  He{\sc ii}/C{\sc iv}              & 0.147 & 0.010 & 0.186 & 0.200 \\
  (O{\sc iii}]+Al{\sc ii})/C{\sc iv}& 0.098 & 0.010 & 0.289 & 0.178 \\
  Al{\sc iii}/C{\sc iv}             & 0.071 & 0.010 & 0.101 & 0.045 \\
  Si{\sc iii}]/C{\sc iv}            & 0.063 & 0.010 & 0.266 & 0.087 \\
  C{\sc iii}]/C{\sc iv}             & 0.351 & 0.018 & 0.474 & 0.258 \\
\hline
\multicolumn{5}{c}{$-26.5 \!\! > \!\! M_B \!\! \geq \!\! -27.5$} \\
\hline
  $Z/Z_\odot$                   &       &       &   5.0 &   5.0 \\
  $\beta$                           &       &       &--1.00 &--1.11 \\
  $\Gamma$                          &       &       &--1.00 &--1.58 \\
  N{\sc v}/C{\sc iv}                & 0.858 & 0.043 & 0.499 & 0.820 \\
  (O{\sc i}+Si{\sc ii})/C{\sc iv}   & 0.095 & 0.010 & 0.163 & 0.081 \\
  C{\sc ii}/C{\sc iv}               & 0.055 & 0.010 & 0.199 & 0.183 \\
  (Si{\sc iv}+O{\sc iv}])/C{\sc iv} & 0.360 & 0.018 & 0.322 & 0.413 \\
  N{\sc iv}]/C{\sc iv}              & 0.022 & 0.010 & 0.152 & 0.153 \\
  He{\sc ii}/C{\sc iv}              & 0.151 & 0.010 & 0.289 & 0.278 \\
  (O{\sc iii}]+Al{\sc ii})/C{\sc iv}& 0.106 & 0.010 & 0.289 & 0.278 \\
  Al{\sc iii}/C{\sc iv}             & 0.101 & 0.010 & 0.101 & 0.096 \\
  Si{\sc iii}]/C{\sc iv}            & 0.108 & 0.010 & 0.266 & 0.155 \\
  C{\sc iii}]/C{\sc iv}             & 0.354 & 0.018 & 0.474 & 0.335 \\
\hline
\end{tabular}
\end{table}

\begin{table}
\addtocounter{table}{-1}
\caption{Observed and model line ratios; $2.5 \leq z < 3.0$ (continued)}
\label{table:13b}
\centering
\begin{tabular}{l c c c c}
\hline\hline
  Parameter                                &
  Obs.                                     &
  Err.$^{\mathrm{a}}$                      &
  \multicolumn{2}{c}{Models$^{\mathrm{b}}$}\\
\hline
\multicolumn{5}{c}{$-27.5 \!\! > \!\! M_B \!\! \geq \!\! -28.5$} \\
\hline
  $Z/Z_\odot$                       &       &       &   5.0 &   5.0 \\
  $\beta$                           &       &       &--1.00 &--1.07 \\
  $\Gamma$                          &       &       &--1.00 &--1.54 \\
  N{\sc v}/C{\sc iv}                & 1.004 & 0.050 & 0.499 & 0.800 \\
  (O{\sc i}+Si{\sc ii})/C{\sc iv}   & 0.119 & 0.010 & 0.163 & 0.086 \\
  C{\sc ii}/C{\sc iv}               & 0.059 & 0.010 & 0.199 & 0.188 \\
  (Si{\sc iv}+O{\sc iv}])/C{\sc iv} & 0.401 & 0.020 & 0.322 & 0.421 \\
  N{\sc iv}]/C{\sc iv}              & 0.019 & 0.010 & 0.152 & 0.144 \\
  He{\sc ii}/C{\sc iv}              & 0.143 & 0.010 & 0.186 & 0.199 \\
  (O{\sc iii}]+Al{\sc ii})/C{\sc iv}& 0.104 & 0.010 & 0.289 & 0.271 \\
  Al{\sc iii}/C{\sc iv}             & 0.109 & 0.010 & 0.101 & 0.104 \\
  Si{\sc iii}]/C{\sc iv}            & 0.152 & 0.010 & 0.266 & 0.155 \\
  C{\sc iii}]/C{\sc iv}             & 0.328 & 0.016 & 0.474 & 0.309 \\
\hline
\multicolumn{5}{c}{$-28.5 \!\! > \!\! M_B \!\! \geq \!\! -29.5$} \\
\hline
  $Z/Z_\odot$                   &       &       &  10.0 &  10.0 \\
  $\beta$                           &       &       &--1.00 &--1.03 \\
  $\Gamma$                          &       &       &--1.00 &--1.53 \\
  N{\sc v}/C{\sc iv}                & 1.248 & 0.062 & 0.837 & 1.191 \\
  (O{\sc i}+Si{\sc ii})/C{\sc iv}   & 0.112 & 0.010 & 0.250 & 0.122 \\
  C{\sc ii}/C{\sc iv}               & 0.074 & 0.010 & 0.317 & 0.290 \\
  (Si{\sc iv}+O{\sc iv}])/C{\sc iv} & 0.462 & 0.023 & 0.426 & 0.586 \\
  N{\sc iv}]/C{\sc iv}              & 0.012 & 0.010 & 0.298 & 0.251 \\
  He{\sc ii}/C{\sc iv}              & 0.125 & 0.010 & 0.178 & 0.180 \\
  (O{\sc iii}]+Al{\sc ii})/C{\sc iv}& 0.096 & 0.010 & 0.373 & 0.370 \\
  Al{\sc iii}/C{\sc iv}             & 0.165 & 0.010 & 0.168 & 0.175 \\
  Si{\sc iii}]/C{\sc iv}            & 0.204 & 0.010 & 0.375 & 0.211 \\
  C{\sc iii}]/C{\sc iv}             & 0.331 & 0.017 & 0.534 & 0.323 \\
\hline
\end{tabular}
\begin{list}{}{}
\item[$^{\mathrm{a}}$]
   Errors adopted for carrying out model fittings.
   Values are changed from those presented in
   Table 4; see text for details.
\item[$^{\mathrm{b}}$]
   Model predictions with fixed $\beta$ and $\Gamma$
   and those with varying $\beta$ and $\Gamma$ are
   given in the left and right columns, respectively.
\end{list}
\end{table}

\clearpage


\begin{table}
\caption{Observed and model line ratios; $3.0 \leq z < 3.5$}
\label{table:14a}
\centering
\begin{tabular}{l c c c c}
\hline\hline
  Parameter                                &
  Obs.                                     &
  Err.$^{\mathrm{a}}$                      &
  \multicolumn{2}{c}{Models$^{\mathrm{b}}$}\\
\hline
\multicolumn{5}{c}{$-25.5 \!\! > \!\! M_B \!\! \geq \!\! -26.5$} \\
\hline
  $Z/Z_\odot$                       &       &       &   5.0 &   2.0 \\
  $\beta$                           &       &       &--1.00 &--1.17 \\
  $\Gamma$                          &       &       &--1.00 &--1.68 \\
  N{\sc v}/C{\sc iv}                & 0.635 & 0.032 & 0.499 & 0.433 \\
  (O{\sc i}+Si{\sc ii})/C{\sc iv}   & 0.090 & 0.010 & 0.163 & 0.040 \\
  C{\sc ii}/C{\sc iv}               & 0.050 & 0.010 & 0.199 & 0.089 \\
  (Si{\sc iv}+O{\sc iv}])/C{\sc iv} & 0.279 & 0.014 & 0.322 & 0.257 \\
  N{\sc iv}]/C{\sc iv}              & 0.030 & 0.010 & 0.152 & 0.066 \\
  He{\sc ii}/C{\sc iv}              & 0.148 & 0.010 & 0.186 & 0.206 \\
  (O{\sc iii}]+Al{\sc ii})/C{\sc iv}& 0.111 & 0.010 & 0.289 & 0.174 \\
  Al{\sc iii}/C{\sc iv}             & 0.072 & 0.010 & 0.101 & 0.047 \\
  Si{\sc iii}]/C{\sc iv}            & 0.068 & 0.010 & 0.266 & 0.083 \\
  C{\sc iii}]/C{\sc iv}             & 0.336 & 0.017 & 0.474 & 0.240 \\
\hline
\multicolumn{5}{c}{$-26.5 \!\! > \!\! M_B \!\! \geq \!\! -27.5$} \\
\hline
  $Z/Z_\odot$                   &       &       &   5.0 &   5.0 \\
  $\beta$                           &       &       &--1.00 &--1.07 \\
  $\Gamma$                          &       &       &--1.00 &--1.54 \\
  N{\sc v}/C{\sc iv}                & 0.819 & 0.041 & 0.499 & 0.795 \\
  (O{\sc i}+Si{\sc ii})/C{\sc iv}   & 0.095 & 0.010 & 0.163 & 0.086 \\
  C{\sc ii}/C{\sc iv}               & 0.049 & 0.010 & 0.199 & 0.188 \\
  (Si{\sc iv}+O{\sc iv}])/C{\sc iv} & 0.329 & 0.016 & 0.322 & 0.418 \\
  N{\sc iv}]/C{\sc iv}              & 0.023 & 0.010 & 0.152 & 0.145 \\
  He{\sc ii}/C{\sc iv}              & 0.148 & 0.010 & 0.186 & 0.199 \\
  (O{\sc iii}]+Al{\sc ii})/C{\sc iv}& 0.111 & 0.010 & 0.289 & 0.272 \\
  Al{\sc iii}/C{\sc iv}             & 0.108 & 0.010 & 0.101 & 0.103 \\
  Si{\sc iii}]/C{\sc iv}            & 0.112 & 0.010 & 0.266 & 0.157 \\
  C{\sc iii}]/C{\sc iv}             & 0.321 & 0.016 & 0.474 & 0.314 \\
\hline
\end{tabular}
\end{table}

\begin{table}
\addtocounter{table}{-1}
\caption{Observed and model line ratios; $3.0 \leq z < 3.5$ (continued)}
\label{table:14b}
\centering
\begin{tabular}{l c c c c}
\hline\hline
  Parameter                                &
  Obs.                                     &
  Err.$^{\mathrm{a}}$                      &
  \multicolumn{2}{c}{Models$^{\mathrm{b}}$}\\
\hline
\multicolumn{5}{c}{$-27.5 \!\! > \!\! M_B \!\! \geq \!\! -28.5$} \\
\hline
  $Z/Z_\odot$                       &       &       &   5.0 &   5.0 \\
  $\beta$                           &       &       &--1.00 &--1.11 \\
  $\Gamma$                          &       &       &--1.00 &--1.61 \\
  N{\sc v}/C{\sc iv}                & 1.034 & 0.052 & 0.499 & 0.840 \\
  (O{\sc i}+Si{\sc ii})/C{\sc iv}   & 0.109 & 0.010 & 0.163 & 0.078 \\
  C{\sc ii}/C{\sc iv}               & 0.055 & 0.010 & 0.199 & 0.183 \\
  (Si{\sc iv}+O{\sc iv}])/C{\sc iv} & 0.402 & 0.020 & 0.322 & 0.420 \\
  N{\sc iv}]/C{\sc iv}              & 0.019 & 0.010 & 0.152 & 0.152 \\
  He{\sc ii}/C{\sc iv}              & 0.148 & 0.010 & 0.186 & 0.200 \\
  (O{\sc iii}]+Al{\sc ii})/C{\sc iv}& 0.108 & 0.010 & 0.289 & 0.277 \\
  Al{\sc iii}/C{\sc iv}             & 0.126 & 0.010 & 0.101 & 0.097 \\
  Si{\sc iii}]/C{\sc iv}            & 0.105 & 0.010 & 0.266 & 0.149 \\
  C{\sc iii}]/C{\sc iv}             & 0.379 & 0.019 & 0.474 & 0.323 \\
\hline
\multicolumn{5}{c}{$-28.5 \!\! > \!\! M_B \!\! \geq \!\! -29.5$} \\
\hline
  $Z/Z_\odot$                       &       &       &   5.0 &   5.0 \\
  $\beta$                           &       &       &--1.00 &--1.06 \\
  $\Gamma$                          &       &       &--1.00 &--1.61 \\
  N{\sc v}/C{\sc iv}                & 0.979 & 0.049 & 0.499 & 0.846 \\
  (O{\sc i}+Si{\sc ii})/C{\sc iv}   & 0.094 & 0.010 & 0.163 & 0.080 \\
  C{\sc ii}/C{\sc iv}               & 0.039 & 0.010 & 0.199 & 0.191 \\
  (Si{\sc iv}+O{\sc iv}])/C{\sc iv} & 0.370 & 0.018 & 0.322 & 0.443 \\
  N{\sc iv}]/C{\sc iv}              & 0.045 & 0.010 & 0.152 & 0.138 \\
  He{\sc ii}/C{\sc iv}              & 0.102 & 0.010 & 0.186 & 0.203 \\
  (O{\sc iii}]+Al{\sc ii})/C{\sc iv}& 0.100 & 0.010 & 0.289 & 0.263 \\
  Al{\sc iii}/C{\sc iv}             & 0.117 & 0.010 & 0.101 & 0.109 \\
  Si{\sc iii}]/C{\sc iv}            & 0.134 & 0.010 & 0.266 & 0.141 \\
  C{\sc iii}]/C{\sc iv}             & 0.281 & 0.014 & 0.474 & 0.269 \\
\hline
\end{tabular}
\begin{list}{}{}
\item[$^{\mathrm{a}}$]
   Errors adopted for carrying out model fittings.
   Values are changed from those presented in
   Table 5; see text for details.
\item[$^{\mathrm{b}}$]
   Model predictions with fixed $\beta$ and $\Gamma$
   and those with varying $\beta$ and $\Gamma$ are
   given in the left and right columns, respectively.
\end{list}
\end{table}

\clearpage


\begin{table}
\caption{Observed and model line ratios; $3.5 \leq z < 4.0$}
\label{table:15a}
\centering
\begin{tabular}{l c c c c}
\hline\hline
  Parameter                                &
  Obs.                                     &
  Err.$^{\mathrm{a}}$                      &
  \multicolumn{2}{c}{Models$^{\mathrm{b}}$}\\
\hline
\multicolumn{5}{c}{$-25.5 \!\! > \!\! M_B \!\! \geq \!\! -26.5$} \\
\hline
  $Z/Z_\odot$                       &       &       &   5.0 &   2.0 \\
  $\beta$                           &       &       &--1.00 &--1.13 \\
  $\Gamma$                          &       &       &--1.00 &--1.60 \\
  N{\sc v}/C{\sc iv}                & 0.618 & 0.031 & 0.499 & 0.406 \\
  (O{\sc i}+Si{\sc ii})/C{\sc iv}   & 0.098 & 0.010 & 0.163 & 0.045 \\
  C{\sc ii}/C{\sc iv}               & 0.030 & 0.010 & 0.199 & 0.091 \\
  (Si{\sc iv}+O{\sc iv}])/C{\sc iv} & 0.261 & 0.013 & 0.322 & 0.255 \\
  N{\sc iv}]/C{\sc iv}              & 0.038 & 0.010 & 0.152 & 0.064 \\
  He{\sc ii}/C{\sc iv}              & 0.153 & 0.010 & 0.186 & 0.202 \\
  (O{\sc iii}]+Al{\sc ii})/C{\sc iv}& 0.097 & 0.010 & 0.289 & 0.175 \\
  Al{\sc iii}/C{\sc iv}             & 0.084 & 0.010 & 0.101 & 0.049 \\
  Si{\sc iii}]/C{\sc iv}            & 0.066 & 0.010 & 0.266 & 0.090 \\
  C{\sc iii}]/C{\sc iv}             & 0.355 & 0.018 & 0.474 & 0.243 \\
\hline
\multicolumn{5}{c}{$-26.5 \!\! > \!\! M_B \!\! \geq \!\! -27.5$} \\
\hline
  $Z/Z_\odot$                       &       &       &   5.0 &   5.0 \\
  $\beta$                           &       &       &--1.00 &--1.08 \\
  $\Gamma$                          &       &       &--1.00 &--1.54 \\
  N{\sc v}/C{\sc iv}                & 0.856 & 0.043 & 0.499 & 0.796 \\
  (O{\sc i}+Si{\sc ii})/C{\sc iv}   & 0.105 & 0.010 & 0.163 & 0.086 \\
  C{\sc ii}/C{\sc iv}               & 0.051 & 0.010 & 0.199 & 0.186 \\
  (Si{\sc iv}+O{\sc iv}])/C{\sc iv} & 0.347 & 0.017 & 0.322 & 0.415 \\
  N{\sc iv}]/C{\sc iv}              & 0.025 & 0.010 & 0.152 & 0.148 \\
  He{\sc ii}/C{\sc iv}              & 0.145 & 0.010 & 0.186 & 0.198 \\
  (O{\sc iii}]+Al{\sc ii})/C{\sc iv}& 0.101 & 0.010 & 0.289 & 0.274 \\
  Al{\sc iii}/C{\sc iv}             & 0.098 & 0.010 & 0.101 & 0.101 \\
  Si{\sc iii}]/C{\sc iv}            & 0.122 & 0.010 & 0.266 & 0.158 \\
  C{\sc iii}]/C{\sc iv}             & 0.332 & 0.017 & 0.474 & 0.323 \\
\hline
\end{tabular}
\end{table}

\begin{table}
\addtocounter{table}{-1}
\caption{Observed and model line ratios; $3.5 \leq z < 4.0$ (continued)}
\label{table:15b}
\centering
\begin{tabular}{l c c c c}
\hline\hline
  Parameter                                &
  Obs.                                     &
  Err.$^{\mathrm{a}}$                      &
  \multicolumn{2}{c}{Models$^{\mathrm{b}}$}\\
\hline
\multicolumn{5}{c}{$-27.5 \!\! > \!\! M_B \!\! \geq \!\! -28.5$} \\
\hline
  $Z/Z_\odot$                       &       &       &   5.0 &   5.0 \\
  $\beta$                           &       &       &--1.00 &--1.07 \\
  $\Gamma$                          &       &       &--1.00 &--1.50 \\
  N{\sc v}/C{\sc iv}                & 1.062 & 0.053 & 0.499 & 0.773 \\
  (O{\sc i}+Si{\sc ii})/C{\sc iv}   & 0.115 & 0.010 & 0.163 & 0.089 \\
  C{\sc ii}/C{\sc iv}               & 0.062 & 0.010 & 0.199 & 0.187 \\
  (Si{\sc iv}+O{\sc iv}])/C{\sc iv} & 0.395 & 0.020 & 0.322 & 0.410 \\
  N{\sc iv}]/C{\sc iv}              & 0.016 & 0.010 & 0.152 & 0.147 \\
  He{\sc ii}/C{\sc iv}              & 0.128 & 0.010 & 0.186 & 0.197 \\
  (O{\sc iii}]+Al{\sc ii})/C{\sc iv}& 0.089 & 0.010 & 0.289 & 0.274 \\
  Al{\sc iii}/C{\sc iv}             & 0.122 & 0.010 & 0.101 & 0.102 \\
  Si{\sc iii}]/C{\sc iv}            & 0.186 & 0.010 & 0.266 & 0.164 \\
  C{\sc iii}]/C{\sc iv}             & 0.343 & 0.017 & 0.474 & 0.329 \\
\hline
\multicolumn{5}{c}{$-28.5 \!\! > \!\! M_B \!\! \geq \!\! -29.5$} \\
\hline
  $Z/Z_\odot$                       &       &       &   5.0 &   5.0 \\
  $\beta$                           &       &       &--1.00 &--1.02 \\
  $\Gamma$                          &       &       &--1.00 &--1.28 \\
  N{\sc v}/C{\sc iv}                & 1.006 & 0.050 & 0.499 & 0.639 \\ 
  (O{\sc i}+Si{\sc ii})/C{\sc iv}   & 0.180 & 0.010 & 0.163 & 0.118 \\
  C{\sc ii}/C{\sc iv}               & 0.105 & 0.010 & 0.199 & 0.194 \\
  (Si{\sc iv}+O{\sc iv}])/C{\sc iv} & 0.421 & 0.021 & 0.322 & 0.377 \\
  N{\sc iv}]/C{\sc iv}              & 0.031 & 0.010 & 0.152 & 0.143 \\
  He{\sc ii}/C{\sc iv}              & 0.108 & 0.010 & 0.186 & 0.194 \\
  (O{\sc iii}]+Al{\sc ii})/C{\sc iv}& 0.056 & 0.010 & 0.289 & 0.276 \\
  Al{\sc iii}/C{\sc iv}             & 0.124 & 0.010 & 0.101 & 0.107 \\
  Si{\sc iii}]/C{\sc iv}            & 0.193 & 0.010 & 0.266 & 0.203 \\
  C{\sc iii}]/C{\sc iv}             & 0.426 & 0.021 & 0.474 & 0.369 \\
\hline
\end{tabular}
\begin{list}{}{}
\item[$^{\mathrm{a}}$]
   Errors adopted for carrying out model fittings.
   Values are changed from those presented in
   Table 6; see text for details.
\item[$^{\mathrm{b}}$]
   Model predictions with fixed $\beta$ and $\Gamma$
   and those with varying $\beta$ and $\Gamma$ are
   given in the left and right columns, respectively.
\end{list}
\end{table}

\clearpage


\begin{table}
\caption{Observed and model line ratios; $4.0 \leq z < 4.5$}
\label{table:16}
\centering
\begin{tabular}{l c c c c}
\hline\hline
  Parameter                                &
  Obs.                                     &
  Err.$^{\mathrm{a}}$                      &
  \multicolumn{2}{c}{Models$^{\mathrm{b}}$}\\
\hline
\multicolumn{5}{c}{$-25.5 \!\! > \!\! M_B \!\! \geq \!\! -26.5$} \\
\hline
  $Z/Z_\odot$                       &       &       &   5.0 &   5.0 \\
  $\beta$                           &       &       &--1.00 &--1.06 \\
  $\Gamma$                          &       &       &--1.00 &--1.24 \\
  N{\sc v}/C{\sc iv}                & 0.724 & 0.036 & 0.499 & 0.614 \\
  (O{\sc i}+Si{\sc ii})/C{\sc iv}   & 0.148 & 0.010 & 0.163 & 0.119 \\
  C{\sc ii}/C{\sc iv}               & 0.067 & 0.010 & 0.199 & 0.188 \\
  (Si{\sc iv}+O{\sc iv}])/C{\sc iv} & 0.349 & 0.017 & 0.322 & 0.352 \\
  N{\sc iv}]/C{\sc iv}              & 0.019 & 0.010 & 0.152 & 0.155 \\
  He{\sc ii}/C{\sc iv}              & 0.132 & 0.010 & 0.186 & 0.185 \\
  (O{\sc iii}]+Al{\sc ii})/C{\sc iv}& 0.049 & 0.010 & 0.289 & 0.288 \\
\hline
\multicolumn{5}{c}{$-26.5 \!\! > \!\! M_B \!\! \geq \!\! -27.5$} \\
\hline
  $Z/Z_\odot$                       &       &       &  10.0 &   5.0 \\
  $\beta$                           &       &       &--1.00 &--1.03 \\
  $\Gamma$                          &       &       &--1.00 &--1.40 \\
  N{\sc v}/C{\sc iv}                & 0.933 & 0.047 & 0.837 & 0.709 \\
  (O{\sc i}+Si{\sc ii})/C{\sc iv}   & 0.146 & 0.010 & 0.250 & 0.103 \\
  C{\sc ii}/C{\sc iv}               & 0.068 & 0.010 & 0.317 & 0.194 \\
  (Si{\sc iv}+O{\sc iv}])/C{\sc iv} & 0.383 & 0.019 & 0.426 & 0.404 \\
  N{\sc iv}]/C{\sc iv}              & 0.010 & 0.010 & 0.298 & 0.139 \\
  He{\sc ii}/C{\sc iv}              & 0.095 & 0.010 & 0.178 & 0.198 \\
  (O{\sc iii}]+Al{\sc ii})/C{\sc iv}& 0.064 & 0.010 & 0.373 & 0.269 \\
\hline
\multicolumn{5}{c}{$-27.5 \!\! > \!\! M_B \!\! \geq \!\! -28.5$} \\
\hline 
  $Z/Z_\odot$                       &       &       &  10.0 &   5.0 \\
  $\beta$                           &       &       &--1.00 &--1.02 \\
  $\Gamma$                          &       &       &--1.00 &--1.54 \\
  N{\sc v}/C{\sc iv}                & 1.075 & 0.054 & 0.837 & 0.805 \\
  (O{\sc i}+Si{\sc ii})/C{\sc iv}   & 0.133 & 0.010 & 0.250 & 0.088 \\
  C{\sc ii}/C{\sc iv}               & 0.064 & 0.010 & 0.317 & 0.197 \\
  (Si{\sc iv}+O{\sc iv}])/C{\sc iv} & 0.409 & 0.020 & 0.426 & 0.445 \\
  N{\sc iv}]/C{\sc iv}              & 0.007 & 0.010 & 0.298 & 0.130 \\
  He{\sc ii}/C{\sc iv}              & 0.095 & 0.010 & 0.178 & 0.204 \\
  (O{\sc iii}]+Al{\sc ii})/C{\sc iv}& 0.075 & 0.010 & 0.373 & 0.257 \\
\hline
\end{tabular}
\begin{list}{}{}
\item[$^{\mathrm{a}}$]
   Errors adopted for carrying out model fittings.
   Values are changed from those presented in
   Table 7; see text for details.
\item[$^{\mathrm{b}}$]
   Model predictions with fixed $\beta$ and $\Gamma$
   and those with varying $\beta$ and $\Gamma$ are
   given in the left and right columns, respectively.
\end{list}
\end{table}


\begin{thebibliography}{}
\bibitem[]{} Abazajian, K., Adelman-McCarthy, J. K., 
             Ag\"{u}eros, M. A., et al. 2004, AJ, 128, 502
\bibitem[]{} Arimoto, N., \& Yoshii, Y. 1987, A\&A, 173, 23
\bibitem[]{} Baldwin, J. A. 1997, ASP Conf. Ser., 113, 80
\bibitem[]{} Baldwin, J. A., Ferland, G. J., Korista, K. T., and
             Verner, D. 1995, ApJ, 455, L119
\bibitem[]{} Baldwin, J. A., Hamann, F., Korista, K. T., et al.
             2003, ApJ, 583, 649
\bibitem[]{} Baldwin, J. A., \& Netzer, H. 1978, ApJ, 226, 1
\bibitem[]{} Baskin, A., \& Laor, A. 2005, MNRAS, 356, 1029
\bibitem[]{} Boyle, B. J. 1990, MNRAS, 243, 231
\bibitem[]{} Cardelli, J. A., Clayton, G. C., \& Mathis, J. S.
             1989, ApJ, 345, 245
\bibitem[]{} Clavel, J., Reichert, G. A., Alloin, D., et al. 
             1991, ApJ, 366, 64
\bibitem[]{} Collin-Souffrin, S., Dumont, S., \& Tully, J.
             1982, A\&A, 106. 302
\bibitem[]{} Collin-Souffrin, S., \& Lasota, J. -P. 1988, PASP, 100, 1041
\bibitem[]{} Corbin, M. R. 1990, ApJ, 357, 346
\bibitem[]{} Corbin, M. R. 1997, ApJS, 113, 245
\bibitem[]{} Croom, S. M., Rhook, K., Corbett, E. A., et al. 
             2002, MNRAS, 337, 275
\bibitem[]{} Croom, S. M., Smith, R. J., Boyle, B. J., et al.
             2004, MNRAS, 349, 1397
\bibitem[]{} Davidson, K. 1977, ApJ, 218, 20
\bibitem[]{} Dietrich, M., Appenzeller, I., Wagner, S. J., et al. 
             1999, A\&A, 352, L1
\bibitem[]{} Dietrich, M., Hamann, F., Shields, J. C., et al. 
             2002, ApJ, 581, 912
\bibitem[]{} Dietrich, M., Hamann, F., Shields, J. C., et al. 
             2003, ApJ, 589, 722
\bibitem[]{} Di Matteo, T., Croft, R. A. C., Springel, V., \&
             Hernquist, L. 2004, ApJ, 610, 80
\bibitem[]{} Espey, B. R., Carswell, R. F., Bailey, J. A., 
             Smith, M. G., \& Ward, M. J. 1989, ApJ, 342, 666
\bibitem[]{} Fan, X., Hennawi, J. F., Richards, G. T., et al.
             2004, AJ, 128, 515
\bibitem[]{} Fan, X., Narayanan, V. K., Lupton, R. H., et al. 
             2001, AJ, 122, 2833
\bibitem[]{} Fan, X., Narayanan, V. K., Strauss, M. A., et al.
             2002, AJ, 123, 1247
\bibitem[]{} Fan, X., Strauss, M. A., Schneider, D. P., et al. 
             2003, AJ, 125, 1649
\bibitem[]{} Ferland, G. J. 1997, Hazy: A Brief Introduction to $Cloudy$
             94.00 (Lexington: Univ. Kentucky Dept. Phys. Astron.)
\bibitem[]{} Ferland, G. J., Baldwin, J. A., Korista, K. T., et al. 
             1996, ApJ, 461, 683
\bibitem[]{} Ferrarese, L., \& Merritt, D. 2000, ApJ, 539, L9
\bibitem[]{} Foltz, C. B., Chaffee, F. H. Jr., Hewett, P. C., et al.
             1987, AJ, 94, 1423
\bibitem[]{} Francis, P. J., \& Koratkar, A. 1995, MNRAS, 274, 504
\bibitem[]{} Gaskell, C. M. 1982, ApJ, 263, 79
\bibitem[]{} Gebhardt, K., Bender, R., Bower, G., et al.
             2000, ApJ, 539, L13
\bibitem[]{} Gnedin, N. Y. 2004, ApJ, 610, 9
\bibitem[]{} Grevesse, N., \& Anders, E. 1989, in AIP Conf. Proc. 183,
             Cosmic Abundance of Matter, ed. C. J. Waddington
             (New York: AIP), 1
\bibitem[]{} Grevesse, N., \& Noels, A. 1993, in Origin and Evolution of 
             the Elements, eds. N. Prantzos, E. Vangioni-Flam, \&
             M. Casse (Cambridge Univ. Press), 15
\bibitem[]{} Hall, P., Anderson, S. F., Strauss, M. A., et al.
             2002, ApJS, 141, 267
\bibitem[]{} Hamann, F., \& Ferland, G. J. 1992, ApJ, 391, L53
\bibitem[]{} Hamann, F., \& Ferland, G. J. 1993, ApJ, 418, 11
\bibitem[]{} Hamann, F., \& Ferland, G. J. 1999, ARA\&A, 37, 487
\bibitem[]{} Hamann, F., Korista, K. T., Ferland, G. J., 
             Warner, C., \& Baldwin, J. A. 2002, ApJ, 564, 592
\bibitem[]{} Kawakatu, N., Umemura, M., \& Mori, M. 2003, ApJ, 583, 85
\bibitem[]{} Kobulnicky, H. A., \& Koo, D. C. 2000, ApJ, 545, 712
\bibitem[]{} Korista, K. T., Alloin, D., Barr, P. et al. 
             1995, ApJS, 97, 285
\bibitem[]{} Korista, K. T., Baldwin, J. A., \& Ferland, G. J.
             1998, ApJ, 507, 24
\bibitem[]{} Korista, K. T., \& Goad, M. R. 2000, ApJ, 536, 284
\bibitem[]{} Laor, A., Bahcall, J. N., Jannuzi, B. T., et al.
             1994, ApJ, 420, 110
\bibitem[]{} Maiolino, R., Juarez, Y., Mujica, R., Nagar, N. M.,
             \& Oliva, E. 2003, ApJ, 596, L155
\bibitem[]{} Marconi, A., \& Hunt, L.~K.\ 2003, \apjl, 589, L21
\bibitem[]{} Marconi, A., Risaliti, G., Gilli, et al. 
             2004, MNRAS, 351, 169
\bibitem[]{} Marziani, P., Sulentic, J. W., Dultzin-Hacyan, D.,
             Calvani, M., \& Moles, M. 1996, ApJS, 104, 37
\bibitem[]{} Matteucci, F., \& Padovani, P. 1993, ApJ, 419, 485
\bibitem[]{} Matteucci, F., \& Tornamb\`{e}, A. 1987, A\&A, 
             185, 51
\bibitem[]{} McIntosh, D. H., Rix, H. -W., Rieke, M. J., \&
             Foltz, C. B. 1999, ApJ, 517, L73
\bibitem[]{} Nagao, T., Maiolino, R., \& Marconi, A. 2005, A\&A,
             submitted (astro-ph/0508652)
\bibitem[]{} Netzer, H., \& Laor, A. 1993, ApJ, 404, L51
\bibitem[]{} Pagel, B. E. J., \& Edmunds, M. G. 1981, ARA\&A, 
             19, 77
\bibitem[]{} Pentericci, L., et al. 2002, AJ, 123, 2151 
\bibitem[]{} Peterson, B. M. 1993, PASP, 105, 1084
\bibitem[]{} Peterson, B. M., \& Wandel, A. 1999, ApJ, 521, L95
\bibitem[]{} Pettini, M., Shapley, A. E., Steidel, C. C., et al. 
             2001, ApJ, 554, 981
\bibitem[]{} Pipino, A., \& Matteucci, F. 2004, MNRAS, 347, 968
\bibitem[]{} Rauch, M. 1998, ARA\&A, 36, 267
\bibitem[]{} Reichard, T. A., Richards, G. T., Hall, P. B., 
             et al. 2003a, AJ, 126, 2594
\bibitem[]{} Reichard, T. A., Richards, G. T., Schneider, D. P., 
             et al. 2003b, AJ, 125, 1711
\bibitem[]{} Richards, G. T., Fan, X., Newberg, H. J., et al. 
             2002a, AJ, 123, 2945
\bibitem[]{} Richards, G. T., Vanden Berk, D. E., Reichard, T. A., 
             et al. 2002b, AJ, 124, 1
\bibitem[]{} Schneider, D. P., Fan, X., Hall, P. B., et al. 
             2003, AJ, 126, 2579
\bibitem[]{} Schneider, D. P., Richards, G. T., Fan, X., et al. 
             2002, AJ, 123, 567
\bibitem[]{} Scott, J. E., Kriss, G. A., Brotherton, M., et al.
             2004, ApJ, 615, 135 
\bibitem[]{} Shemmer, O., Netzer, H., Maiolino, R., et al.
             2004, ApJ, 614, 547
\bibitem[]{} Stoughton, C., Lupton, R. H., Bernardi, M., et al. 
             2002, AJ, 123, 485
\bibitem[]{} Sulentic, J. W., Marziani, P., Zamanov, R., et al.
             2002, ApJ, 566, L71
\bibitem[]{} Telfer, R. C., Zheng, W., Kriss, G. A., \&
             Davidson, A. F. 2002, ApJ, 565, 773
\bibitem[]{} Teplitz, H. I., McLean, I. S., Becklin, E. E., 
             et al. 2000, ApJ, 533, L65
\bibitem[]{} Tremonti, C. A., Heckman, T. M., Kauffmann, G., 
             et al. 2004, ApJ, 613, 898
\bibitem[]{} Vanden Berk, D. E., Richards, G. T., Bauer, A., et al. 
             2001, AJ, 122, 549
\bibitem[]{} Venkatesan, A., Schneider, R., \& Ferrara, A.
             2004, MNRAS, 349, L43
\bibitem[]{} V\'{e}ron-Cetty, M. -P., V\'{e}ron, P., \&
             Tarenghi, M. 1983, A\&A, 119, 69
\bibitem[]{} Vestergaard, M., \& Peterson, B. M. 2005, ApJ, 
             625, 688
\bibitem[]{} Vestergaard, M., \& Wilkes, B. J. 2001, ApJS, 134, 1
\bibitem[]{} Warner, C., Hamann, F., \& Dietrich, M. 2004, ApJ, 
             608, 136
\bibitem[]{} Weymann, R. J., Morris, S. L., Foltz, C. B., \&
             Hewett, P. C. 1991, ApJ, 373, 23
\bibitem[]{} Wilkes, B. J. 1984, MNRAS, 207, 73
\bibitem[]{} Wilkes, B. J. 1986, MNRAS, 218, 331
\bibitem[]{} York, D. G., Adelman, J., Anderson, J. E., et al. 
             2000, AJ, 120, 1579
\bibitem[]{} Zamorani, G., Henry, J. P., Maccacaro, T., 
             et al. 1981, ApJ, 245, 357
\bibitem[]{} Zheng, W., Kriss, G. A., Telfer, R. C., 
             Grimes, J. P., \& Davidson, A. F. 1997, ApJ, 475, 469
\bibitem[]{} Zheng, W., Kriss, G. A., Telfer, R. C., 
             Grimes, J. P., \& Davidson, A. F. 1998, ApJ, 492, 855
\end{thebibliography}
\end{document}